%% file: main.tex
\documentclass[twocolumn,aps,pra,letter,amsmath,amssymb,superscriptaddress]{revtex4-2}
\usepackage{amsmath}
\usepackage{amssymb}
\usepackage{hyperref,mathtools,graphicx}
\usepackage{bm,color,bbm}
\usepackage{amsfonts}    
\usepackage{verbatim}   
\usepackage{subfigure}  
\usepackage{natbib}
\usepackage{booktabs}
\usepackage{threeparttable}
\usepackage{dcolumn}
\usepackage{multirow}
\usepackage{array}
\usepackage{ulem}
\usepackage{diagbox}
\usepackage{braket}
\usepackage{extarrows}
\DeclareMathOperator{\Tr}{Tr}
\DeclareMathOperator{\sinc}{sinc}

\definecolor{maroon}{rgb}{0.7,0,0}

\definecolor{ngreen}{rgb}{0.3,0.7,0.3}

\definecolor{golden}{rgb}{0.8,0.6,0.1}

\graphicspath{ {./figures/} }

\def\Re{\mathop{\text Re}}
\def\Im{\mathop{\text Im}}

\begin{document}
\title{Topological resilience of optical skyrmions in local decoherence}
\author{Li-Wen Wang}
\affiliation{Laboratory of Quantum Information, University of Science and Technology of China, Hefei, 230026, China}
\affiliation{Anhui Province Key Laboratory of Quantum Network, University of Science and Technology of China, Hefei, 230026, China}
\affiliation{Synergetic Innovation Center of Quantum Information and Quantum Physics, University of Science and Technology of China, Hefei, 230026, China}

\author{Sheng Liu}
\email{shengliu@ustc.edu.cn}
\affiliation{Laboratory of Quantum Information, University of Science and Technology of China, Hefei, 230026, China}
\affiliation{Anhui Province Key Laboratory of Quantum Network, University of Science and Technology of China, Hefei, 230026, China}
\affiliation{Synergetic Innovation Center of Quantum Information and Quantum Physics, University of Science and Technology of China, Hefei, 230026, China}

\author{Cheng-Jie Zhang}
\affiliation{School of Physical Science and Technology, Ningbo University, Ningbo, 315211, China}
\affiliation{Hefei National Laboratory, University of Science and Technology of China, Hefei, 230088, China}

\author{Geng Chen}
\affiliation{Laboratory of Quantum Information, University of Science and Technology of China, Hefei, 230026, China}
\affiliation{Anhui Province Key Laboratory of Quantum Network, University of Science and Technology of China, Hefei, 230026, China}
\affiliation{Synergetic Innovation Center of Quantum Information and Quantum Physics, University of Science and Technology of China, Hefei, 230026, China}
\affiliation{Hefei National Laboratory, University of Science and Technology of China, Hefei, 230088, China}

\author{Yong-Sheng Zhang}
\email{yshzhang@ustc.edu.cn}
\affiliation{Laboratory of Quantum Information, University of Science and Technology of China, Hefei, 230026, China}
\affiliation{Anhui Province Key Laboratory of Quantum Network, University of Science and Technology of China, Hefei, 230026, China}
\affiliation{Synergetic Innovation Center of Quantum Information and Quantum Physics, University of Science and Technology of China, Hefei, 230026, China}
\affiliation{Hefei National Laboratory, University of Science and Technology of China, Hefei, 230088, China}

\author{Chuan-Feng Li}
\affiliation{Laboratory of Quantum Information, University of Science and Technology of China, Hefei, 230026, China}
\affiliation{Anhui Province Key Laboratory of Quantum Network, University of Science and Technology of China, Hefei, 230026, China}
\affiliation{Synergetic Innovation Center of Quantum Information and Quantum Physics, University of Science and Technology of China, Hefei, 230026, China}
\affiliation{Hefei National Laboratory, University of Science and Technology of China, Hefei, 230088, China}

\author{Guang-Can Guo}
\affiliation{Laboratory of Quantum Information, University of Science and Technology of China, Hefei, 230026, China}
\affiliation{Anhui Province Key Laboratory of Quantum Network, University of Science and Technology of China, Hefei, 230026, China}
\affiliation{Synergetic Innovation Center of Quantum Information and Quantum Physics, University of Science and Technology of China, Hefei, 230026, China}
\affiliation{Hefei National Laboratory, University of Science and Technology of China, Hefei, 230088, China}
\date{\today}

\begin{abstract}
The topologically protected configuration embedded in skyrmions has prompted some investigations into their fundamental properties and versatile applications, sparking interest and guiding ongoing development. The topological protection associated with skyrmions was initially observed in systems with interactions. It is widely believed that skyrmions are stable yet relevant confirmation and empirical research remain limited. A pertinent question is whether skyrmion configurations formed by a single classical beam with two coupled degrees of freedom also exhibit topological stability. In this study, we affirm this hypothesis by investigating the effects of local decoherence. We analytically and numerically demonstrate the topological resilience of skyrmions and the occurrence of transition points of skyrmion numbers in local decoherence across three typical decoherence channels. On the other hand, we show that these qualities are independent of the initial state. From the numerical results, we find that inhomogeneous but continuous decoherence channels also have the same behaviors and maintain topological stability of skyrmions as homogeneous decoherence channels do. These properties of skyrmions contribute to further applications in various areas, including communication and imaging.
\end{abstract}
\maketitle

\section{Introduction}
In 1960s, Skyrme introduced the concept of skyrmions primarily to describe the structure of nucleons \cite{skyrme1961non,skyrme1962unified}. The skyrmion model proposes that nucleons are topological structures embedded in space, providing a theoretical framework to elucidate their interactions and dynamic behaviors. Due to their topological excitation properties, skyrmions have wide-ranging applications in fields such as condensed matter physics \cite{bogdanov2020physical,wang2010spin}, liquid crystals \cite{nagase2019smectic}, and quantum Hall systems \cite{taguchi2001spin,neubauer2009topological,pfleiderer2010single,ritz2013formation,mochizuki2014thermally}. Not surprisingly, skyrmion textures can emerge in different physical systems with various mechanisms. Extensive studies have been conducted on the theoretical aspects of magnetic skyrmions, and their spin configurations \cite{fert2017magnetic,zhang2020skyrmion} have been experimentally observed in specific material environments \cite{bogdanov1989thermodynamically,bogdanov1994thermodynamically,roessler2006spontaneous,mühlbauer2009skyrmion,yu2010real,garel1982phase,malozemoff2013magnetic,heinze2011spontaneous,romming2013writing}. \par
Given the potential and significant application prospects exhibited by skyrmions, their counterpart in optics is now being developed to explore analogous topological properties \cite{shen2024optical} and it has been first realized by an evanescent electromagnetic field on an interface \cite{tsesses2018optical,du2019deep}. Very recently, optical skyrmionic structures in free space based on vectorial optical fields can be generated and observed by a superposition of two structured light modes \cite{rosales2018review,nape2022revealing,peters2023invariance} with two mutually orthogonal polarizations \cite{gao2020paraxial,gutierrez2021optical,shen2022generation}. Meanwhile, more advanced and sophisticated topological textures are revealed, such as momentum space skyrmions \cite{guo2020meron}, space-time skyrmions with toroidal pulses \cite{zdagkas2022observation} and even 3-dimensional (3D) hopfions \cite{shen2023topological}.\par
Skyrmions have topologically protected configurations characterized by quantized topological invariants, known as skyrmion numbers. It is commonly believed that skyrmions are stable. Owing to their inherent topological stability, skyrmions are generally considered to be robust quasi-particles, making them promising candidates for diverse applications in spintronics, information technology, and other imaging systems. Recent studies have initiated investigations into the robustness of skyrmions against various perturbations {\cite{liu2022disorder,ornelas2024non,wang2024topological,ornelas2025topological}}. For instance, Liu \textit{et al.} examined the stability of optical skyrmions against disorder induced by random fluctuations in amplitude and phase of the field \cite{liu2022disorder}. Similarly, Ornelas \textit{et al.} experimentally demonstrated the topological resilience of optical skyrmions to entanglement decay by amplitude damping operations \cite{ornelas2024non}. However, these studies remain incomplete and further research is essential to address the general case, where amplitude, polarization, and phase perturbations should be considered. And all these impressive advances do not thoroughly explore the robustness of skyrmions in the presence of local decoherence. A comprehensive understanding of skyrmions in local decoherence is crucial and indispensable for their practical implementation in communication, information encoding and imaging applications.\par
Here, we consider three typical decoherence channels and an optical skyrmion field constructed by structured light carrying orbital angular momentum (OAM) with two orthogonal polarizations \cite{gao2020paraxial,zhu2021synthesis} to study different topological properties and stability of skyrmions while the light field propagates through different noisy channels. We explore more general scenarios beyond merely considering pure states and report the topological resilience of skyrmions in local decoherence.

\section{Background}
A 2-dimensional (2D) skyrmion can be regarded as a mapping from the 2D transverse spatial plane onto a Bloch sphere or a Poincar{\'e} sphere with $4\pi$ solid angle \cite{ornelas2024non}. In general, a skyrmion is expressed simply as a state $\ket{\Psi(\mathbf{r})} = \psi_1(\mathbf{r}) \ket{0}+\psi_2(\mathbf{r}) \ket{1}$ \cite{gao2020paraxial,gordon2000pmd}, and $\psi_{1,2}(\mathbf{r})$ are transverse spatial modes, $\ket{0} = \begin{pmatrix}
    1 \\ 
    0
\end{pmatrix}$ and $\ket{1} = \begin{pmatrix}
    0 \\ 
    1
\end{pmatrix}$ represent two mutually orthogonal vectors. And this statement is applicable to two-level systems. In this work, we consider optical skyrmion fields in free space, starting with the classical light field, and take paraxial Laguerre-Gaussian (LG) beams $\psi_{m,l}(\mathbf{r})$ carrying OAM as transverse spatial modes of two orthogonal polarization states. This type of skyrmionic beams can be described by \cite{gao2020paraxial}
\begin{equation}
    \ket{\Psi(\mathbf{r})} = a\psi_{l_1}(\mathbf{r})\ket{0}+b\psi_{l_2}(\mathbf{r})\ket{1},
    \label{skyrmionic state classically}
\end{equation}
where the radial index $m=0$, $|a|^2+|b|^2=1$, and $\ket{0}$ and $\ket{1}$ can be regarded as horizontal polarization $\ket{H}$ and vertical polarization $\ket{V}$ in the optical field. Indeed, even a classical light beam can be described in quantum terms \cite{gordon2000pmd}. In linear optical systems without frequency conversion, identical single photons exhibit equivalence with classical light beam. For convenience and conciseness, the state in Eq.~(\ref{skyrmionic state classically}) can be written in a general form
\begin{equation}
    \ket{\Psi} = a\ket{l_1}\ket{0}+b\ket{l_2}\ket{1},
    \label{skyrmionic state quantum}
\end{equation}involving two degrees of freedom (DOFs) of a classical beam, where $\ket{l} = \int_{\mathcal{R}^2} |\psi_l(\mathbf{r})|e^{il\phi}\ket{\mathbf{r}}d\mathbf{r}$. \par
For a skyrmionic beam, we need to introduce the normalized local Stokes parameters \cite{seki2016skyrmions}, a three-component vector $\mathbf{S(\mathbf{r})} = \langle S_x(\mathbf{r}),S_y(\mathbf{r}),S_z(\mathbf{r}) \rangle (S_x^2+S_y^2+S_z^2=1)$, to define the spin direction of the skyrmionic field. We should note that the normalization of Stokes parameters is necessary in general cases. The three components of spatial Stokes parameters correspond to the expected values of three Pauli operators in the local normalized state, i.e. $S_j(\mathbf{r}) = \langle \sigma_j \rangle (\mathbf{r}) = \frac{I_j^+(\mathbf{r})-I_j^-(\mathbf{r})}{I_j^+(\mathbf{r})+I_j^-(\mathbf{r})}$, where $j = x,y,z$ and $I_j^\pm(\mathbf{r}) = |\langle\lambda_j^\pm|\Psi(\mathbf{r})\rangle|^2$ are the projection intensities of the state $\ket{\Psi(\mathbf{r})}$ on the eigenstates $\ket{\lambda_j^\pm}$ related to the eigenvalues $\lambda_j^\pm = \pm1$ of the Pauli operator $\sigma_j$ ($\sigma_x = \ket{0}\bra{1}+\ket{1}\bra{0}$, $\sigma_y = -i\ket{0}\bra{1}+i\ket{1}\bra{0}$ and $\sigma_z = \ket{0}\bra{0}-\ket{1}\bra{1}$). Here, we consider the density matrix $\rho$ of the overall system with two DOFs in a general form of Eq.~(\ref{skyrmionic state quantum}), i.e., $\rho=\ket{\Psi}\bra{\Psi}$, therefore we calculate the Stokes parameters as $S_j(\mathbf{r}) = \left< \ket{\mathbf{r}}\bra{\mathbf{r}}\otimes\sigma_j\right> = \Tr(\ket{\mathbf{r}}\bra{\mathbf{r}}\otimes\sigma_j\rho)$, where $\mathrm{Tr}(\cdot)$ is the trace operator.\par
To characterize the topological nature of the optical skyrmion field, a topological number, also known as the skyrmion number, is required. The skyrmion number indicates the number of times that the 2D transverse plane $\mathcal{R}^2$ wraps around a unit sphere $\mathcal{S}^2$. The expression of the skyrmion number is 
\begin{equation}
        N_z  = \frac{1}{4\pi}\int_{\mathcal{R}^2} \mathbf{S}\cdot \frac{\partial \mathbf{S}}{\partial x}\times \frac{\partial \mathbf{S}}{\partial y}{\rm d}x{\rm d}y.
\end{equation}

To understand the topological resilience of skyrmions in local decoherence, we employ three typical decoherence channels acting on a single qubit, i.e., phase damping channel, depolarizing channel and amplitude damping channel \cite{preskill1998lecture,cuevas2017experimental}. These channels are non-unitary processes for systems and lead to loss of information and degradation of input state \cite{narang2020comparative}, both of which allow specific representations via Kraus operators. We assume that this channel map is described by the notation $E$ and the output density matrix of this system is \cite{preskill1998lecture}
\begin{equation}
    \rho_{out} = E(\rho) = \sum _{\nu}K_\nu \rho K_\nu^\dagger, 
\end{equation}
where $K$ is the Kraus operator satisfying $\sum_\nu K_\nu^\dagger K_\nu = \mathbb{I}$ and $\rho$ is the input density matrix of the system.\par
This decoherence effect is applied to the polarization DOF, while the spatial DOF remains unaffected (or equivalently, subjected to an identity operation).
When an input state in Eq.~(\ref{skyrmionic state quantum}) passes through a phase damping channel, its output form of the density matrix is 
\begin{equation}
    \rho_{out} = \left(1-\frac{p_1}{2}\right)\rho + \frac{p_1}{2}(\mathbb{I}\otimes \sigma_z)\rho (\mathbb{I}\otimes \sigma_z),
    \label{PDC rho}
\end{equation}
where $p_1 \in [0,1]$. $p_1=0$ denotes that the system remains unaffected. Here, $\mathbb{I}$ is identity operator on the spatial wave function. Phase damping channel can destroy the off-diagonal term of the state while the diagonal term remains unchanged. \par
The depolarizing channel represents maximally unbiased decoherence, with uniform error probabilities giving it unique state-degradation symmetry. The output density matrix is 
\begin{equation}
\begin{split}
    \rho_{out} &= \left(1-p_2\right)\rho + \frac{p_2}{3}[(\mathbb{I}\otimes \sigma_x)\rho (\mathbb{I}\otimes \sigma_x) \\
    &+ (\mathbb{I}\otimes \sigma_y)\rho (\mathbb{I}\otimes \sigma_y) + (\mathbb{I}\otimes \sigma_z)\rho (\mathbb{I}\otimes \sigma_z)].
    \label{DC rho}
\end{split}
\end{equation}
where $p_2\in[0,\frac{3}{4}]$. $p_2=0$ suggests that the error probability is $0$, while $p_2\in (\frac{3}{4},1]$ is non-physical. \par
The Kraus operators of amplitude damping channel \cite{cuevas2017experimental,cuevas2017cut,yu2003qubit} are 
\begin{equation}
    K_0 = \begin{pmatrix}
    1 & 0 \\
    0 & \sqrt{\eta}
\end{pmatrix}, K_1 = \begin{pmatrix}
    0 & \sqrt{1-\eta} \\
    0 & 0
\end{pmatrix}, \eta \in [0, 1].
\end{equation} Here $\eta$ represents the transmission coefficient, showing perfect transmission in $\eta=1$ and complete amplitude damping in $\eta=0$. The diagonal and off-diagonal terms of the state will be both affected by this channel, so it can cause the dissipation and decoherence of the state. The output state is expressed as
\begin{equation}
    \rho_{out} = (\mathbb{I}\otimes K_0)\rho (\mathbb{I}\otimes K_0^\dagger) + (\mathbb{I}\otimes K_1)\rho (\mathbb{I}\otimes K_1^\dagger).
    \label{ADC rho}
\end{equation} 
More details of the above three channels are in Sec.~\uppercase\expandafter{\romannumeral3} of the Supplemental Material \cite{supp}. \par

\section{Results}
We analytically and numerically calculate skyrmion numbers of optical skyrmion fields that propagate in different local decoherence channels. Intriguingly, the interplay between topology and coherence is remarkably subtle, showing the topological resilience of skyrmions. As an optical skyrmion field traverses a decoherence channel, we derive the density matrix and Stokes parameters of the system. By assuming beam propagation in the $z$-direction, we compute the $z$-component of the skyrmion field, thereby reveal the corresponding skyrmion number upon 2D planar integration. It is noteworthy that the Stokes parameters equivalently indicate the direction of the effective magnetization and are solely related to the pointing angle. Therefore, it is essential to normalize them so that the normalized Stokes vector lies on the unit sphere \cite{seki2016skyrmions}.\par
Considering LG spatial modes with radial index $m=0$, for simplicity, we express the complicated formulas $\psi_{l_1}(\mathbf{r})$ and $\psi_{l_2}(\mathbf{r})$ in a concise form
\begin{equation}
    v(\mathbf{r}) = v(r,\phi,z) = \frac{b\psi_{l_2}(\mathbf{r})}{a\psi_{l_1}(\mathbf{r})} = A(z) r^{|l_2|-|l_1|}e^{i\Delta l\phi},
    \label{vr}
\end{equation}
where $A(z)$ is a complex expression containing $z$ and $\Delta l = |l_2-l_1|$ with $l_1 \neq -l_2$. When the optical skyrmion field propagates without noise, the corresponding Stokes parameters are expressed compactly as $(S_x^2+S_y^2+S_z^2=1)$:
\begin{align}
    S_x(\mathbf{r}) &= \frac{v(\mathbf{r})+v^\ast(\mathbf{r})}{1+|v(\mathbf{r})|^2},\\
    S_y(\mathbf{r}) &= -i\frac{v(\mathbf{r})-v^\ast(\mathbf{r})}{1+|v(\mathbf{r})|^2},\\
    S_z(\mathbf{r}) &= \frac{1-|v(\mathbf{r})|^2}{1+|v(\mathbf{r})|^2}.
\end{align}
Thus, the $z$-component of the skyrmion field takes the form
\begin{equation}
    \Sigma_z = \frac{4\Delta l (|l_2|-|l_1|)|v|^2}{r^2(1+|v|^2)^2},
    \label{Sigma_z}
\end{equation}
and yields the skyrmion number $N_z = \Delta l$ when $\Sigma_z$ is integrated over the entire plane \cite{gao2020paraxial}.
When the optical skyrmion field propagates in different local decoherence channels, analogously, we calculate the $z$-components of the three output optical skyrmion fields, i.e.,
\begin{align}
    \Sigma_z^{\text{PDC}} &= \frac{4\Delta l (|l_2|-|l_1|)|v|^2(1-p_1)^2(1+|v|^2)}{r^2[(1+|v|^2)^2-4p_1(2-p_1)|v|^2]^{3/2}}, \\
    \Sigma_z^{\text{DC}} &= \frac{4\Delta l (|l_2|-|l_1|)|v|^2}{r^2(1+|v|^2)^2},\label{Sigma_zDC}  \\
    \Sigma_z^{\text{ADC}} &= \frac{4\Delta l (|l_2|-|l_1|)|v|^2\eta[1+(2\eta-1)|v|^2]}{r^2[(1+|v|^2)^2-4\eta(1-\eta)|v|^4]^{3/2}}.
\end{align}

And we can obtain the analytical solutions of corresponding skyrmion numbers as follows.
\begin{align}
\centering
    &\text{Phase damping channel}:\notag \\
    &N_z^{\text{PDC}} = \frac{1}{4\pi}\iint \Sigma_z^{\text{PDC}} {\rm d}x{\rm d}y
    = \left\{ \begin{matrix}
        \Delta l, & 0\leq p_1 < 1 \\
        0, & p_1 = 1
    \end{matrix} \right.. \label{PDC Nz} 
\end{align}

\begin{align}
\centering
    &\text{Depolarizing channel}: \notag \\
    &N_z^{\text{DC}} = \frac{1}{4\pi}\iint \Sigma_z^{\text{DC}} {\rm d}x{\rm d}y
    = \left\{ \begin{matrix}
        \Delta l, & 0 \leq p_2 <\frac{3}{4} \\
        0, & p_2 = \frac{3}{4}
    \end{matrix} \right.. \label{DC Nz}
\end{align}

\begin{align}
\centering
    &\text{Amplitude damping channel}: \notag \\
    &N_z^{\text{ADC}} = \frac{1}{4\pi}\iint \Sigma_z^{\text{ADC}} {\rm d}x{\rm d}y 
    = \left\{ \begin{matrix}
        \Delta l, & \frac{1}{2}<\eta \leq 1 \\
        0, & 0\leq \eta < \frac{1}{2}
    \end{matrix} \right., \label{ADC Nz}
\end{align}
where $\eta \neq \frac{1}{2}$ in Eq.~(\ref{ADC Nz}). More detailed calculations are in Sec.~\uppercase\expandafter{\romannumeral3} of the Supplemental Material \cite{supp}. \par

\begin{figure}[ht!]
    \centering
    \subfigbottomskip = -5pt
    \subfigcapskip = -1pt    
    \subfigure[Phase damping channel]{
    \begin{minipage}[t]{0.99\linewidth}
        \centering
        \includegraphics[width=2.5in]{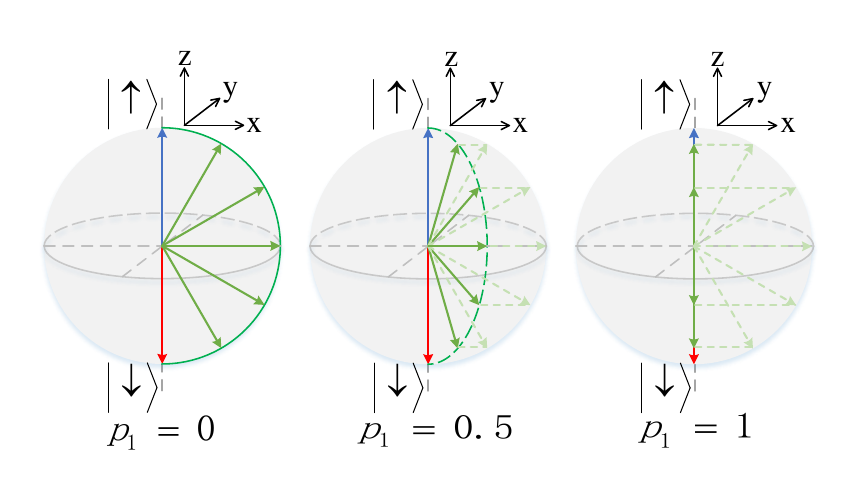}
        \label{fig:Blochsphere_PDC}
    \end{minipage}
    }
    
    \subfigure[Depolarizing channel]{
    \begin{minipage}[t]{0.99\linewidth}
        \centering
        \includegraphics[width=2.5in]{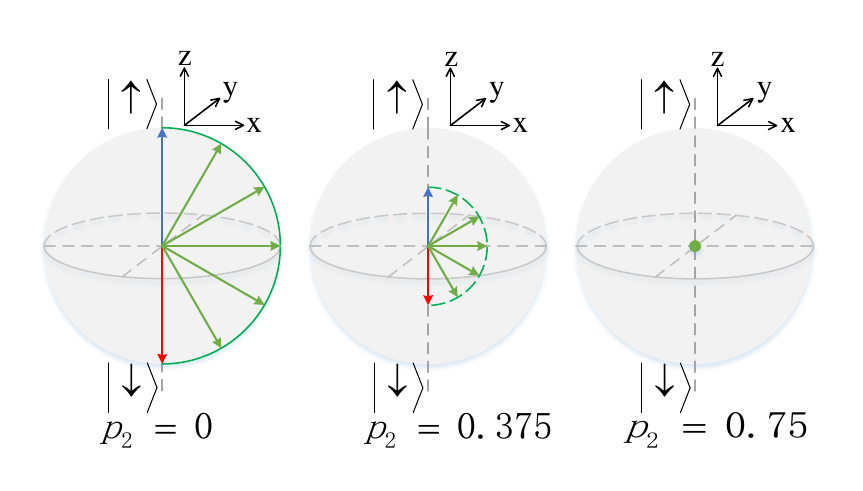}
        \label{fig:Blochsphere_DC}
    \end{minipage}
    }

    \subfigure[Amplitude damping channel]{
    \begin{minipage}[t]{0.99\linewidth}
        \centering
        \includegraphics[width=2.5in]{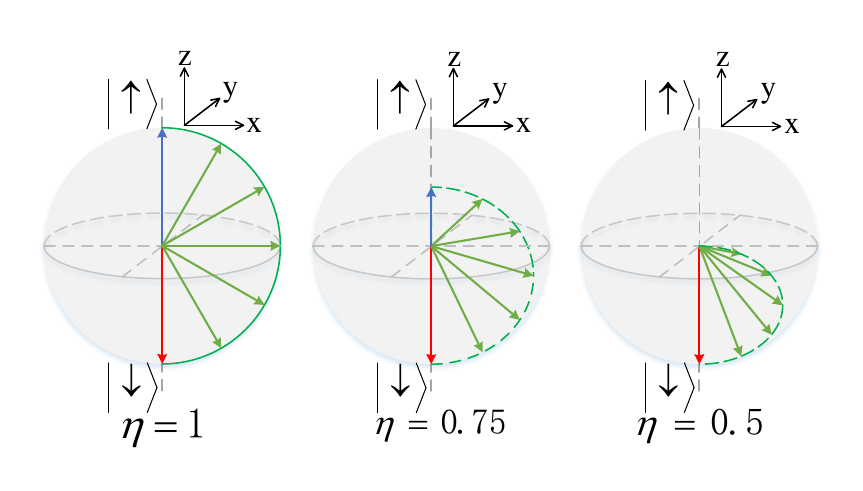}
        \label{fig:Blochsphere_ADC}
    \end{minipage}
    }
    \caption{Changes of the spin vector on the Bloch sphere after passing through the decoherence channel. The normalized Stokes parameters $\mathbf{S}(\mathbf{r}) = \langle S_x(\mathbf{r}),S_y(\mathbf{r}),S_z(\mathbf{r})\rangle = (\sin{\theta}\cos{\phi},\sin{\theta}\sin{\phi},\cos{\theta})$ (polar angle: $\theta \in [0,\pi]$; azimuthal angle: $\phi \in [0,2\pi]$), denote the unit vector on the Bloch sphere. We take a slice ($S_y = 0$ i.e. azimuth $\phi=0$) on the Bloch sphere as an example to observe the changes of spin vectors. \textbf{(a)} Phase damping channel: With the damping factor $p_1$ increases, $S_x$ and $S_y$ gradually decreases, whereas $S_z$ remains unchanged. Once $p_1=1$, all the spin vectors will fall on the central axis and they are maximal mixed states without coherence. \textbf{(b)} Depolarizing channel: As the damping factor $p_2$ decreases, the three components of the Stokes parameters will proportionally contract towards the center of the sphere, eventually converging to a single point. \textbf{(c)} Amplitude damping channel: When $\eta=1$, spin vectors can cover the whole Bloch sphere. When $\eta\in (\frac{1}{2},1)$ (like $\eta=0.75$), $S_x$ decreases and the magnitude of the blue spin vector pointing to the north pole is decreasing. When $\eta=0.5$, the blue spin vector vanishes and there are only spin vectors for the southern hemisphere. }
    \label{fig:Blochsphere}
\end{figure}

In Eq.~(\ref{PDC Nz}), we can see that skyrmion numbers are not affected by phase local decoherence unless $p_1=1$, i.e., the phase information is destroyed totally. In this case, the corresponding unnormalized Stokes parameters are $\mathbf{S'}(\mathbf{r}) = \langle (1-p_1)S_x(\mathbf{r}),(1-p_1)S_y(\mathbf{r}),S_z(\mathbf{r})\rangle$. As the damping factor $p_1$ is close to $1$, the two components $S_x(\mathbf{r})$ and $S_y(\mathbf{r})$ approach $0$ while $S_z(\mathbf{r})$ holds unchanged, producing the effect that the coherence between the two components of the optical skyrmion field gradually disappears and finally the output state is a maximal mixed state. Therefore, topological stability is maintained as long as coherence is preserved, as shown in Fig.~\ref{fig:Blochsphere_PDC}. There is another decoherence noisy channel called bit flip channel, characterized by the $\sigma_x$ operator instead of the $\sigma_z$ operator used in the phase damping channel. For this channel, the unnormalized Stokes parameters transform as $\mathbf{S'}(\mathbf{r}) = \langle S_x(\mathbf{r}),(1-p_1')S_y(\mathbf{r}),(1-p_1')S_z(\mathbf{r})\rangle$, where only $S_x(\mathbf{r})$ remains unchanged. More importantly, the bit flip channel's effect is equivalent to that of a phase damping channel, which acts on the state in the diagonal basis. Thus, we do not discuss the bit flip noise alone in this work, with further details available in Sec.~\uppercase\expandafter{\romannumeral9} of the Supplemental Material \cite{supp}. \par

For depolarizing channel, due to the final Stokes parameters are in proportion to the initial values, i.e. $\mathbf{S'}(\mathbf{r}) = (1-\frac{4p_2}{3})\langle S_x(\mathbf{r}),S_y(\mathbf{r}),S_z(\mathbf{r})\rangle$, the Stokes parameters after being normalized in local decoherence are identical and invariant except for the case of $p_2=\frac{3}{4}$. Therefore, the $z$-component of the skyrmion field in Eq.~(\ref{Sigma_zDC}) is equivalent to that in Eq.~(\ref{Sigma_z}), with the analytical solution in Eq.~(\ref{DC Nz}) demonstrating that the skyrmion number remains constant and is determined by $\Delta l$ (for $p_2\neq \frac{3}{4}$). The illustration is in Fig.~\ref{fig:Blochsphere_DC}. \par

We can see that the emergence of transition points of the first two channels are similar, showing the absence of non-trivial topology of skyrmions when completely decohered. For amplitude damping channel, the skyrmion number vanishes when the system is not fully decohered and $\eta=\frac{1}{2}$ is its transition point. Skyrmion numbers hold invariant as long as the damping factor satisfies $\eta\in (\frac{1}{2},1]$. We explain this phenomenon as follows. The amplitude damping channel demonstrates a process of the decay of a two-level (atom) system due to the spontaneous emission. \{$\ket{1}$, $\ket{0}$\} can be seen as the upper energy level (the excited state) and the lower energy level (the ground state) of this system, and initial populations are $|b|^2$ and $|a|^2$, respectively. There is a probability $1-\eta$ of decaying from the excited state to the ground state so that final populations of excited state and ground state become $\eta|b|^2$ and $|a|^2+(1-\eta)|b|^2$, respectively. As $\eta$ decreases, the population of the upper level gradually decreases. When $\eta\leq \frac{1}{2}$, the inequality $\eta|b|^2\leq|a|^2+(1-\eta)|b|^2$ is always true. It is well-known that skyrmions is a mapping from $\mathcal{R}^2$ to $\mathcal{S}^2$ containing the whole $4\pi$ solid angle, however, it will only map half of the Bloch sphere and topological spin textures will also be destroyed once $\eta\leq \frac{1}{2}$, as shown in Fig.~\ref{fig:Blochsphere_ADC}. We mention $\eta\neq \frac{1}{2}$ in Eq.~(\ref{ADC Nz}) because the full mapping cannot be completed under the condition of this parameter, even when the skyrmion number is equal to $\frac{\Delta l}{2}$ (not reach $0$). \par
\begin{figure}[htbp!]
    \centering
    \vspace{-5pt}
    \subfigbottomskip = -5pt
    \subfigcapskip = -1pt
    \subfigure[Phase damping channel]{
    \begin{minipage}[t]{0.49\linewidth}
        \centering
        \includegraphics[width=1.73in]{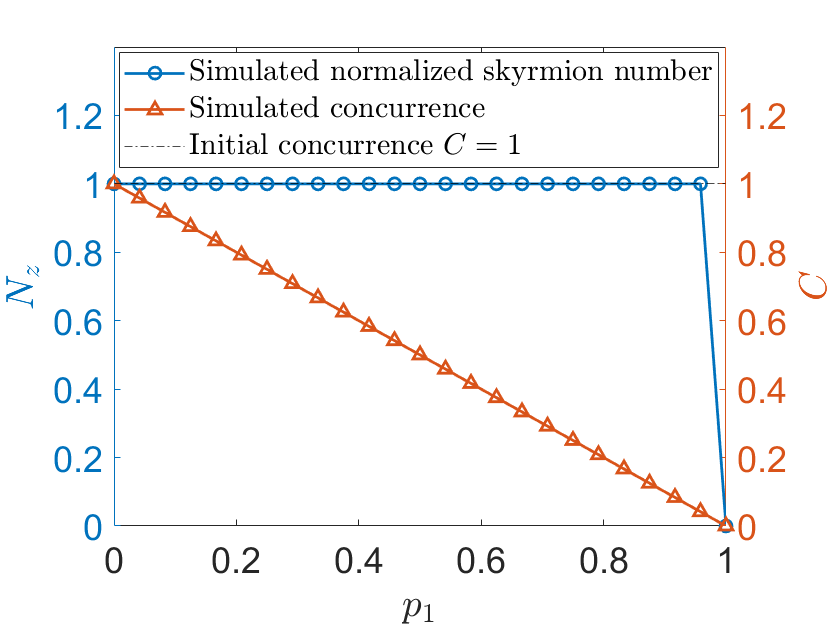}
        \label{figure20_PDC}
    \end{minipage}
    }\subfigure[Depolarizing channel]{
    \begin{minipage}[t]{0.49\linewidth}
        \centering
        \includegraphics[width=1.73in]{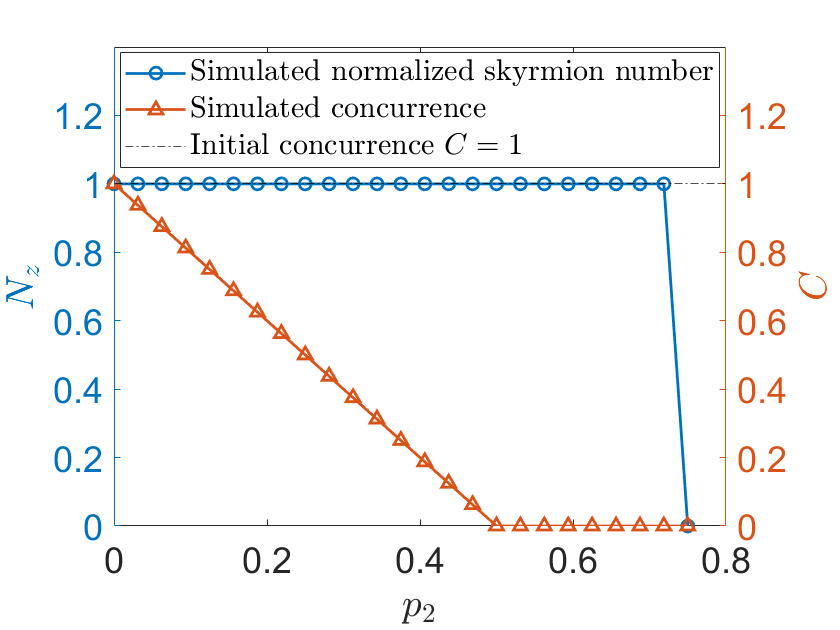}
        \label{figure30_DC}
    \end{minipage}
    }

    \subfigure[Amplitude damping channel]{
    \begin{minipage}[t]{0.49\linewidth}
        \centering
        \includegraphics[width=1.73in]{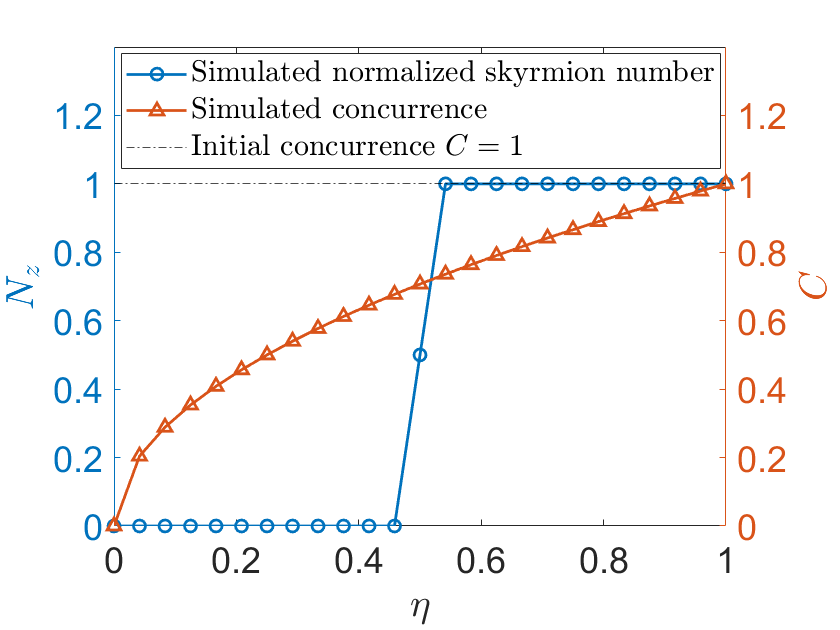}
        \label{figure10_ADC}
    \end{minipage}
    }
    \caption{Normalized skyrmion numbers and relevant concurrence. Blue curves with hollow circles represent simulated normalized skyrmion numbers. Orange curves with hollow triangles represent simulated concurrence values. Black dotted lines are initial concurrence values (here it is $C=1$ with $a=b=\frac{1}{\sqrt{2}}$). \textbf{(a)} Phase damping channel: the damping factor is $p_1 \in [0,1]$. \textbf{(b)} Depolarizing channel: the damping factor is $p_2 \in [0,\frac{3}{4}]$. \textbf{(c)} Amplitude damping channel: the damping factor is $\eta \in [0,1]$. }
    \label{fig:skyrmion number and concurr=1}
\end{figure}

Meanwhile, we numerically demonstrate the results that the normalized skyrmion numbers vary with damping factors in three local decoherence channels in Fig.~\ref{fig:skyrmion number and concurr=1}. Here, we consider that initial coefficients are $a = b = \frac{1}{\sqrt{2}}$ and the azimuthal indices of two OAM modes are $l_1 = 8$ and $l_2=0$. Appropriate truncation of the integration domain is essential in this simulation, as residual low-intensity contributions from peripheral regions—though physically negligible—can distort the gradient of the effective magnetization and destabilize the calculation of the skyrmion number \cite{zhu2021synthesis,wang2025131568}. It is obvious that these numerical results in Fig.~\ref{figure20_PDC},~\ref{figure30_DC} and~\ref{figure10_ADC} (represented by curves with hollow circles in blue) are in good agreement with the analytic solutions in Eq.~(\ref{PDC Nz}),~(\ref{DC Nz}) and~(\ref{ADC Nz}). Additionally, Eq.~(\ref{vr}) shows that the occurrence of these transition points is independent of the initial state. Apart from the case of maximum entanglement (i.e. $a=b=\frac{1}{\sqrt{2}}$), as long as $a\neq0$ and $b\neq0$, all trends hold. And the selection of initial spatial modes also does not affect these properties.\par

Comparing the skyrmion number with the concurrence-quantified entanglement between two DOFs \cite{hill1997entanglement,wootters1998entanglement,yu2004finite}, we find that the skyrmion number does not change synchronously with entanglement during local decoherence. Indeed, this behavior is physically justified because the skyrmion number emerges from the spatially varying polarization and field amplitude but the concurrence is independent of the spatial modes. Entanglement quantified by the concurrence is an important resource and has various applications in information processing and transmission, yet it is not resilient to decoherence noises. We obtain the analytical formulas about concurrence in three channels in the following
\begin{align}
    &C^{\text{PDC}}(\rho_{out}) = 2|a||b|(1-p_1), p_1 \in [0,1],\label{PDC C}\\
    &C^{\text{DC}}(\rho_{out}) = \max\{2|a||b|(1-2p_2),0\}, p_2 \in [0,\frac{3}{4}],\label{DC C} \\
    &C^{\text{ADC}}(\rho_{out}) = 2|a||b|\sqrt{\eta}, \eta\in [0,1].\label{ADC C}
\end{align}
And the numerical curves of concurrence in Fig.~\ref{fig:skyrmion number and concurr=1} are perfectly fitted with theoretical values. Notably, the concurrence exhibits an approximately continuous variation, demonstrating a gradual decay with the increasing damping strength. Moreover, in phase damping channel and amplitude damping channel, the concurrence approaches zero when complete coherence damping occurs, as shown in Fig.~\ref{figure20_PDC} and~\ref{figure10_ADC}. For depolarizing channel in Fig.~\ref{figure30_DC}, however, it behaves very differently. The concurrence goes to zero prematurely, a phenomenon known as entanglement sudden death \cite{yu2006quantum,almeida2007environment,cui2008entanglement,yu2009sudden,meng2020environment,yashodamma2014effectiveness,bavontaweepanya2018effect}, which describes the complete loss of quantum correlation (or entanglement) in a finite time rather than an asymptotic decay over an infinite duration. More details are in Sec.~\uppercase\expandafter{\romannumeral4} and \uppercase\expandafter{\romannumeral6} of the Supplemental Material \cite{supp}. Therefore, as shown in Fig.~\ref{fig:skyrmion number and concurr=1}, due to the presence of transition points of skyrmion numbers, the skyrmion number is more robust than concurrence in the first two cases, establishing its potential as a superior information carrier. In particular, we can see that the skyrmion number changes in a stepwise manner so that it has certain advantages in discrete encoding. \par
Intensity loss of two polarization components is also an aspect to be considered and we can characterize the loss by two loss parameters $T_a\in [0,1]$ and $T_b\in [0,1]$. In fact, the variations of loss parameters $T_a$ and $T_b$ are equivalent to the variations of the coefficients $a$ and $b$ of the initial state. When the two components have the same loss, the state is unchanged after normalization. When the loss parameters are unequal, Eq.~(\ref{skyrmionic state quantum}) is transformed into $\ket{\Psi'} = a'\ket{l_1}\ket{0}+b'\ket{l_2}\ket{1} (|a'|^2+|b'|^2=1)$. As mentioned before, the skyrmion number remains invariant under variations of the system coefficients, provided that the light field retains its vectorial character without collapsing into a scalar field configuration. Consequently, loss parameters alone cannot alter the skyrmion number unless at least one component is entirely lost--similar to the pure state scenario discussed in Ref. \cite{ornelas2024non}. More details are in Sec.~\uppercase\expandafter{\romannumeral7} of the Supplemental Material \cite{supp}. It should be noted that we only consider the effect from propagating in noisy channels and the effect from detection is not taken into account here. 

To analyze the effects of decoherence noises on the skyrmion number, we proceed from the fundamental model and all of the above studies are based on a homogeneous decoherence process, where the damping factors are uniformly distributed in space (i.e. a constant). In some realistic systems, decoherence noise can exhibit spatial variations, implying that different locations experience distinct levels of decoherence effects. The inhomogeneous decoherence noise primarily manifests in photon polarization within inhomogeneous media, exemplified by stress-induced birefringence in fiber and non-uniform refractive index in atmospheric turbulence. Extending this investigation, we find that the topological resilience of an optical skyrmion field propagating through an inhomogeneous yet continuous local decoherence channel still persists in some conditions. We numerically generate an inhomogeneous local decoherence channel, where the damping factor varies randomly across the spatial domain and the correlation of damping factors between any two positions is characterized by a specific correlation length $\epsilon$ \cite{goodman2007speckle}. The correlation length $\epsilon$ contains the possibility that the decoherence effect of any two positions may exist a certain correlated feature. The continuity of the channel is supposed to be guaranteed by increasing the correlation length $\epsilon$ because the skyrmion numbers are unaffected by the operations of smooth deformations \cite{ornelas2024non}. In our simulation, the numerical grid size is $1024\times1024$. Consequently, when $\epsilon=1024$, the inhomogeneous channel deteriorates into a homogeneous one. When $\epsilon=1$, the noise is completely random without correlation. We consider that the initial state is $\ket{\Psi(\mathbf{r})} = \frac{1}{\sqrt{2}}(\ket{l_1=8}\ket{0}+\ket{l_2=0}\ket{1})$ and then passes through the local decoherence channel, whose damping factor distributes in $x-y$ plane. The correlation character of decoherence effects at different positions can be described by several types of correlation functions and we take the usual $\sinc$ function as an example to generate the inhomogeneous noise. We perform 50 realizations for the ensemble average on account of the existence of random process and this averaging operation is conducted on the final results, denoting multiple trials. \par
The results are shown in Tab.~\ref{tab:vary damping}. Similarly to the uniform case, we confirm numerically that skyrmion numbers remain stable and the topology is preserved in these instances: (1) $p_1(\mathbf{r})\in [0,1)$ in phase damping channel; (2) $p_2(\mathbf{r})\in [0,3/4)$ in depolarizing channel; (3) $\eta(\mathbf{r})\in (1/2,1]$ and $\epsilon > 2$ (the continuity of the noisy channel is better) in amplitude damping channel. Therefore, the properties in homogeneous scenarios can also be generalized in inhomogeneous conditions. More details are in Sec.~\uppercase\expandafter{\romannumeral8} of the Supplemental Material \cite{supp}. It should be noted that optical skyrmions are not robust against all perturbations and we only demonstrate their topological structures in these noisy models. In the inhomogeneous noisy model, the probabilities $p_1(\mathbf{r})=1$ and $p_2(\mathbf{r})=3/4$ occur very rarely, thus the skyrmion number remains stable even when $p_1(\mathbf{r})\in [0,1]$ or $p_2(\mathbf{r})\in [0,3/4]$. In contrast, choosing $\eta(\mathbf{r})\in [0,1]$ results in significantly more sites with damping factor in $[0,1/2]$ which destabilizes the skyrmion number. If generating an inhomogeneous noise with $\eta(\mathbf{r})\in [0,1]$ instead of $\eta(\mathbf{r})\in (1/2,1]$, the corresponding skyrmion number will not be maintained. \par

\begin{table}[htbp!]
    \centering
    \setlength{\abovecaptionskip}{0.17cm}
    \setlength{\belowcaptionskip}{-0.46cm}
    \begin{tabular}{c|ccc}
    \hline\hline
    \diagbox[]{$\epsilon$}{$N_z^{\text{sim}}$/$N_z^{\text{the}}$}{Type} & PDC & DC & ADC\\
         \hline
       1 & 0.9986 & 1.0000 & 0.9186 \\
       2 & 0.9999 & 1.0000 & 0.9589 \\
       4 & 1.0000 & 1.0000 & 0.9849 \\
       8 & 1.0000 & 1.0000 & 0.9964 \\
       16 & 1.0000 & 1.0000 & 0.9993 \\
       32 & 1.0000 & 1.0000 & 1.0000 \\
       64 & 1.0000 & 1.0000 & 1.0000 \\
       128 & 0.9998 & 1.0000 & 1.0000 \\
       256 & 0.9995 & 1.0000 & 1.0000 \\
       512 & 0.9966 & 1.0000 & 1.0000 \\
         \hline\hline
    \end{tabular}
    \caption{Normalized skyrmion numbers in the inhomogeneous local decoherence channels with different types and different correlation lengths $\epsilon$ in units of a single grid point. $N_z^{\text{sim}}$ is the simulated skyrmion number and $N_z^{\text{the}}$ is the theoretical skyrmion number. For phase damping channel, the damping factor $p_1(\mathbf{r})\in [0,1)$; for depolarizing channel, the damping factor $p_2(\mathbf{r})\in[0,3/4)$; for amplitude damping channel, the damping factor $\eta(\mathbf{r})\in (1/2,1]$.}
    \label{tab:vary damping}
\end{table}

\section{Discussions}
We have investigated the topological resilience of skyrmions in three local decoherence scenarios. Both analytically and numerically, we find the validity of this topological property. By considering a 2D optical skyrmion field constructed by paraxial LG spatial modes with two orthogonal polarization states, we have found that the two skyrmion numbers $N_z^{\text{PDC}}$ and $N_z^{\text{DC}}$ in phasing noise and depolarizing noise maintain invariant unless their damping factors $p_1$ and $p_2$ reach their respective maximum values indicating completely damping. Moreover, as the decoherence strength increasing, the skyrmion number $N_z^{\text{ADC}}$ remains topologically stable until the factor $\eta$ of amplitude noise is less than or equal to $\frac{1}{2}$, which means the destruction of topological spin texture and the incompleteness of mapping. The immunity against local decoherence endowed by skyrmions' topology is independent of the coefficients $(a,b)$ and spatial modes' azimuthal indices $(l_1,l_2)$ of the initial state. Due to the presence of transition points and stepwise variations in skyrmion numbers under local decoherence, skyrmion numbers exhibit distinct advantages over concurrence for discrete encoding applications. We also demonstrate numerically that these features are still valid for generalized inhomogeneous yet continuous decoherence channels.\par
In this paper, we consider two DOFs of a single classical beam to create the optical skyrmion field. The analysis is not restricted by this condition, and our results can be applicable to common systems (other two-level or two-level ensemble systems), inspiring additional generalization among skyrmions in optics and magnetism. Meanwhile, compared with the pure states and unitary channels in Refs.~\cite{nape2022revealing,ornelas2024non}, we employ more general expressions and non-unitary channels to study the stability of skyrmion numbers. The robustness of skyrmion numbers can be used to guide the transmission of classical discrete signals. It is expected that topological resilience in local decoherence has greater application prospects in more fields, including communication, information encoding and processing, metrology and imaging.\par
In addition, in our work we mainly talk about the stability of an optical skyrmion field in transmission noise. For the influence of the measurement noise  relevant to intensities of optical field to be observed, we have a brief discussion and simulation in Appendix \ref{meas noise}. Indeed, it may weaken the ability of extracting sufficient information and even lead to non-smooth process, thus it is an important issue worthy of further investigations. Moreover, for the effect of spatial resolution on the skyrmion number, Appendix \ref{sec resolution} shows that spatial resolution acts as a fundamental limiting factor governing the observable topological properties.

\textit{Note added.}---We notice that there is a similar work \cite{koch2024quantum} when our paper is in preparation. Our work differs from the work in Ref \cite{koch2024quantum}: we consider a state from a classical beam with two coupled degrees of freedom (polarization and spatial modes) rather than entangled bi-photon states. Our findings are applicable to classical optical fields and can be generalized to other physical systems, as long as the noise in those systems is localized. \par

\section*{Acknowledgments}
This work was supported by the National Natural Science Foundation of China (No.~92065113), Innovation Program for Quantum Science and Technology (No.~2021ZD0301201) and the University Synergy Innovation Program of Anhui Province (No.~GXXT-2022-039). \par

\appendix
\renewcommand{\thesection}{\Alph{section}}
\numberwithin{figure}{section}  
\numberwithin{table}{section} 
\numberwithin{equation}{section}

\section{Discussions about the measurement noise}
\label{meas noise}
In this work, we primarily focus on these investigations about the topological resilience of optical skyrmion fields under different local decoherence noises. However, the measurement noise remains an issue worthy of further consideration and exists not only in optical skyrmion fields propagating through local decoherence channels but also in unperturbed skyrmion fields. When the decoherence noise is stronger, more factors should be taken into account for the process of a practical detection. Here, we proceed a brief discussion about the measurement noise. In general, we need to detect full Stokes parameters to calculate the corresponding skyrmion number of an optical skyrmion field. As the previous section mentioned, the three components of spatial Stokes parameters correspond to the expected values of three Pauli operators in the local normalized state, thus the intensity distributions in six polarization bases ($\{\ket{D}=(\ket{0}+\ket{1})/\sqrt{2},\ket{A}=(\ket{0}-\ket{1})/\sqrt{2},\ket{R}=(\ket{0}+i\ket{1})/\sqrt{2},\ket{L}=(\ket{0}-i\ket{1})/\sqrt{2},\ket{H}=\ket{0},\ket{V}=\ket{1}\}$) are required in experiment. We consider a noise model of the detector \cite{xu2020approaching}. In this model, the main noise sources are Poisson-distributed quantum noise, dark noise ($K_d\sim \mathcal{N}(\mu_d,\sigma_d^2)$), and an extra classical noise ($K_a\sim \mathcal{N}(0,\sigma_a^2)$). Assuming the total average number of photons of the skyrmionic beam is $\overline{n}_t$ per exposure and the number of photons at the $j$th pixel of the detector is $\overline{n}_j$, the photon-electrons received at the $j$th pixel is $N_j$, following the Poisson distribution $P(N_j|\eta \overline{n}_j)$ ($\eta$ is the detection efficiency). Due to the existence of detector noise, the readout $k_j$ at the $j$th pixel includes not only $N_j$, but also $K_d$ and $K_a$. Therefore, the readout of intensity distribution in each polarization basis should include these measurement noises. Some parameters related to the detection are referred to in Ref. \cite{xu2020approaching} ($\eta=0.125, \mu_d=511.38, \sigma_d=7.05, \ln{(\sigma_a^2)}=1.19\ln{(\overline{n}_j)}-4.39$). It is noted that we do not consider the saturated threshold of the detector because of the choice of single-photon working mode of detector, which only aims to demonstrate the effect of the measurement noise.\par
\begin{figure}[htbp!]
    \centering
    \vspace{-5pt}
    \subfigbottomskip = -5pt
    \subfigcapskip = -1pt
    \subfigure[Phase damping channel]{
    \begin{minipage}[t]{0.99\linewidth}
        \centering
        \includegraphics[width=2.5in]{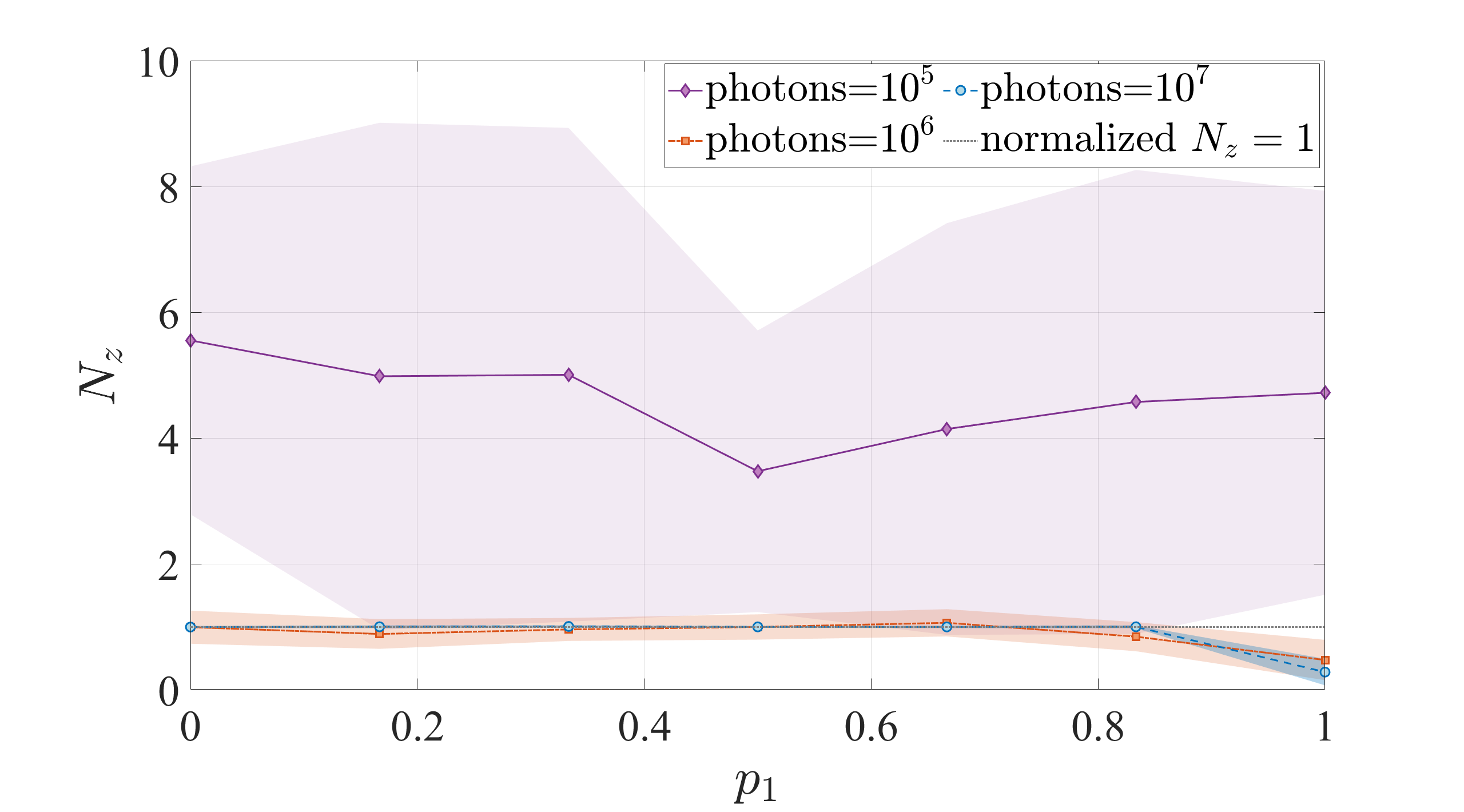}
        \label{PDC_1024_photons}
    \end{minipage}
    }
    
    \subfigure[Depolarizing channel]{
    \begin{minipage}[t]{0.99\linewidth}
        \centering
        \includegraphics[width=2.5in]{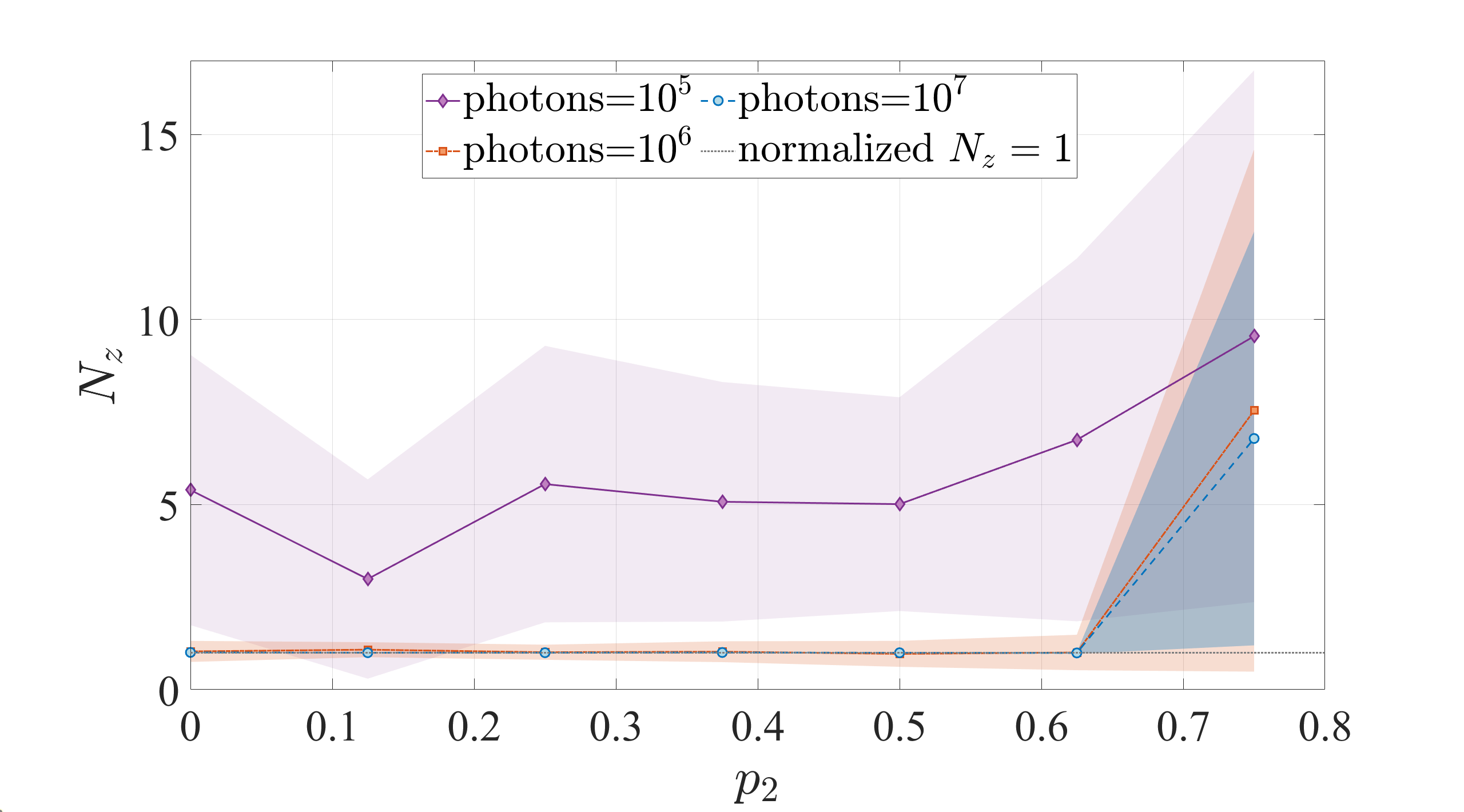}
        \label{DC_1024_photons}
    \end{minipage}
    }

    \subfigure[Amplitude damping channel]{
    \begin{minipage}[t]{0.99\linewidth}
        \centering
        \includegraphics[width=2.5in]{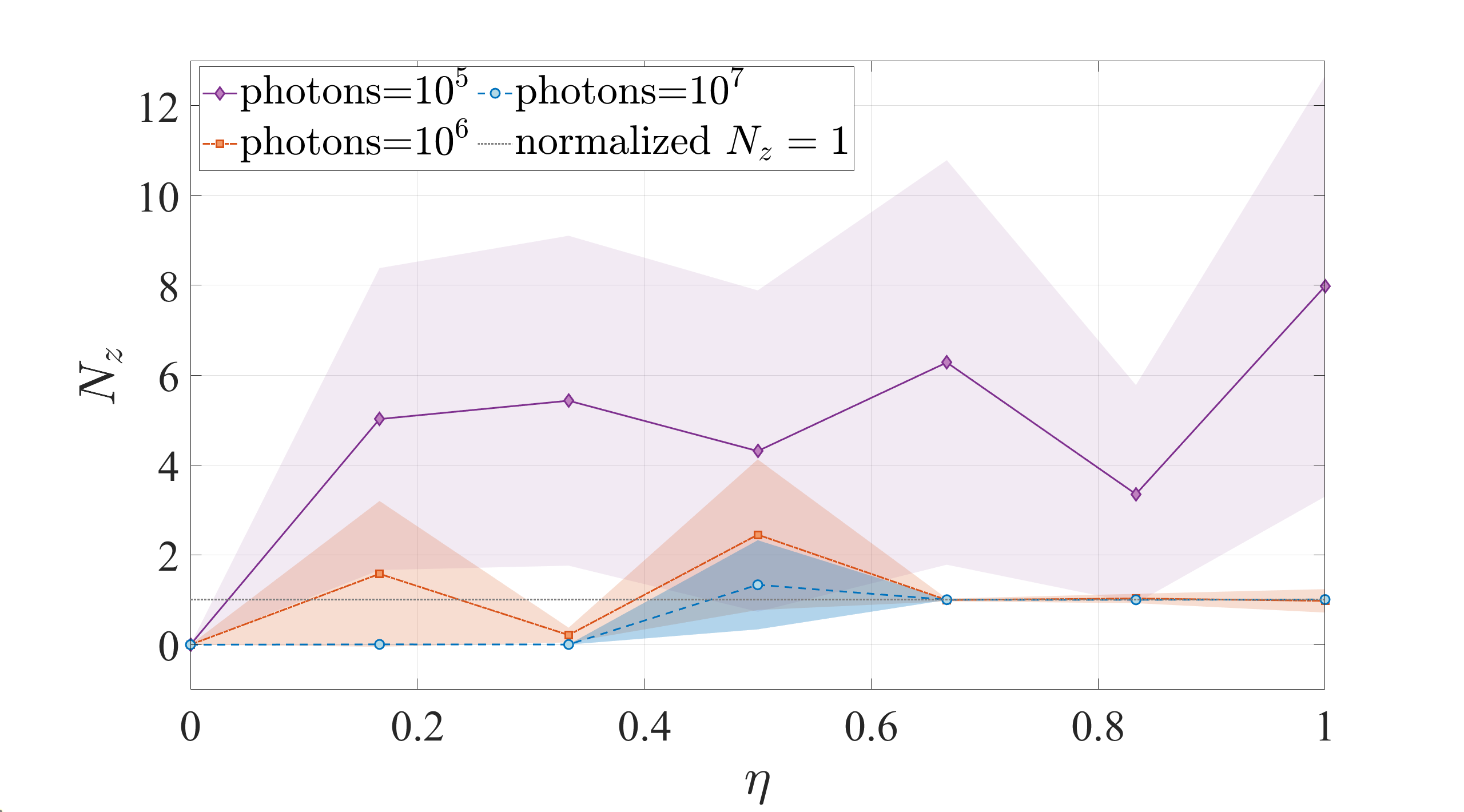}
        \label{ADC_1024_photons}
    \end{minipage}
    }
    \caption{Normalized skyrmion numbers in three local decoherence channels with varying total photon numbers ($\overline{n}_t= \{10^5, 10^6, 10^7\}$). The gray dot-dash line represents the value of $N_z=1$. The shadow part indicates the variance of the skyrmion number in 20 realizations.}
    \label{photons}
\end{figure}

When the number of incident photons is sufficiently large, the influence of measurement noise on the experimental outcomes can be reduced. We examine varying total average photon counts, denoted by different values of $\overline{n}_t$, and analyze how measurement noise affects final changing trends of skyrmion numbers in these three local decoherence channels (here $l_1=8,l_2=0$). Since the Poisson statistics governing photoelectron distribution at each pixel, we perform 20 realizations and average the final results. Relevant results are in Fig.~\ref{photons}. For the phase damping channel, the normalized skyrmion number $N_z$ exhibits a certain stability depending on the total average photon number $\overline{n}_t$. When $p_1 \neq 1$, $N_z$ remains robust for $\overline{n}_t=1\times 10^7$, whereas a slight deviation occurs when $\overline{n}_t=1\times 10^6$. In the case of the depolarizing channel, provided $p_2\neq 3/4$, $N_z$ closely aligns with theoretical value $1$ for both $\overline{n}_t=\{1\times 10^7, 1\times 10^6\}$. For the amplitude damping channel, when $\overline{n}_t=\{1\times 10^7, 1\times 10^6\}$, the skyrmion number remains stable within the damping factor range $1/2 < \eta \leq 1$. However, near the critical point $\eta=1/2$ where $N_z$ should originally be equal to $1/2$ (and in the range of $0\leq\eta <1/2$ where $N_z$ should be $0$), $N_z$ becomes highly sensitive to measurement noise -- a behavior similarly observed in the first two channels ($p_1=1$ and $p_2=3/4$ where $N_z$ both should be $0$). Notably, when $\overline{n}_t=1\times 10^5$, the skyrmion numbers in all three channels deteriorates completely. These results demonstrate that lower photon numbers exacerbate the detrimental influence of the detector noise on $N_z$, leading to greater fluctuations and larger variances. It is reasonable. With adequate photon numbers, the skyrmion number maintains notable resilience against local decoherence in noisy channels.

\section{Discussions about the spatial resolution}
\label{sec resolution}

\begin{figure}[htbp!]
    \centering
    \vspace{-5pt}
    \subfigbottomskip = -4pt
    \subfigcapskip = 0pt
    \subfigure[Phase damping channel]{
    \begin{minipage}[t]{0.99\linewidth}
        \centering
        \includegraphics[width=2.7in]{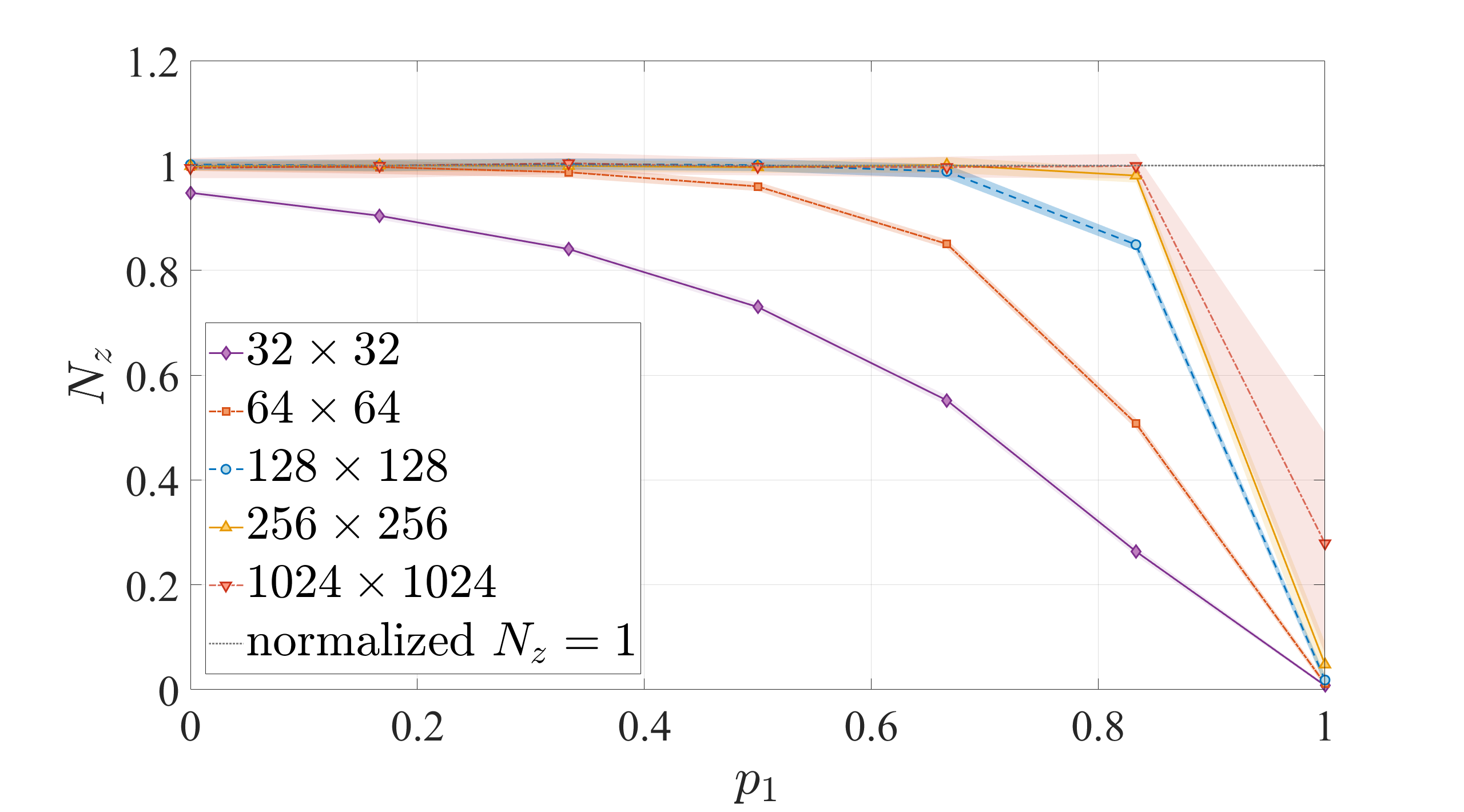}
        \label{PDC_resolution}
    \end{minipage}
    }
    
    \subfigure[Depolarizing channel]{
    \begin{minipage}[t]{0.99\linewidth}
        \centering
        \includegraphics[width=2.7in]{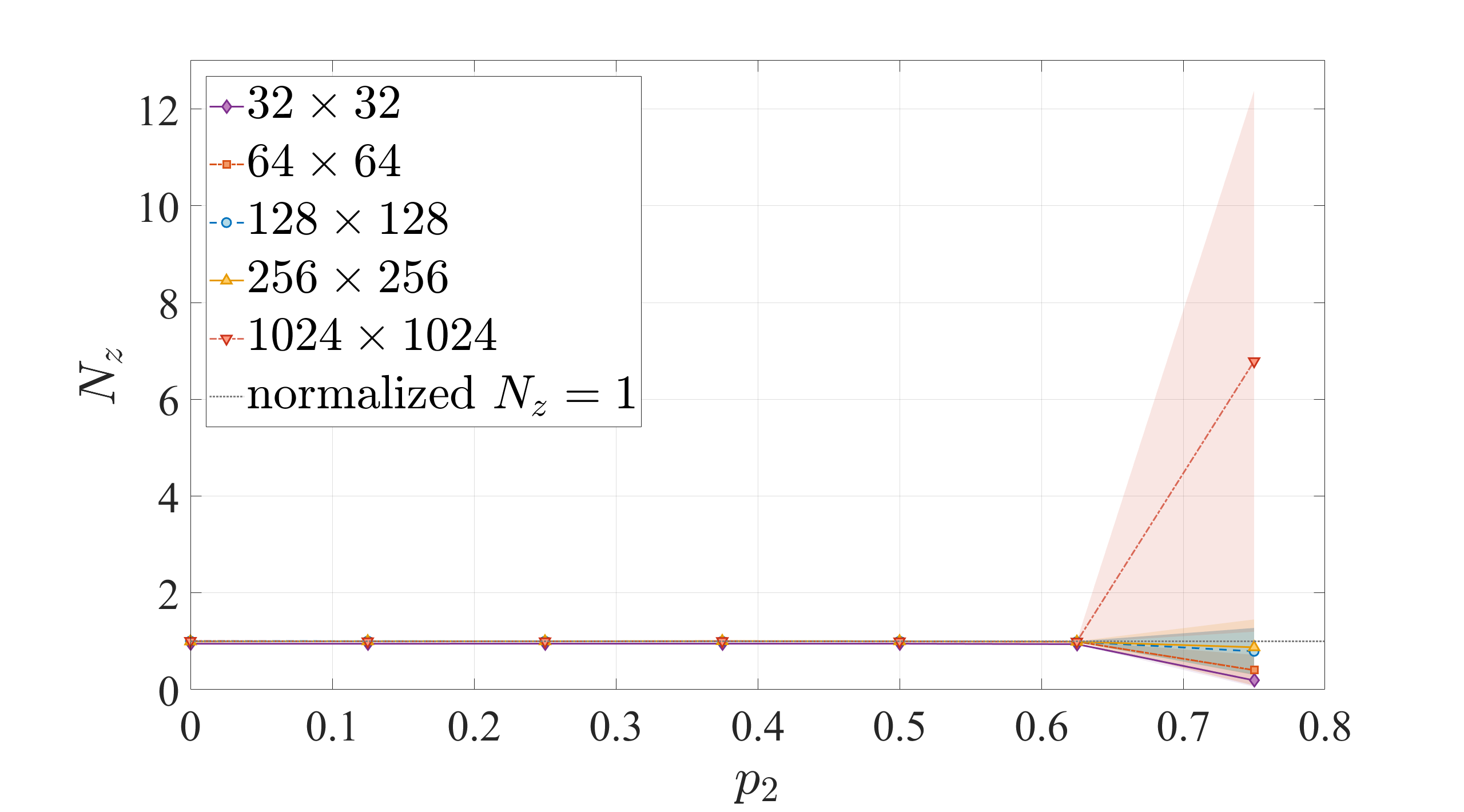}
        \label{DC_resolution}
    \end{minipage}
    }

    \subfigure[Amplitude damping channel]{
    \begin{minipage}[t]{0.99\linewidth}
        \centering
        \includegraphics[width=2.7in]{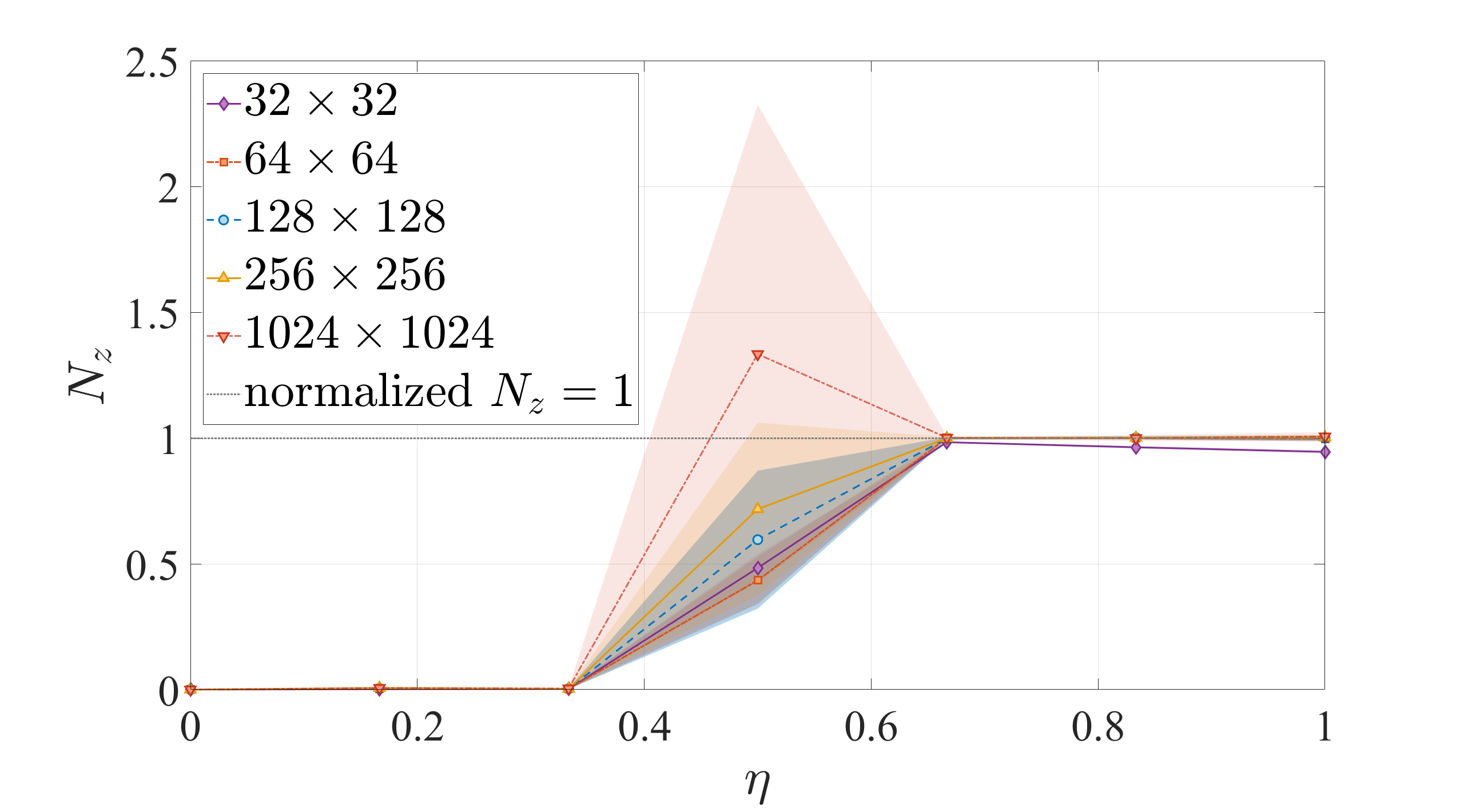}
        \label{ADC_resolution}
    \end{minipage}
    }
    \caption{Normalized skyrmion numbers in three local decoherence channels with varying grid sizes ($\{32\times 32,64\times 64, 128\times 128, 256\times 256, 1024\times 1024\}$). The gray dot-dash line represents the value of $N_z=1$. The shadow part indicates the variance of the skyrmion number in 20 realizations.}
    \label{resolution}
\end{figure}

We further discuss the spatial resolution considerations for the detection of the skyrmion number. In actual measurements or theoretical simulations, the spatial resolution is a crucial factor and inadequate spatial resolution adversely affects the precision of the obtained results. In our work, we numerically study the skyrmion number under noise conditions using a grid size of $1024\times 1024$ with pixel size $\Delta x = \Delta y = 0.0125$mm and an input Gaussian beam's radius of $w_0=1.5$mm. The spatial resolution is related to the pixel size, and the smaller $\Delta x$ $(\Delta y)$ means the higher resolution. On the other hand, selection of enough spatial resolution also depends on the beam spot size (related to $w_0$), because we need to have a smaller pixel size for a smaller beam spot size. Thus, we can define a relative coefficient $\delta_R = \frac{w_0}{\Delta x}$ (here $\Delta x = \Delta y$). When we reduce $\Delta x$ with the same $w_0$, the spatial resolution will increase. The initial coefficient is $\delta_R = \frac{1.5\text{mm}}{0.0125\text{mm}}=120$. When fixing the screen size, we change the grid size to obtain different pixel sizes (or different spatial resolutions). As shown in Fig.~\ref{resolution}, we study the effect of spatial resolutions with five grid sizes ($\{32\times 32,64\times 64, 128\times 128, 256\times 256, 1024\times 1024\}$) corresponding to five relative coefficients ($\delta_R = 3.75, 7.5, 15, 30, 120$) and we also take the measurement noise into account. In the phase damping channel, the skyrmion number $N_z$ gradually loses its characteristic stepwise behavior as the grid size decreases. In contrast, both the depolarizing and amplitude damping channels preserve this stepwise feature, although their transition points remain particularly sensitive to noise. Furthermore, in Fig.~B.\ref{DC_resolution} and Fig.~B.\ref{ADC_resolution}, reduced spatial resolution leads to minor deviations in $N_z$ when the grid size is $32\times 32$.\par
In our work, the skyrmion number is normalized by its theoretical maximum value of 8, corresponding to the initial pure state ($l_1=8,l_2=0$). This normalization quantifies the relative decay under local decoherence, rendering the results independent of the initial spatial modes and facilitating a more general interpretation of the stability.


\normalem
\bibliography{main.bbl}

\clearpage
\appendix
\setcounter{table}{0}
\setcounter{figure}{0}
\setcounter{section}{0}
\setcounter{equation}{0}
\renewcommand{\thetable}{S\arabic{table}}
\renewcommand{\thefigure}{S\arabic{figure}}
\renewcommand{\theequation}{S\arabic{equation}}
\renewcommand{\thesection}{\Roman{section}}
\renewcommand{\appendixname}{} 

\begin{widetext}
    \section*{Supplemental Materials} 
    \input{supplemental.tex}
\end{widetext}


\end{document}

%% file: supplemental.tex
\section*{\expandafter{\romannumeral1}. Skyrmion numbers $N_z$}
Skyrmions are one kind of topological spin textures characterized by corresponding topological numbers, i.e. skyrmion numbers. For the electron spin, the normalized local magnetization $\mathbf{m}(\mathbf{r})$ defines the relevant skyrmion field. For the light beam, a normalized Stokes vector is introduced to calculate the skyrmion number. An optical skyrmion can be regarded as a mapping from a transverse spatial plane $\mathcal{R}^2$ to a Poincar{\'e} sphere (or a Bloch sphere in spin) $\mathcal{S}^2$, which has a two-dimensional (2D) surface of a three-deimensional (3D) ball and completely covers a $4\pi$ solid angle, and the skyrmion number denotes the number of times of $\mathcal{R}^2$ wrapping around $\mathcal{S}^2$. The $i$th component of the skyrmion field can be expressed by \cite{gao2020paraxial}
\begin{equation}
    \Sigma_i = \frac{1}{2}\epsilon_{ijk}\epsilon_{pqr}S_p\frac{\partial S_q}{\partial x_j}\frac{\partial S_r}{\partial x_k},
\end{equation}
where the notations $\epsilon_{ijk}$ and $\epsilon_{pqr}$ are both Levi-civita symbols and the subscript $i=x,y,z$. Nevertheless, we only need to exploit the $z$th component $\Sigma_z$
\begin{align}
     \Sigma_z(x,y) =& \frac{1}{2}\epsilon_{pqr}S_p\frac{\partial S_q}{\partial x}\frac{\partial S_r}{\partial y}
     - \frac{1}{2}\epsilon_{pqr}S_p\frac{\partial S_q}{\partial y}\frac{\partial S_r}{\partial x}
     \notag
     \\ =& S_x(\frac{\partial S_y}{\partial x} \frac{\partial S_z}{\partial y} - \frac{\partial S_z}{\partial x} \frac{\partial S_y}{\partial y}) 
     \notag
     \\ &+S_y(\frac{\partial S_z}{\partial x} \frac{\partial S_x}{\partial y} - \frac{\partial S_x}{\partial x} \frac{\partial S_z}{\partial y})
     \notag
     \\ &+S_z(\frac{\partial S_x}{\partial x} \frac{\partial S_y}{\partial y} - \frac{\partial S_y}{\partial x} \frac{\partial S_x}{\partial y})
     \notag
     \\ =& \mathbf{S}\cdot \left( \frac{\partial \mathbf{S}}{\partial x}\times \frac{\partial \mathbf{S}}{\partial y} \right), \label{SigmaZ}
\end{align}
where $S_i$ is spatially distributed, i.e. $S_i(x,y)$. Therefore, the skyrmion number is defined as \cite{gao2020paraxial}
\begin{equation}
    N_z = \frac{1}{4\pi}\iint \Sigma_z(x,y) {\rm d}x{\rm d}y.
    \label{skyrme number}
\end{equation}

\section*{\expandafter{\romannumeral2}. Stokes parameters $\mathbf{S}(\mathbf{r})$}
To calculate skyrmion numbers, we need to obtain relevant Stokes parameters and this operation is also called Stokes measurements. The input state of this system is written in the general form $\ket{\Psi} = a\ket{l_1}\ket{0}+b\ket{l_2}\ket{1}$, where $\ket{l} = \int_{\mathcal{R}^2} |\psi_l(\mathbf{r})|e^{il\phi}\ket{\mathbf{r}}d\mathbf{r}$ and \{$\ket{0},\ket{1}$\} are two mutually orthogonal states. Here, we consider paraxial Laguerre-Gaussian (LG) modes with no radial index ($m=0$) and the $\psi_{m,l}(\mathbf{r})$ is described as
\begin{align}
    \psi_{m,l}(r,\phi,z) = &\frac{C_{ml}}{w(z)}\left [\frac{\sqrt{2}r}{w(z)} \right]^{\mid l \mid}L_m^{\mid l \mid }\left[\frac{2r^2}{w^2(z)}\right]\exp\left[-\frac{r^2}{w^2(z)}\right] \exp(il\phi)
    \notag
    \\& \times \exp\left[\frac{-ikr^2z}{2(z^2+z_R^2)}\right]\exp\left[i(2m+\mid l \mid +1)\arctan\left(\frac{z}{z_R}\right)\right],
    \label{LGmode}
\end{align}
where $C_{ml} = \sqrt{\frac{2m!}{\pi(m+\mid l \mid)!}}$  is a normalization constant, $L_m^{\mid l \mid}(x)$ is the generalized Laguerre polynomial, $z_R = \frac{\pi w_0^2}{\lambda}$ is the Rayleigh distance, $\lambda$ represents the beam wavelength, $w_0$ is the beam waist of the fundamental mode, and $w(z) = w_0 \sqrt{1+\left(\frac{z}{z_R}\right)^2}$ is the beam radius at the $z$ plane.\par
The initial density matrix is 
\begin{align}
    \rho &= \ket{\Psi}\bra{\Psi}\notag \\
    &= |a|^2\ket{l_1}\bra{l_1}\otimes\ket{0}\bra{0}+ab^\ast\ket{l_1}\bra{l_2}\otimes\ket{0}\bra{1} \notag \\
    &+ a^\ast b \ket{l_2}\bra{l_1}\otimes\ket{1}\bra{0}+|b|^2\ket{l_2}\bra{l_2}\otimes\ket{1}\bra{1} \notag\\
    &= \sum _{pqst=1}^2 \mu_{pqst}\ket{l_p}\ket{l_q}\otimes\ket{e_s}\bra{e_t},
    \label{intial rho}
\end{align}
where $\ket{e_1}=\ket{0}$, $\ket{e_2}=\ket{1}$, $\mu_{1111} = |a|^2$, $\mu_{1212}=ab^\ast$, $\mu_{2121}=a^\ast b$, $\mu_{2222}=|b|^2$, and other coefficients are zero. The Stokes parameters correspond to the expected values of measurement operators of the overall system, i.e. 
\begin{equation}
    S_{j=x,y,z}(\mathbf{r}) = \left< \ket{\mathbf{r}}\bra{\mathbf{r}}\otimes\sigma_j\right> = \Tr(\ket{\mathbf{r}}\bra{\mathbf{r}}\otimes\sigma_j\rho),
    \label{Stokes para1}
\end{equation}
where $\Tr(\cdot)$ is the trace operator, the position basis of orbital angular momentum (OAM) degree of freedom (DOF) $\{\ket{\mathbf{r}}, \mathbf{r}\in \mathcal{R}^2\}$ satisfies the orthogonality $\langle\mathbf{r}_1 | \mathbf{r}_2\rangle = \delta (\mathbf{r}_1-\mathbf{r}_2)$ and the completeness relation $\int _{\mathcal{R}^2} \ket{\mathbf{r}}\bra{\mathbf{r}}d\mathbf{r}=\mathbb{I}$. By employing Eq. (\ref{intial rho}) into Eq. (\ref{Stokes para1}), we can get 
\begin{align}
    S_{j=x,y,z}(\mathbf{r}) &= \sum _{pqst=1}^2 \mu_{pqst} \Tr(\ket{\mathbf{r}}\langle \mathbf{r} | l_p\rangle \bra{l_q})\Tr(\sigma_j\ket{e_s}\bra{e_t}) \notag \\
    &= \sum_{pqst=1}^2 \mu_{pqst} \langle\mathbf{r}|l_p\rangle \langle l_q|\mathbf{r}\rangle \Tr(\sigma_j\ket{e_s}\bra{e_t}) \notag \\ 
    &= \sum _{pqst=1}^2 \mu_{pqst} \psi_{l_p}(\mathbf{r})\psi_{l_q}^\ast (\mathbf{r}) \Tr(\sigma_j\ket{e_s}\bra{e_t}),
    \label{Stokes para2}
\end{align}
where $\langle \mathbf{r}|l\rangle = \psi_l(\mathbf{r})$.
For the term $\Tr(\sigma_j\ket{e_s}\bra{e_t})$, we can rewrite it by the spectral decomposition of Pauli matrices, i.e. $\sigma_j = \lambda_j^+P_j^++\lambda_j^-P_j^-$. So Eq. (\ref{Stokes para2}) can be expressed by
\begin{align}
    S_{j=x,y,z}(\mathbf{r}) =& \lambda_j^+\sum_{pqst=1}^2\mu_{pqst}\psi_{l_p}(\mathbf{r})\psi_{l_q}^\ast(\mathbf{r})\langle e_t|P_j^+ |e_s\rangle \notag \\
    &+ \lambda_j^-\sum_{pqst=1}^2\mu_{pqst}\psi_{l_p}(\mathbf{r})\psi_{l_q}^\ast(\mathbf{r})\langle e_t|P_j^- |e_s\rangle,
    \label{Stokes prara3}
\end{align}
where $\lambda_j^{\pm}=\pm 1$ and $P_j^{\pm}$ are the projection operators. Therefore, we can calculate three locally normalized Stokes components of the initial density matrix in detail.\par
\begin{align}
    \Tilde{S}_x(\mathbf{r}) &= \frac{I_D(\mathbf{r}) - I_A(\mathbf{r})}{I_D(\mathbf{r}) + I_A(\mathbf{r})},\label{Sx1}\\
    \Tilde{S}_y(\mathbf{r}) &= \frac{I_R(\mathbf{r}) - I_L(\mathbf{r})}{I_R(\mathbf{r}) + I_L(\mathbf{r})},\label{Sy1}\\
    \Tilde{S}_z(\mathbf{r}) &= \frac{I_H(\mathbf{r}) - I_V(\mathbf{r})}{I_H(\mathbf{r}) + I_V(\mathbf{r})},\label{Sz1}
\end{align}
where $I_k(\mathbf{r}) = I|\langle k| \langle \mathbf{r}| \Psi\rangle|^2 = I\left[\Tr(\ket{\mathbf{r}}\bra{\mathbf{r}}\otimes \ket{k}\bra{k}\rho)\right] (k=D,A,R,L,H,V)$, and $I$ is the total intensity in the skyrmionic beam. In fact, $I_k(\mathbf{r})$ is the first or second term of Eq. (\ref{Stokes prara3}) after the eigenvalues are removed. These states $\{\ket{D},\ket{A}\},\{\ket{R},\ket{L}\},\{\ket{H},\ket{V}\}$ are eigenvectors of three Pauli operators $\sigma_x$, $\sigma_y$ and $\sigma_z$, respectively.
($\ket{D}=1/\sqrt{2}(\ket{0}+\ket{1})$, $\ket{A}=1/\sqrt{2}(\ket{0}-\ket{1})$, $\ket{R}=1/\sqrt{2}(\ket{0}+i\ket{1})$, $\ket{L}=1/\sqrt{2}(\ket{0}-i\ket{1})$, $\ket{H}=\ket{0}$, $\ket{V}=\ket{1}$.)
Each $I_k(\mathbf{r})$ is 
\begin{align}
    I_D(\mathbf{r}) =& (1/2)\mu_{1111}\psi_{l_1}(\mathbf{r})\psi_{l_1}^\ast(\mathbf{r})+(1/2)\mu_{1212}\psi_{l_1}(\mathbf{r})\psi_{l_2}^\ast(\mathbf{r}) \notag \\ 
    &+(1/2)\mu_{2121}\psi_{l_2}(\mathbf{r})\psi_{l_1}^\ast(\mathbf{r})+(1/2)\mu_{2222}\psi_{l_2}(\mathbf{r})\psi_{l_2}^\ast(\mathbf{r}),\label{ID}
\end{align}

\begin{align}
    I_A(\mathbf{r}) =& (1/2)\mu_{1111}\psi_{l_1}(\mathbf{r})\psi_{l_1}^\ast(\mathbf{r})+(-1/2)\mu_{1212}\psi_{l_1}(\mathbf{r})\psi_{l_2}^\ast(\mathbf{r}) \notag \\ 
    &+(-1/2)\mu_{2121}\psi_{l_2}(\mathbf{r})\psi_{l_1}^\ast(\mathbf{r})+(1/2)\mu_{2222}\psi_{l_2}(\mathbf{r})\psi_{l_2}^\ast(\mathbf{r}),\label{IA}
\end{align}

\begin{align}
    I_R(\mathbf{r}) =& (1/2)\mu_{1111}\psi_{l_1}(\mathbf{r})\psi_{l_1}^\ast(\mathbf{r})+(i1/2)\mu_{1212}\psi_{l_1}(\mathbf{r})\psi_{l_2}^\ast(\mathbf{r}) \notag \\ 
    &+(-i1/2)\mu_{2121}\psi_{l_2}(\mathbf{r})\psi_{l_1}^\ast(\mathbf{r})+(1/2)\mu_{2222}\psi_{l_2}(\mathbf{r})\psi_{l_2}^\ast(\mathbf{r}),\label{IR}
\end{align}

\begin{align}
    I_L(\mathbf{r}) =& (1/2)\mu_{1111}\psi_{l_1}(\mathbf{r})\psi_{l_1}^\ast(\mathbf{r})+(-i1/2)\mu_{1212}\psi_{l_1}(\mathbf{r})\psi_{l_2}^\ast(\mathbf{r}) \notag \\ 
    &+(i1/2)\mu_{2121}\psi_{l_2}(\mathbf{r})\psi_{l_1}^\ast(\mathbf{r})+(1/2)\mu_{2222}\psi_{l_2}(\mathbf{r})\psi_{l_2}^\ast(\mathbf{r}),\label{IL}
\end{align}

\begin{align}
    I_H(\mathbf{r}) =& \mu_{1111}\psi_{l_1}(\mathbf{r})\psi_{l_1}^\ast(\mathbf{r}),\label{IH}
\end{align}

\begin{align}
    I_V(\mathbf{r}) =& \mu_{2222}\psi_{l_2}(\mathbf{r})\psi_{l_2}^\ast(\mathbf{r}).\label{IV}
\end{align}
Substituting Eq. (\ref{ID})-(\ref{IV}) into Eq. (\ref{Sx1})-(\ref{Sz1}), we can obtain the final forms of locally normalized Stokes parameters
\begin{align}
    \Tilde{S}_x(\mathbf{r}) &= \frac{2\Re[ab^\ast \psi_{l_1}(\mathbf{r})\psi_{l_2}^\ast(\mathbf{r})]}{|a|^2|\psi_{l_1}(\mathbf{r})|^2+|b|^2|\psi_{l_2}(\mathbf{r})|^2},\label{initial Sx}\\
    \Tilde{S}_y(\mathbf{r}) &= \frac{-2\Im[ab^\ast \psi_{l_1}(\mathbf{r})\psi_{l_2}^\ast(\mathbf{r})]}{|a|^2|\psi_{l_1}(\mathbf{r})|^2+|b|^2|\psi_{l_2}(\mathbf{r})|^2},\label{initial Sy}\\
    \Tilde{S}_z(\mathbf{r}) &= \frac{|a|^2|\psi_{l_1}(\mathbf{r})|^2-|b|^2|\psi_{l_2}(\mathbf{r})|^2}{|a|^2|\psi_{l_1}(\mathbf{r})|^2+|b|^2|\psi_{l_2}(\mathbf{r})|^2}\label{initial Sz}.
\end{align}
And above three elements satisfy the vector normalization relation $\Tilde{S}_x^2 + \Tilde{S}_y^2+\Tilde{S}_z^2=1$.

\section*{\expandafter{\romannumeral3}. Analytical solutions of skyrmion numbers in local decoherence}
We primarily consider three decoherence channels, i.e., phase damping channel, depolarizing channel and amplitude damping channel. The decoherence channels solely act on the polarization DOF and OAM DOF maintains unaffected by imposing on an identity operator $\mathbb{I}$. We use Kraus operators $\{K_\nu\}$ to characterize these channels.\par

\subsection{Phase damping channel}
For the phase damping channel, the output state is 
\begin{equation}
    \rho_{\text{PDC}} = E(\rho) = \left(1-\frac{p_1}{2}\right)\rho + \frac{p_1}{2}(\mathbb{I}\otimes \sigma_z)\rho (\mathbb{I}\otimes \sigma_z),\label{pdc_rho}
\end{equation}
 and this channel can be related by the Kraus operators $K_0 = \sqrt{1-p_1}\mathbb{I}, K_1 = \sqrt{p_1}\begin{pmatrix}
    1 & 0 \\
    0 & 0
\end{pmatrix}, K_2 = \sqrt{p_1}\begin{pmatrix}
    0 & 0 \\
    0 & 1
\end{pmatrix}, p_1 \in [0, 1]$. Kraus operators have the property of $\sum_\nu K_\nu^\dagger K_\nu=\mathbb{I}$. If we input a density matrix of a single qubit $\rho_{i}  = \begin{pmatrix}
    \rho_{00} & \rho_{01} \\
    \rho_{10} & \rho_{11}
\end{pmatrix}$,
the output density matrix after passing through the channel is
\begin{equation}
    \rho_{o} = \begin{pmatrix}
    \rho_{00} & (1-p_1)\rho_{01} \\
    (1-p_1)\rho_{10} & \rho_{11}
\end{pmatrix}.\label{rhoo_pdc}
\end{equation}
The final density matrix of the phase damping channel is  (from the initial state in Eq. (\ref{intial rho}))
\begin{equation}
    \rho_{\text{PDC}} = \sum _{pqst=1}^2 \mu'_{pqst} \ket{l_p}\bra{l_q}\otimes \ket{e_s}\bra{e_t},
    \label{supp:PDC rho}
\end{equation}
where $\mu'_{1111} = |a|^2$, $\mu'_{1212}=(1-p_1)ab^\ast$, $\mu'_{2121}=(1-p_1)a^\ast b$, $\mu'_{2222}=|b|^2$, and other coefficients are zero. The locally normalized Stokes parameters are 
\begin{align}
    \Tilde{S}^{\text{PDC}}_x(\mathbf{r}) &= \frac{(1-p_1)\cdot 2\Re[ab^\ast \psi_{l_1}(\mathbf{r})\psi_{l_2}^\ast(\mathbf{r})]}{\sqrt{[|a|^2|\psi_{l_1}(\mathbf{r})|^2+|b|^2|\psi_{l_2}(\mathbf{r})|^2]^2 - 4p_1(2-p_1)|a|^2|b|^2|\psi_{l_1}(\mathbf{r})|^2|\psi_{l_2}(\mathbf{r})|^2}}\notag \\
    &= \frac{(1-p_1)[v(\mathbf{r})+v^\ast(\mathbf{r})]}{\sqrt{[1+|v(\mathbf{r})|^2]^2-4p_1(2-p_1)|v(\mathbf{r})|^2}},
    \label{Sxpdc2}\\
    \Tilde{S}^{\text{PDC}}_y(\mathbf{r}) &= \frac{-(1-p_1) \cdot 2\Im[ab^\ast \psi_{l_1}(\mathbf{r})\psi_{l_2}^\ast(\mathbf{r})]}{\sqrt{[|a|^2|\psi_{l_1}(\mathbf{r})|^2+|b|^2|\psi_{l_2}(\mathbf{r})|^2]^2 - 4p_1(2-p_1)|a|^2|b|^2|\psi_{l_1}(\mathbf{r})|^2|\psi_{l_2}(\mathbf{r})|^2}}\notag \\
    &= \frac{-i(1-p_1)[v(\mathbf{r})-v^\ast(\mathbf{r})]}{\sqrt{[1+|v(\mathbf{r})|^2]^2-4p_1(2-p_1)|v(\mathbf{r})|^2}},
    \label{Sypdc2}\\
    \Tilde{S}^{\text{PDC}}_z(\mathbf{r}) &= \frac{|a|^2|\psi_{l_1}(\mathbf{r})|^2-|b|^2|\psi_{l_2}(\mathbf{r})|^2}{\sqrt{[|a|^2|\psi_{l_1}(\mathbf{r})|^2+|b|^2|\psi_{l_2}(\mathbf{r})|^2]^2 - 4p_1(2-p_1)|a|^2|b|^2|\psi_{l_1}(\mathbf{r})|^2|\psi_{l_2}(\mathbf{r})|^2}}\notag \\
    &= \frac{1-|v(\mathbf{r})|^2}{\sqrt{[1+|v(\mathbf{r})|^2]^2-4p_1(2-p_1)|v(\mathbf{r})|^2}}.
    \label{Szpdc2}
\end{align}
The equality $\Tilde{S}_x^2 + \Tilde{S}_y^2+\Tilde{S}_z^2=1$ must be fulfilled and we simplify two LG spatial modes ($\psi_{l_1}(\mathbf{r})$ and $\psi_{l_2}(\mathbf{r})$) into the following form 
\begin{equation}
    v(\mathbf{r}) = v(r,\phi,z) = \frac{b\psi_{l_2}(\mathbf{r})}{a\psi_{l_1}(\mathbf{r})} = A(z) r^{|l_2|-|l_1|}e^{i\Delta l\phi},
\end{equation}
where $\Delta l = |l_2-l_1|$.\par
In the polar coordinate $\{r,\phi,z\}$, we can get  
\begin{align}
    \frac{\partial \Tilde{S}^{\text{PDC}}_x}{\partial x} &= \cos{\phi}\frac{\partial \Tilde{S}^{\text{PDC}}_x}{\partial r} - \frac{1}{r}\sin{\phi}\frac{\partial \Tilde{S}^{\text{PDC}}_x}{\partial \phi} \notag \\ 
    &= \frac{(1-p_1)}{rB_1}\{ \cos{\phi}(|l_2|-|l_1|)(v+v^\ast)[B_1^2 \notag \\
    &- 2|v|^2(1+|v|^2-2p_1(2-p_1))]/B_1^2-i \sin{\phi}\Delta l (v-v^\ast)\},
\end{align}

\begin{align}
    \frac{\partial \Tilde{S}^{\text{PDC}}_x}{\partial y} &= \sin{\phi}\frac{\partial \Tilde{S}^{\text{PDC}}_x}{\partial r} + \frac{1}{r}\cos{\phi}\frac{\partial \Tilde{S}^{\text{PDC}}_x}{\partial \phi} \notag \\ 
    &= \frac{(1-p_1)}{rB_1}\{ \sin{\phi}(|l_2|-|l_1|)(v+v^\ast)[B_1^2 \notag \\
    &- 2|v|^2(1+|v|^2-2p_1(2-p_1))]/B_1^2+i \cos{\phi}\Delta l (v-v^\ast)\},
\end{align}

\begin{align}
    \frac{\partial \Tilde{S}^{\text{PDC}}_y}{\partial x} &= \cos{\phi}\frac{\partial \Tilde{S}^{\text{PDC}}_y}{\partial r} - \frac{1}{r}\sin{\phi}\frac{\partial \Tilde{S}^{\text{PDC}}_y}{\partial \phi} \notag \\ 
    &= -\frac{i(1-p_1)}{rB_1}\{ \cos{\phi}(|l_2|-|l_1|)(v-v^\ast)[B_1^2 \notag \\
    &- 2|v|^2(1+|v|^2-2p_1(2-p_1))]/B_1^2-i \sin{\phi}\Delta l (v+v^\ast)\},
\end{align}

\begin{align}
    \frac{\partial \Tilde{S}^{\text{PDC}}_y}{\partial y} &= \sin{\phi}\frac{\partial \Tilde{S}^{\text{PDC}}_y}{\partial r} + \frac{1}{r}\cos{\phi}\frac{\partial \Tilde{S}^{\text{PDC}}_y}{\partial \phi} \notag \\ 
    &= -\frac{i(1-p_1)}{rB_1}\{ \sin{\phi}(|l_2|-|l_1|)(v-v^\ast)[B_1^2 \notag \\
    &- 2|v|^2(1+|v|^2-2p_1(2-p_1))]/B_1^2+i \cos{\phi}\Delta l (v+v^\ast)\},
\end{align}

\begin{align}
    \frac{\partial \Tilde{S}^{\text{PDC}}_z}{\partial x} &= \cos{\phi}\frac{\partial \Tilde{S}^{\text{PDC}}_z}{\partial r} - \frac{1}{r}\sin{\phi}\frac{\partial \Tilde{S}^{\text{PDC}}_z}{\partial \phi} \notag \\ 
    &= -\frac{2(|l_2|-|l_1|)|v|^2}{rB_1^3}\{ \cos{\phi}[B_1^2 + (1-|v|^2)(1+|v|^2-2p_1(2-p_1))]\},
\end{align}

\begin{align}
    \frac{\partial \Tilde{S}^{\text{PDC}}_z}{\partial y} &= \sin{\phi}\frac{\partial \Tilde{S}^{\text{PDC}}_z}{\partial r} + \frac{1}{r}\cos{\phi}\frac{\partial \Tilde{S}^{\text{PDC}}_z}{\partial \phi} \notag \\ 
    &= -\frac{2(|l_2|-|l_1|)|v|^2}{rB_1^3}\{ \sin{\phi}[B_1^2+ (1-|v|^2)(1+|v|^2-2p_1(2-p_1))]\},
\end{align}
where $B_1^2 = (1+|v|^2)^2-4p_1(2-p_1)|v|^2$. The relevant $z$th component of the field is
\begin{equation}
    \Sigma_z^{\text{PDC}} = \frac{4\Delta l (|l_2|-|l_1|)|v|^2}{r^2}\cdot \frac{(1-p_1)^2(1+|v|^2)}{[(1+|v|^2)^2-4p_1(2-p_1)|v|^2]^{3/2}}.
    \label{KernelPDC}
\end{equation}

We integrate Eq.~(\ref{KernelPDC}) over the whole $x-y$ plane, and for convenience, we proceed this process in the polar coordinate, i.e., 
\begin{align}
    N_z^{\text{PDC}} &= \frac{1}{4\pi}\int \Sigma_z^{\text{PDC}}{\rm d}x{\rm d}y \notag \\
    &= \frac{1}{4\pi}\int_0^\infty \int_0^{2\pi} \Sigma_z^{\text{PDC}}r{\rm d}r{\rm d}\phi \notag \\
    &= \frac{1}{4\pi}2\pi \int_0^\infty \frac{4\Delta l (|l_2|-|l_1|)|v|^2}{r} \cdot \frac{(1-p_1)^2(1+|v|^2)}{[(1+|v|^2)^2-4p_1(2-p_1)|v|^2]^{3/2}} {\rm d}r \notag \\
    &= 2\Delta l (|l_2|-|l_1|)(1-p_1)^2\int_0^\infty \frac{|A(z)|^2r^{2(|l_2|-|l_1|)-1}(1+|A(z)|^2r^{2(|l_2|-|l_1|)})}{[(1+|A(z)|^2r^{2(|l_2|-|l_1|)})^2-4p_1(2-p_1)|A(z)|^2r^{2(|l_2|-|l_1|)}]^{3/2}}{\rm d}r \notag \\
    &\xlongequal{R=r^{2(|l_2|-|l_1|)}} \Delta l (1-p_1)^2\int_0^ \infty \frac{|A(z)|^2(1+|A(z)|^2R)}{[(1+|A(z)|^2R)^2-4p_1(2-p_1)|A(z)|^2R]^{3/2}}{\rm d}R \notag \\ 
    &\xlongequal{R'=|A(z)|^2R} \Delta l (1-p_1)^2 \int_0^\infty \frac{(1+R')}{[(1+R')^2-4p_1(2-p_1)R']^{3/2}}{\rm d}R'.
\end{align}

Thus, the corresponding skyrmion number is
\begin{align}
    N_z^{\text{PDC}} &= \left\{ \begin{matrix}
        \Delta l, & 0\leq p_1 < 1 \\
        0, & p_1=1
    \end{matrix} \right. \label{supp:PDC Nz}
\end{align}

\subsection{Depolarizing channel}
For the depolarizing channel, the output state is 
\begin{equation}
    \rho_{\text{DC}} = E(\rho) = \left(1-p_2\right)\rho + \frac{p_2}{3}\left[(\mathbb{I}\otimes \sigma_x)\rho (\mathbb{I}\otimes \sigma_x) + (\mathbb{I}\otimes \sigma_y)\rho (\mathbb{I}\otimes \sigma_y) + (\mathbb{I}\otimes \sigma_z)\rho (\mathbb{I}\otimes \sigma_z) \right],
\end{equation}
and the Kraus operators of this channel are 
\begin{equation}
   K_0 = \sqrt{1-p_2}\mathbb{I}, K_1 = \sqrt{\frac{p_2}{3}}\begin{pmatrix}
    0 & 1 \\
    1 & 0
\end{pmatrix}, K_2 = \sqrt{\frac{p_2}{3}}\begin{pmatrix}
    0 & -i \\
    i & 0
\end{pmatrix}, K_3 = \sqrt{\frac{p_2}{3}}\begin{pmatrix}
    1 & 0 \\
    0 & -1
\end{pmatrix},
p_2 \in [0, \frac{3}{4}]. 
\end{equation}
 If we input a density matrix of a single qubit $\rho_{i}  = \begin{pmatrix}
    \rho_{00} & \rho_{01} \\
    \rho_{10} & \rho_{11}
\end{pmatrix}$,
the output density matrix after passing through the channel is 
\begin{equation}
    \rho_{o} = \begin{pmatrix}
    (1-\frac{2p_2}{3})\rho_{00}+\frac{2p_2}{3}\rho_{11} & (1-\frac{4p_2}{3})\rho_{01} \\
    (1-\frac{4p_2}{3})\rho_{10} & (1-\frac{2p_2}{3})\rho_{11}+\frac{2p_2}{3}\rho_{00}
\end{pmatrix}.\label{rhoo_dc}
\end{equation}
The final density matrix of the depolarizing channel is 
\begin{equation}
    \rho_{\text{DC}} = \sum _{pqst=1}^2 \mu''_{pqst} \ket{l_p}\bra{l_q}\otimes \ket{e_s}\bra{e_t},
    \label{supp:DC rho}
\end{equation}
where $\mu''_{1111} = (1-\frac{2p_2}{3})|a|^2$, $\mu''_{1212}=(1-\frac{4p_2}{3})ab^\ast$, $\mu''_{2121}=(1-\frac{4p_2}{3})a^\ast b$, $\mu''_{2222}=(1-\frac{2p_2}{3})|b|^2$, $\mu''_{2211}=\frac{2p_2}{3}|b|^2$, $\mu''_{1122}=\frac{2p_2}{3}|a|^2$, and other coefficients are zero. The Stokes parameters in depolarizing channel differ only by a factor $(1-\frac{4p_2}{3})$ from the Stokes parameters in the input state, and this factor has no effect after vector normalization. Thereby the locally normalized Stokes parameters in depolarizing channel are the same as Eq. (\ref{initial Sx}), (\ref{initial Sy}) and (\ref{initial Sz}):
\begin{align}
    \Tilde{S}^{\text{DC}}_x(\mathbf{r}) &= \frac{2\Re[ab^\ast \psi_{l_1}(\mathbf{r})\psi_{l_2}^\ast(\mathbf{r})]}{|a|^2|\psi_{l_1}(\mathbf{r})|^2+|b|^2|\psi_{l_2}(\mathbf{r})|^2}\notag \\
    &= \frac{v(\mathbf{r})+v^\ast(\mathbf{r})}{1+|v(\mathbf{r})|^2},
    \label{Sxdc2}\\
    \Tilde{S}^{\text{DC}}_y(\mathbf{r}) &= \frac{-2\Im[ab^\ast \psi_{l_1}(\mathbf{r})\psi_{l_2}^\ast(\mathbf{r})]}{|a|^2|\psi_{l_1}(\mathbf{r})|^2+|b|^2|\psi_{l_2}(\mathbf{r})|^2}\notag \\
    &= \frac{-i[v(\mathbf{r})-v^\ast(\mathbf{r})]}{1+|v(\mathbf{r})|^2},
    \label{Sydc2}\\
    \Tilde{S}^{\text{DC}}_z(\mathbf{r}) &= \frac{|a|^2|\psi_{l_1}(\mathbf{r})|^2-|b|^2|\psi_{l_2}(\mathbf{r})|^2}{|a|^2|\psi_{l_1}(\mathbf{r})|^2+|b|^2|\psi_{l_2}(\mathbf{r})|^2}\notag \\
    &= \frac{1-|v(\mathbf{r})|^2}{1+|v(\mathbf{r})|^2}.
    \label{Szdc2}
\end{align}
Here, $\Tilde{S}_x^2 + \Tilde{S}_y^2+\Tilde{S}_z^2=1$ and 
$
    v(\mathbf{r}) = v(r,\phi,z) = \frac{b\psi_{l_2}(\mathbf{r})}{a\psi_{l_1}(\mathbf{r})} = A(z) r^{|l_2|-|l_1|}e^{i\Delta l\phi},
$
where $\Delta l = |l_2-l_1|$.\par
Then
\begin{align}
    \frac{\partial \Tilde{S}^{\text{DC}}_x}{\partial x} &= \cos{\phi}\frac{\partial \Tilde{S}^{\text{DC}}_x}{\partial r} - \frac{1}{r}\sin{\phi}\frac{\partial \Tilde{S}^{\text{DC}}_x}{\partial \phi} \notag \\ 
    &= \frac{1}{r(1+|v|^2)}\{ \cos{\phi}(|l_2|-|l_1|)(v+v^\ast)(1-|v|^2)/(1+|v|^2)-i \sin{\phi}\Delta l (v-v^\ast)\},
\end{align}

\begin{align}
    \frac{\partial \Tilde{S}^{\text{DC}}_x}{\partial y} &= \sin{\phi}\frac{\partial \Tilde{S}^{\text{DC}}_x}{\partial r} + \frac{1}{r}\cos{\phi}\frac{\partial \Tilde{S}^{\text{DC}}_x}{\partial \phi} \notag \\ 
    &= \frac{1}{r(1+|v|^2)}\{ \sin{\phi}(|l_2|-|l_1|)(v+v^\ast)(1-|v|^2)/(1+|v|^2)+i \cos{\phi}\Delta l (v-v^\ast)\},
\end{align}

\begin{align}
    \frac{\partial \Tilde{S}^{\text{DC}}_y}{\partial x} &= \cos{\phi}\frac{\partial \Tilde{S}^{\text{DC}}_y}{\partial r} - \frac{1}{r}\sin{\phi}\frac{\partial \Tilde{S}^{\text{DC}}_y}{\partial \phi} \notag \\ 
    &= -\frac{i}{r(1+|v|^2)}\{ \cos{\phi}(|l_2|-|l_1|)(v-v^\ast)(1-|v|^2)/(1+|v|^2)-i \sin{\phi}\Delta l (v+v^\ast)\},
\end{align}

\begin{align}
    \frac{\partial \Tilde{S}^{\text{DC}}_y}{\partial y} &= \sin{\phi}\frac{\partial \Tilde{S}^{\text{DC}}_y}{\partial r} + \frac{1}{r}\cos{\phi}\frac{\partial \Tilde{S}^{\text{DC}}_y}{\partial \phi} \notag \\ 
    &= -\frac{i}{r(1+|v|^2)}\{ \sin{\phi}(|l_2|-|l_1|)(v-v^\ast)(1-|v|^2)/(1+|v|^2)+i \cos{\phi}\Delta l (v+v^\ast)\},
\end{align}

\begin{align}
    \frac{\partial \Tilde{S}^{\text{DC}}_z}{\partial x} &= \cos{\phi}\frac{\partial \Tilde{S}^{\text{DC}}_z}{\partial r} - \frac{1}{r}\sin{\phi}\frac{\partial \Tilde{S}^{\text{DC}}_z}{\partial \phi} \notag \\ 
    &= -\frac{4}{r(1+|v|^2)}\{\cos{\phi}(|l_2|-|l_1|)|v|^2/(1+|v|^2)\},
\end{align}

\begin{align}
    \frac{\partial \Tilde{S}^{\text{DC}}_z}{\partial y} &= \sin{\phi}\frac{\partial \Tilde{S}^{\text{DC}}_z}{\partial r} + \frac{1}{r}\cos{\phi}\frac{\partial \Tilde{S}^{\text{DC}}_z}{\partial \phi} \notag \\ 
    &= -\frac{4}{r(1+|v|^2)}\{\sin{\phi}(|l_2|-|l_1|)|v|^2/(1+|v|^2)\}.
\end{align}
The relevant $z$th component of the field is
\begin{equation}
    \Sigma_z^{\text{DC}} = \frac{4\Delta l (|l_2|-|l_1|)|v|^2}{r^2}\cdot \frac{1}{(1+|v|^2)^2},
    \label{KernelDC}
\end{equation}

We integrate Eq.~(\ref{KernelDC}) over the whole $x-y$ plane, and for convenience, we proceed this process in the polar coordinate, i.e., 
\begin{align}
    N_z^{\text{DC}} &= \frac{1}{4\pi}\int \Sigma_z^{\text{DC}}{\rm d}x{\rm d}y \notag \\
    &= \frac{1}{4\pi}\int_0^\infty \int_0^{2\pi} \Sigma_z^{\text{DC}}r{\rm d}r{\rm d}\phi \notag \\
    &= \frac{1}{4\pi}2\pi \int_0^\infty \frac{4\Delta l (|l_2|-|l_1|)|v|^2}{r} \cdot \frac{1}{(1+|v|^2)^2} {\rm d}r \notag \\
    &= 2\Delta l (|l_2|-|l_1|)\int_0^\infty \frac{|A(z)|^2r^{2(|l_2|-|l_1|)-1}}{(1+|A(z)|^2r^{2(|l_2|-|l_1|)})^2}{\rm d}r \notag \\
    &\xlongequal{R=r^{2(|l_2|-|l_1|)}} \Delta l \int_0^ \infty \frac{|A(z)|^2}{(1+|A(z)|^2R)^2}{\rm d}R \notag \\ 
    &\xlongequal{R'=1+|A(z)|^2R} \Delta l \int_0^\infty \frac{1}{R'^2}{\rm d}R'.
\end{align}

Thus, if $p_2 \neq 3/4$, the corresponding skyrmion number remains unchanged, i.e., 
\begin{align}
    N_z^{\text{DC}} &= \left\{ \begin{matrix}
        \Delta l, & 0\leq p_2 < \frac{3}{4} \\
        0, & p_2=\frac{3}{4}
    \end{matrix} \right. \label{supp:DC Nz}
\end{align}
In fact, the skyrmion number in this scenario is the same as Ref. \cite{gao2020paraxial} after re-normalization of Stokes parameters.

\subsection{Amplitude damping channel}
For the amplitude damping channel, the Kraus operators are 
\begin{equation}
    K_0 = \begin{pmatrix}
    1 & 0 \\
    0 & \sqrt{\eta}
\end{pmatrix}, K_1 = \begin{pmatrix}
    0 & \sqrt{1-\eta} \\
    0 & 0
\end{pmatrix}, \eta \in [0, 1].
\end{equation}
Here, $\eta$ is the damping factor. If we input a density matrix of a single qubit $\rho_{i}  = \begin{pmatrix}
    \rho_{00} & \rho_{01} \\
    \rho_{10} & \rho_{11}
\end{pmatrix}$,
the output density matrix after passing through the channel is 
\begin{equation}
    \rho_{o} = \begin{pmatrix}
    \rho_{00}+(1-\eta)\rho_{11} & \sqrt{\eta}\rho_{01} \\
    \sqrt{\eta}\rho_{10} & \eta \rho_{11}
\end{pmatrix}.\label{rhoo_adc}
\end{equation}
The initial state in Eq. (\ref{intial rho}) goes through an amplitude damping channel, the final state can be presented as
\begin{equation}
    \rho_{\text{ADC}} = \sum _{pqst=1}^2 \mu'''_{pqst} \ket{l_p}\bra{l_q}\otimes \ket{e_s}\bra{e_t},
    \label{supp:ADC rho}
\end{equation}
where $\mu'''_{1111} = |a|^2$, $\mu'''_{2211}=(1-\eta)|b|^2$, $\mu'''_{1212}=\sqrt{\eta}ab^\ast$, $\mu'''_{2121}=\sqrt{\eta}a^\ast b$, $\mu'''_{2222}=\eta|b|^2$, and other coefficients are zero. The locally normalized Stokes parameters are 
\begin{align}
    \Tilde{S}^{\text{ADC}}_x(\mathbf{r}) &= \frac{\sqrt{\eta}\cdot 2\Re[ab^\ast \psi_{l_1}(\mathbf{r})\psi_{l_2}^\ast(\mathbf{r})]}{\sqrt{[|a|^2|\psi_{l_1}(\mathbf{r})|^2+|b|^2|\psi_{l_2}(\mathbf{r})|^2]^2 - 4\eta(1-\eta)|b|^4|\psi_{l_2}(\mathbf{r})|^4}},\label{Sxadc}\\
    \Tilde{S}^{\text{ADC}}_y(\mathbf{r}) &= \frac{-\sqrt{\eta} \cdot 2\Im[ab^\ast \psi_{l_1}(\mathbf{r})\psi_{l_2}^\ast(\mathbf{r})]}{\sqrt{[|a|^2|\psi_{l_1}(\mathbf{r})|^2+|b|^2|\psi_{l_2}(\mathbf{r})|^2]^2 - 4\eta(1-\eta)|b|^4|\psi_{l_2}(\mathbf{r})|^4}},\label{Syadc}\\
    \Tilde{S}^{\text{ADC}}_z(\mathbf{r}) &= \frac{|a|^2|\psi_{l_1}(\mathbf{r})|^2-(2\eta-1)|b|^2|\psi_{l_2}(\mathbf{r})|^2}{\sqrt{[|a|^2|\psi_{l_1}(\mathbf{r})|^2+|b|^2|\psi_{l_2}(\mathbf{r})|^2]^2 - 4\eta(1-\eta)|b|^4|\psi_{l_2}(\mathbf{r})|^4}}.\label{Szadc}
\end{align}
Similarly, the equality $\Tilde{S}_x^2 + \Tilde{S}_y^2+\Tilde{S}_z^2=1$ must be fulfilled.
Furthermore, we simplify two LG spatial modes ($\psi_{l_1}(\mathbf{r})$ and $\psi_{l_2}(\mathbf{r})$) into 
$
    v(\mathbf{r}) = v(r,\phi,z) = \frac{b\psi_{l_2}(\mathbf{r})}{a\psi_{l_1}(\mathbf{r})} = A(z) r^{|l_2|-|l_1|}e^{i\Delta l\phi},
    \label{supp:vr}
$
where $\Delta l = |l_2-l_1|$. From it, we know that the specific values of coefficients $a$ and $b$ have no impact on the calculation of skyrmion numbers ($a\neq0, b\neq0$). Based on the above results, Eq. (\ref{Sxadc})-(\ref{Szadc}) can be simplified to
\begin{align}
    \Tilde{S}^{\text{ADC}}_x(\mathbf{r}) &= \frac{\sqrt{\eta}[v(\mathbf{r})+v^\ast(\mathbf{r})]}{\sqrt{[1+|v(\mathbf{r})|^2]^2-4\eta(1-\eta)|v(\mathbf{r})|^4}},\label{Sxadc2}\\
    \Tilde{S}^{\text{ADC}}_y(\mathbf{r}) &= \frac{-i\sqrt{\eta}[v(\mathbf{r})-v^\ast(\mathbf{r})]}{\sqrt{[1+|v(\mathbf{r})|^2]^2-4\eta(1-\eta)|v(\mathbf{r})|^4}},\label{Syadc2}\\
    \Tilde{S}^{\text{ADC}}_z(\mathbf{r}) &= \frac{1-(2\eta-1)|v(\mathbf{r})|^2}{\sqrt{[1+|v(\mathbf{r})|^2]^2-4\eta(1-\eta)|v(\mathbf{r})|^4}}.\label{Szadc2}
\end{align}
Then
\begin{align}
    \frac{\partial \Tilde{S}^{\text{ADC}}_x}{\partial x} &= \cos{\phi}\frac{\partial \Tilde{S}^{\text{ADC}}_x}{\partial r} - \frac{1}{r}\sin{\phi}\frac{\partial \Tilde{S}^{\text{ADC}}_x}{\partial \phi} \notag \\ 
    &= \frac{\sqrt{\eta}}{rB_2}\{ \cos{\phi}(|l_2|-|l_1|)(v+v^\ast)[B_2^2 \notag \\
    &- 2|v|^2(1+(2\eta-1)^2|v|^2)]/B_2^2-i \sin{\phi}\Delta l (v-v^\ast)\},
\end{align}

\begin{align}
    \frac{\partial \Tilde{S}^{\text{ADC}}_x}{\partial y} &= \sin{\phi}\frac{\partial \Tilde{S}^{\text{ADC}}_x}{\partial r} + \frac{1}{r}\cos{\phi}\frac{\partial \Tilde{S}^{\text{ADC}}_x}{\partial \phi} \notag \\ 
    &= \frac{\sqrt{\eta}}{rB_2}\{ \sin{\phi}(|l_2|-|l_1|)(v+v^\ast)[B_2^2 \notag \\
    &- 2|v|^2(1+(2\eta-1)^2|v|^2)]/B_2^2+i \cos{\phi}\Delta l (v-v^\ast)\},
\end{align}

\begin{align}
    \frac{\partial \Tilde{S}^{\text{ADC}}_y}{\partial x} &= \cos{\phi}\frac{\partial \Tilde{S}^{\text{ADC}}_y}{\partial r} - \frac{1}{r}\sin{\phi}\frac{\partial \Tilde{S}^{\text{ADC}}_y}{\partial \phi} \notag \\ 
    &= -\frac{i\sqrt{\eta}}{rB_2}\{ \cos{\phi}(|l_2|-|l_1|)(v-v^\ast)[B_2^2 \notag \\
    &- 2|v|^2(1+(2\eta-1)^2|v|^2)]/B_2^2-i \sin{\phi}\Delta l (v+v^\ast)\},
\end{align}

\begin{align}
    \frac{\partial \Tilde{S}^{\text{ADC}}_y}{\partial y} &= \sin{\phi}\frac{\partial \Tilde{S}^{\text{ADC}}_y}{\partial r} + \frac{1}{r}\cos{\phi}\frac{\partial \Tilde{S}^{\text{ADC}}_y}{\partial \phi} \notag \\ 
    &= -\frac{i\sqrt{\eta}}{rB_2}\{ \sin{\phi}(|l_2|-|l_1|)(v-v^\ast)[B_2^2 \notag \\
    &- 2|v|^2(1+(2\eta-1)^2|v|^2)]/B_2^2+i \sin{\phi}\Delta l (v+v^\ast)\},
\end{align}

\begin{align}
    \frac{\partial \Tilde{S}^{\text{ADC}}_z}{\partial x} &= \cos{\phi}\frac{\partial \Tilde{S}^{\text{ADC}}_z}{\partial r} - \frac{1}{r}\sin{\phi}\frac{\partial \Tilde{S}^{\text{ADC}}_z}{\partial \phi} \notag \\ 
    &= \cos{\phi}\frac{2(|l_2|-|l_1|)|v|^2}{rB_2^3} \left[  (1-2\eta)B_2^2+(2\eta-1)^3|v|^4-2(2\eta-1)(\eta-1)|v|^2-1 \right],
\end{align}

\begin{align}
    \frac{\partial \Tilde{S}^{\text{ADC}}_z}{\partial y} &= \sin{\phi}\frac{\partial \Tilde{S}^{\text{ADC}}_z}{\partial r} + \frac{1}{r}\cos{\phi}\frac{\partial \Tilde{S}^{\text{ADC}}_z}{\partial \phi} \notag \\ 
    &= \sin{\phi}\frac{2(|l_2|-|l_1|)|v|^2}{rB_2^3} \left[  (1-2\eta)B_2^2+(2\eta-1)^3|v|^4-2(2\eta-1)(\eta-1)|v|^2-1 \right],
\end{align}
where $B_2^2 = (1+|v|^2)^2-4\eta(1-\eta)|v|^4$. By using Eq. (\ref{SigmaZ}), we get the $z$th component of the field 
\begin{equation}
    \Sigma_z^{\text{ADC}} = \frac{4\Delta l (|l_2|-|l_1|)|v|^2}{r^2}\cdot \frac{\eta[1+(2\eta-1)|v|^2]}{[(1+|v|^2)^2-4\eta(1-\eta)|v|^4]^{3/2}},
    \label{KernelADC}
\end{equation}

We integrate Eq.~(\ref{KernelADC}) over the whole $x-y$ plane, and for convenience, we proceed this process in the polar coordinate, i.e., 
\begin{align}
    N_z^{\text{ADC}} &= \frac{1}{4\pi}\int \Sigma_z^{\text{ADC}}{\rm d}x{\rm d}y \notag \\
    &= \frac{1}{4\pi}\int_0^\infty \int_0^{2\pi} \Sigma_z^{\text{ADC}}r{\rm d}r{\rm d}\phi \notag \\
    &= \frac{1}{4\pi}2\pi \int_0^\infty \frac{4\Delta l (|l_2|-|l_1|)|v|^2}{r}\cdot \frac{\eta[1+(2\eta-1)|v|^2]}{[(1+|v|^2)^2-4\eta(1-\eta)|v|^4]^{3/2}} {\rm d}r \notag \\
    &= 2\Delta l (|l_2|-|l_1|) \eta \int_0^\infty \frac{[1+(2\eta-1)|A(z)|^2r^{2(|l_2|-|l_1|)}]|A(z)|^2r^{2(|l_2|-|l_1|)-1}}{[(1+|A(z)|^2r^{2(|l_2|-|l_1|)})^2-4\eta(1-\eta)|A(z)|^4r^{4(|l_2|-|l_1|)}]^{3/2}} {\rm d}r \notag \\
    &\xlongequal{R=r^{2(|l_2|-|l_1|)}} \Delta l \eta \int_0^ \infty \frac{[1+(2\eta-1)|A(z)|^2R]|A(z)|^2}{[(1+|A(z)|^2R)^2-4\eta(1-\eta)|A(z)|^4R^2]^{3/2}}{\rm d}R \notag \\ 
    &\xlongequal{R'=|A(z)|^2R} \Delta l \eta \int_0^\infty \frac{[1+(2\eta-1)R']}{[(1+R')^2-4\eta(1-\eta)R'^2]^{3/2}}{\rm d}R',
\end{align}

and then the skyrmion number is
\begin{align}
    N_z^{\text{ADC}} &= \left\{ \begin{matrix}
        \Delta l, & \frac{1}{2}<\eta \leq 1 \\
        0, & 0\leq \eta < \frac{1}{2}
    \end{matrix} \right. \label{supp:ADC Nz}
\end{align}
where $\eta\neq \frac{1}{2}$ in Eq. (\ref{supp:ADC Nz}). When $\eta = \frac{1}{2}$, $N_z^{\text{ADC}} = \frac{\Delta l}{2}$. It is well-known that skyrmions is a mapping from the two-dimensional transverse plane $\mathcal{R}^2$ to the unit sphere $\mathcal{S}^2$ containing the whole $4\pi$ solid angle. Once $\eta\leq\frac{1}{2}$, it will only map half of the unit sphere and the spin configuration will be destroyed. Therefore, a full mapping can no longer be completed when $\eta$ is set to $\frac{1}{2}$, even when the skyrmion number does not reach $0$. \par

\section*{\expandafter{\romannumeral4}. Different initial coefficients $(a,b)$ and corresponding skyrmion numbers}\label{different ab}
An initial state of the system is $\ket{\Psi} = a\ket{l_1}\ket{0}+b\ket{l_2}\ket{1}, (a\neq0, b\neq0)$ and coefficients $(a,b)$ can vary prompting different initial concurrence values under the condition of normalization (i.e. $|a|^2+|b|^2=1$). However, different values of $a$ and $b$ do not influence the trend of skyrmion numbers, regardless of the presence or absence of local decoherence.\par
\begin{figure}[ht!]
    \centering
    \includegraphics[width=3in]{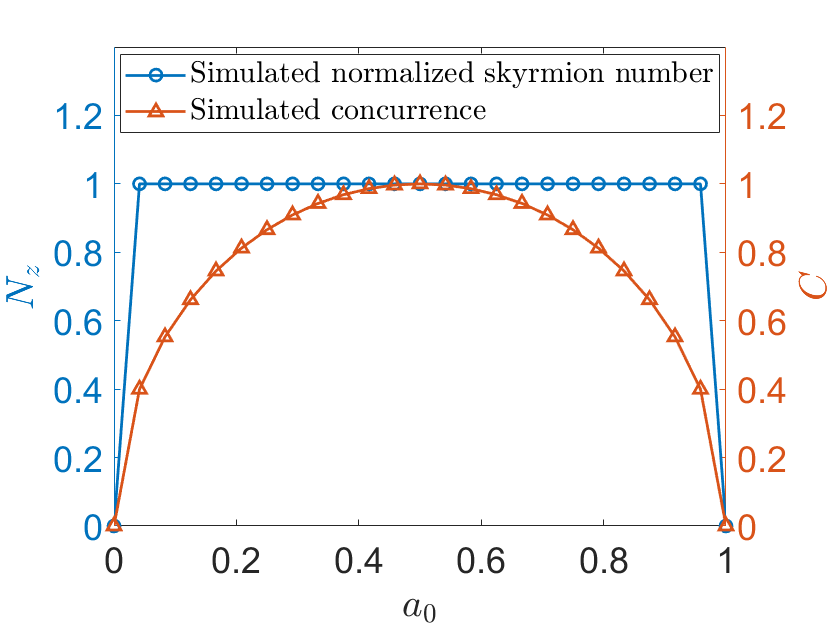}
    \caption{The simulated normalized skyrmion numbers and concurrence values vary with coefficient $a_0$ in the case of no decoherence.}
    \label{fig:figure40_purestate}
\end{figure}

Firstly, we examine the case without decoherence in Fig. \ref{fig:figure40_purestate}. Here, $a=\sqrt{a_0}$ and $b=\sqrt{1-a_0}$. By varying $a_0$, we obtain different initial states. However, skyrmion numbers remain topologically stable and unchanged unless the degree of entanglement vanishes. This result is consistent with the one in Ref.~\cite{ornelas2024non}.

\begin{figure*}[ht!]
    \centering
    \subfigure[Phase damping channel]{
    \begin{minipage}[t]{0.33\linewidth}
        \centering
        \includegraphics[width=1.65in]{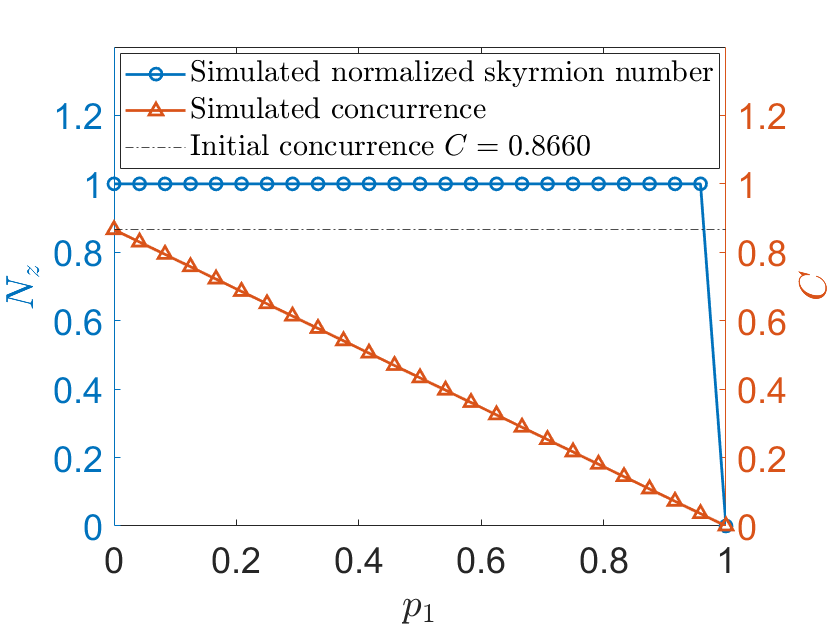}
        \label{figure21_PDC}
    \end{minipage}
    }\subfigure[Depolarizing channel]{
    \begin{minipage}[t]{0.33\linewidth}
        \centering
        \includegraphics[width=1.65in]{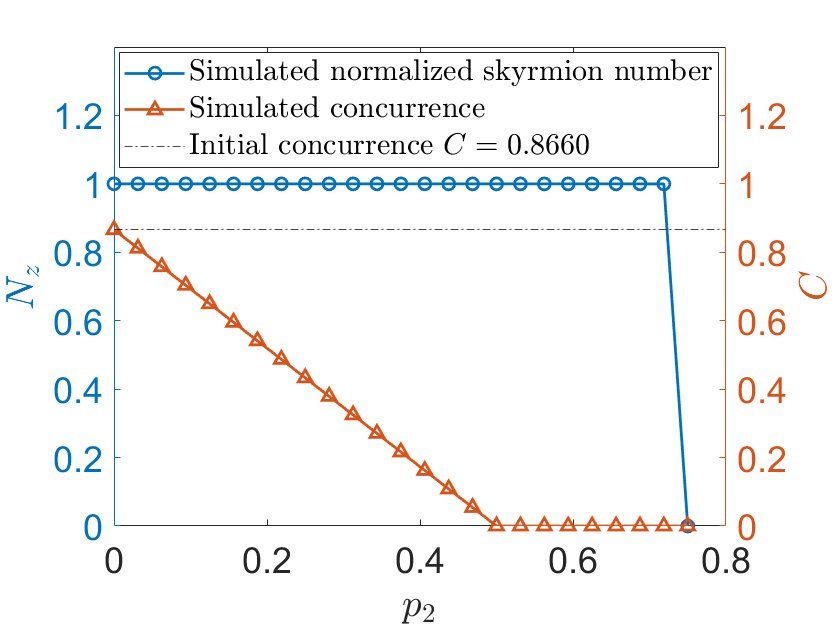}
        \label{figure31_DC}
    \end{minipage}
    }\subfigure[Amplitude damping channel]{
    \begin{minipage}[t]{0.33\linewidth}
        \centering
        \includegraphics[width=1.65in]{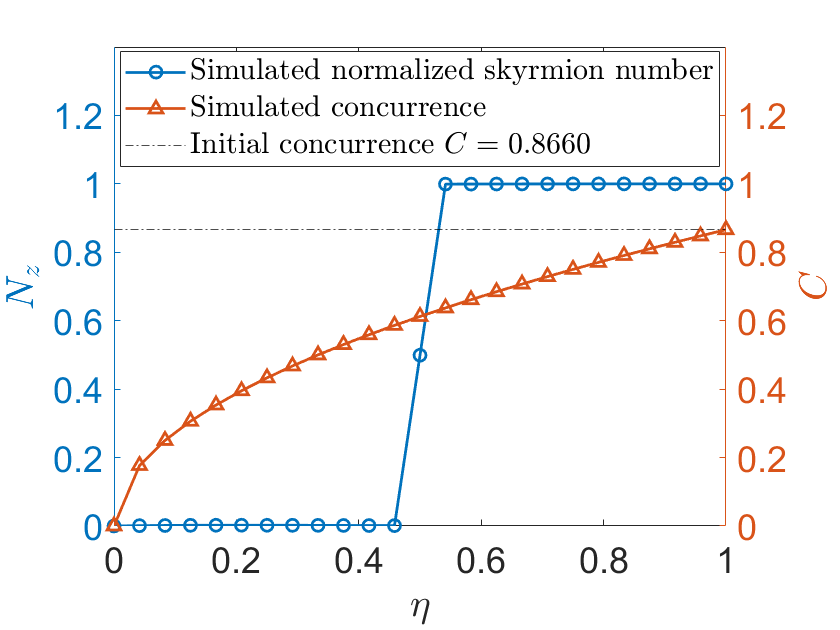}
        \label{figure11_ADC}
    \end{minipage}
    }
    \caption{The simulated normalized skyrmion numbers and concurrence values in different local decoherence channels. Coefficients $a_0=1/4$, $a=1/2$ and $b=\sqrt{3}/2$ correspond to an initial concurrence of $C=0.8660$.}
    \label{fig:skyrmion number and concurr=0.8660}
\end{figure*}

\begin{figure*}[ht!]
    \centering
    \subfigure[Phase damping channel]{
    \begin{minipage}[t]{0.33\linewidth}
        \centering
        \includegraphics[width=1.65in]{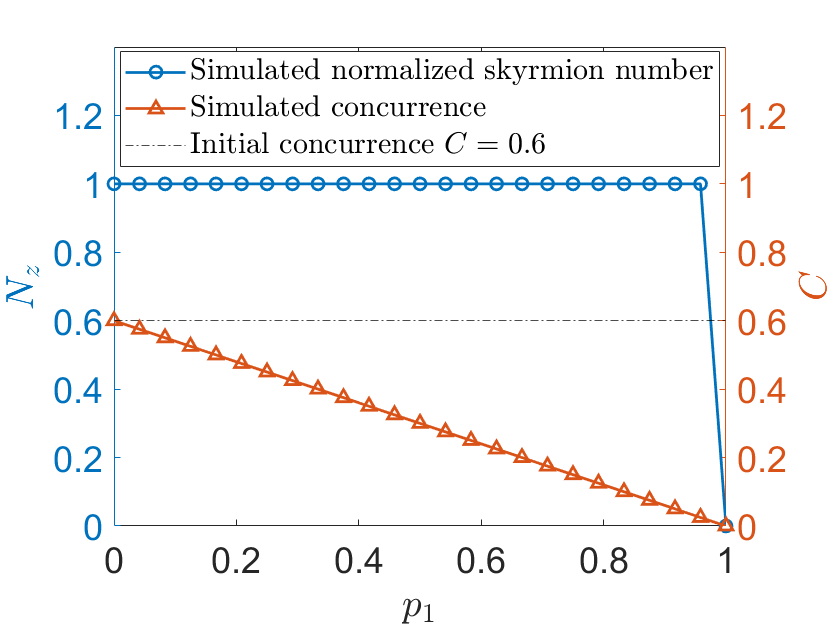}
        \label{figure22_PDC}
    \end{minipage}
    }\subfigure[Depolarizing channel]{
    \begin{minipage}[t]{0.33\linewidth}
        \centering
        \includegraphics[width=1.65in]{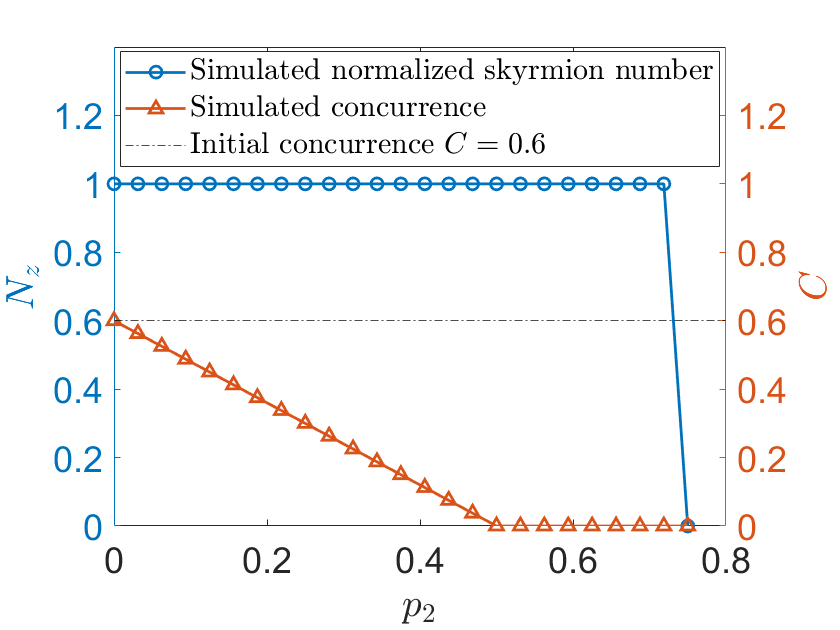}
        \label{figure32_DC}
    \end{minipage}
    }\subfigure[Amplitude damping channel]{
    \begin{minipage}[t]{0.33\linewidth}
        \centering
        \includegraphics[width=1.65in]{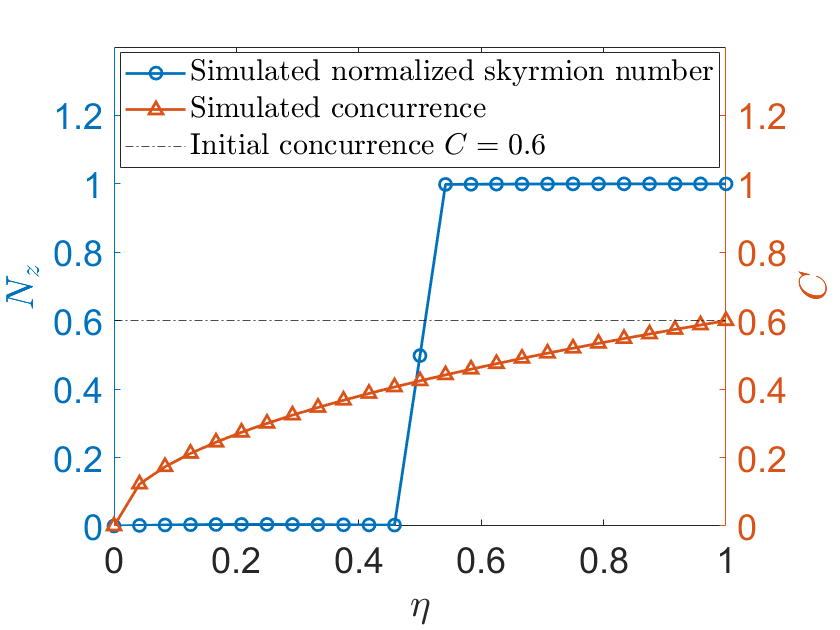}
        \label{figure12_ADC}
    \end{minipage}
    }
    \caption{The simulated normalized skyrmion numbers and concurrence values in different local decoherence channels. Coefficients $a_0=1/10$, $a=1/\sqrt{10}$ and $b=3/\sqrt{10}$ correspond to an initial concurrence of $C=0.6$.}
    \label{fig:skyrmion number and concurr=0.6}
\end{figure*}

Secondly, in the main text, we study the case of maximum entanglement degree (concurrence $C=1$ corresponds to coefficient $a_0 = 1/2$) with presence of local decoherence channels. In Fig. \ref{fig:skyrmion number and concurr=0.8660} and \ref{fig:skyrmion number and concurr=0.6}, coefficients are $a_0=1/4$ and $a_0=1/10$, respectively. But different initial states have the similar results so that each transition point of skyrmion numbers in three channels is impervious to local decoherence.\par

\begin{figure*}[ht!]
    \centering
    \subfigure[Phase damping channel]{
    \begin{minipage}[t]{0.33\linewidth}
        \centering
        \includegraphics[width=1.65in]{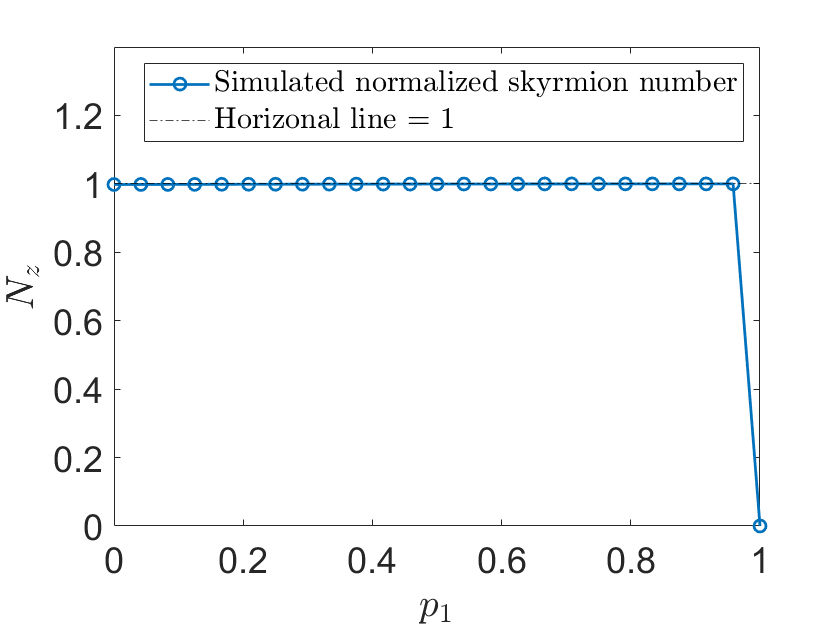}
        \label{PDC_L1=2&L2=0_a0=0.5}
    \end{minipage}
    }\subfigure[Depolarizing channel]{
    \begin{minipage}[t]{0.33\linewidth}
        \centering
        \includegraphics[width=1.65in]{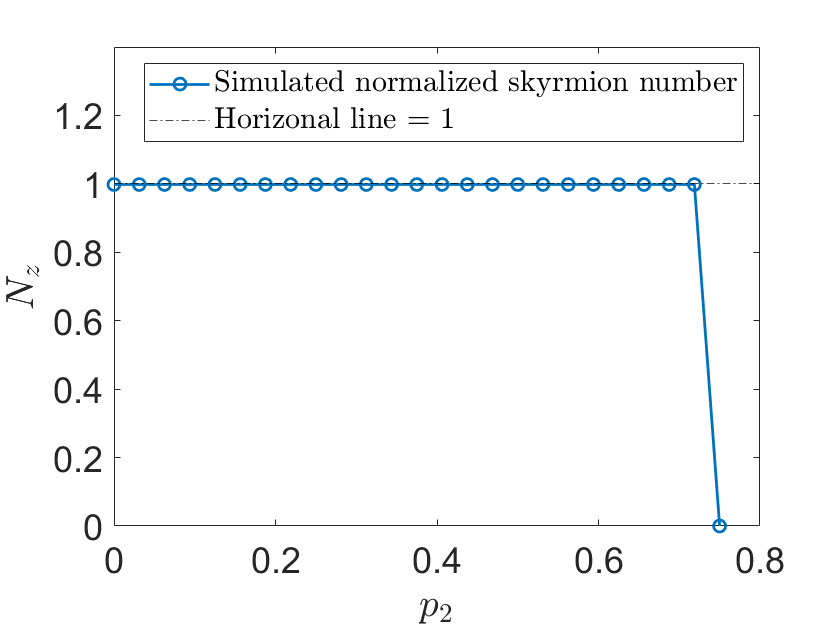}
        \label{DC_L1=2&L2=0_a0=0.5}
    \end{minipage}
    }\subfigure[Amplitude damping channel]{
    \begin{minipage}[t]{0.33\linewidth}
        \centering
        \includegraphics[width=1.65in]{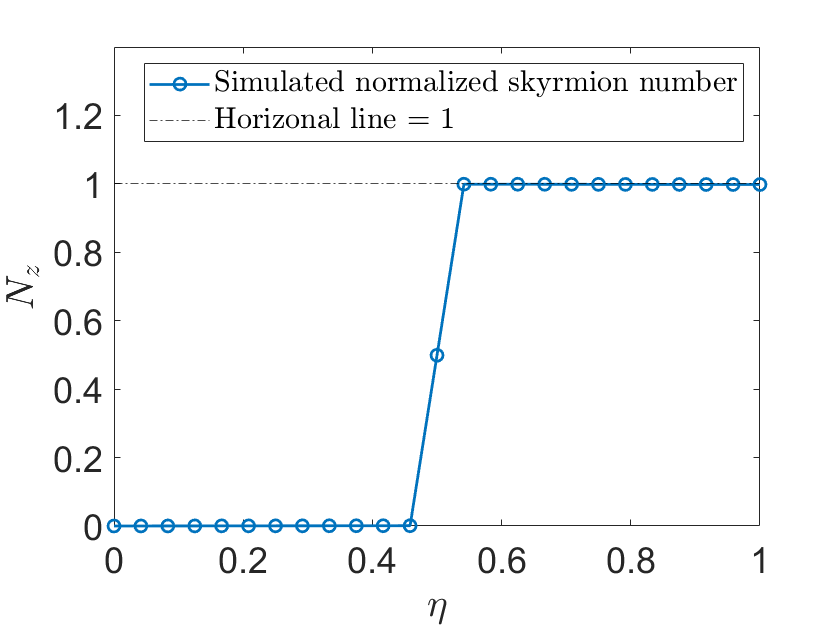}
        \label{ADC_L1=2&L2=0_a0=0.5}
    \end{minipage}
    }
    \caption{The simulated normalized skyrmion numbers in different local decoherence channels. The azimuthal indices of two spatial modes are $l_1=2$ and $l_2=0$. Coefficients $a_0=1/2$, $a=1/\sqrt{2}$ and $b=1/\sqrt{2}$ correspond to an initial concurrence of $C=1$.}
    \label{fig:skyrmion number and concurr=1_L1=2&L2=0}
\end{figure*}

In addition, in the main text, we take two LG modes with azimuthal indices $(l_1=8,l_2=0)$ as an example to interpret skyrmions' properties. In Fig. \ref{fig:skyrmion number and concurr=1_L1=2&L2=0}, we take a smaller value $N_z=2$ with $(l_1=2,l_2=0)$, and the features of skyrmions are all the same.

\section*{\expandafter{\romannumeral5}. The Stokes parameters vary with the decoherence strength}
\begin{figure*}[ht!]
    \centering
    \subfigure[$p_1=0$]{
    \begin{minipage}[t]{0.33\linewidth}
        \centering
        \includegraphics[width=1.65in]{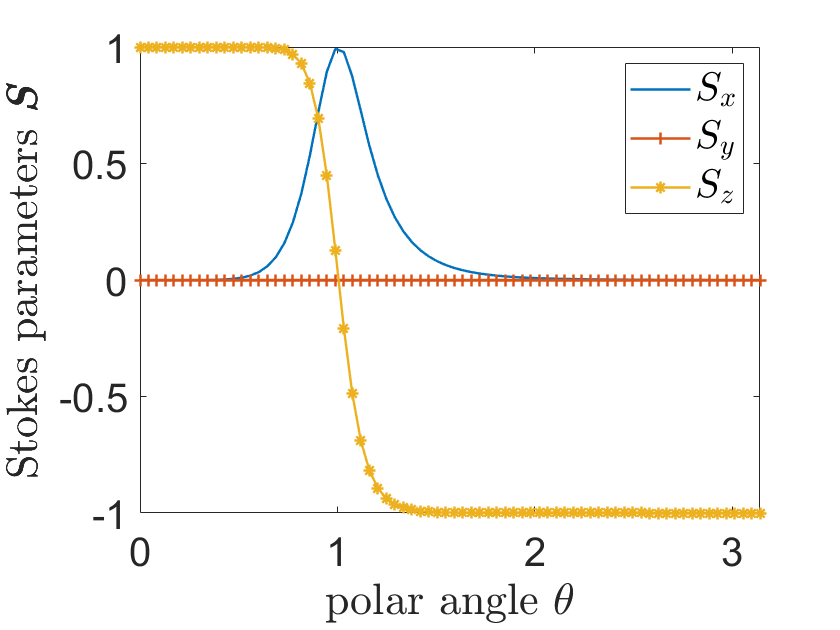}
        \label{figure52_PDC_Sxyz_p=0}
    \end{minipage}
    }\subfigure[$p_1=0.5$]{
    \begin{minipage}[t]{0.33\linewidth}
        \centering
        \includegraphics[width=1.65in]{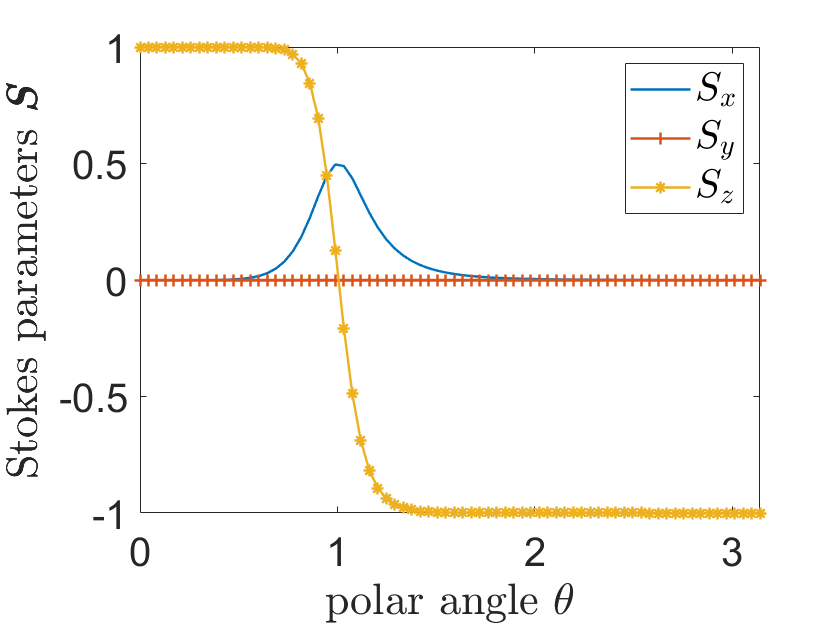}
        \label{figure52_PDC_Sxyz_p=0.5}
    \end{minipage}
    }\subfigure[$p_1=1$]{
    \begin{minipage}[t]{0.33\linewidth}
        \centering
        \includegraphics[width=1.65in]{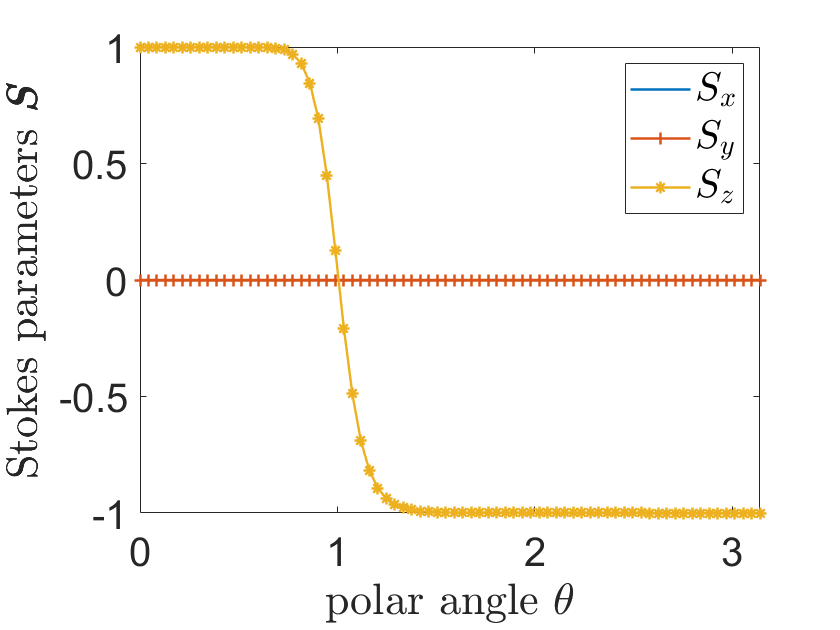}
        \label{figure52_PDC_Sxyz_p=1}
    \end{minipage}
    }
    \caption{The Stokes parameters $(S_x,S_y,S_z)(\mathbf{r})$ vary with different decoherence strength ($p_1=0, p_1=0.5, p_1=1$) of phase damping channel. We choose a slice of $S_y=0$ as an example.}
    \label{fig:Bloch S vector PDC}
\end{figure*}

\begin{figure*}[ht!]
    \centering
    \subfigure[$p_2=0$]{
    \begin{minipage}[t]{0.33\linewidth}
        \centering
        \includegraphics[width=1.65in]{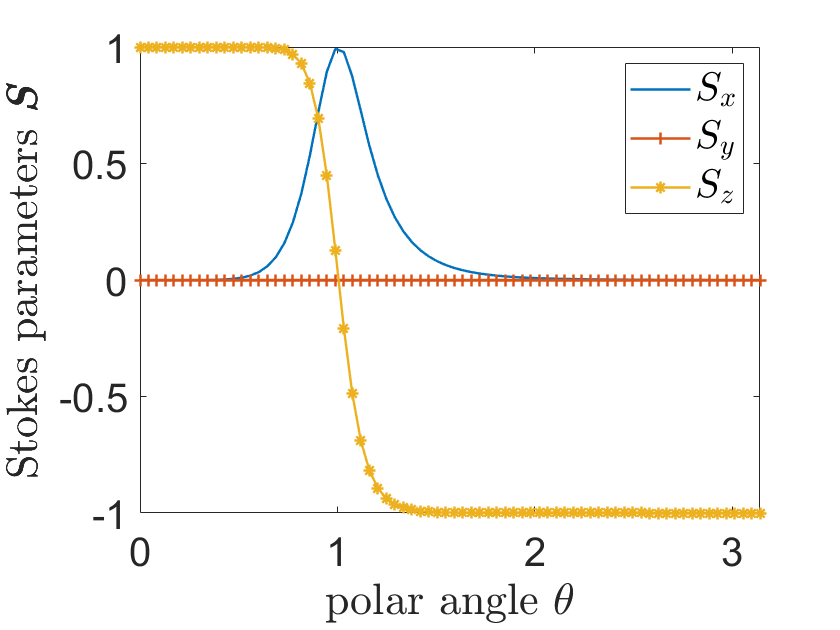}
        \label{figure53_DC_Sxyz_p=0}
    \end{minipage}
    }\subfigure[$p_2=0.375$]{
    \begin{minipage}[t]{0.33\linewidth}
        \centering
        \includegraphics[width=1.65in]{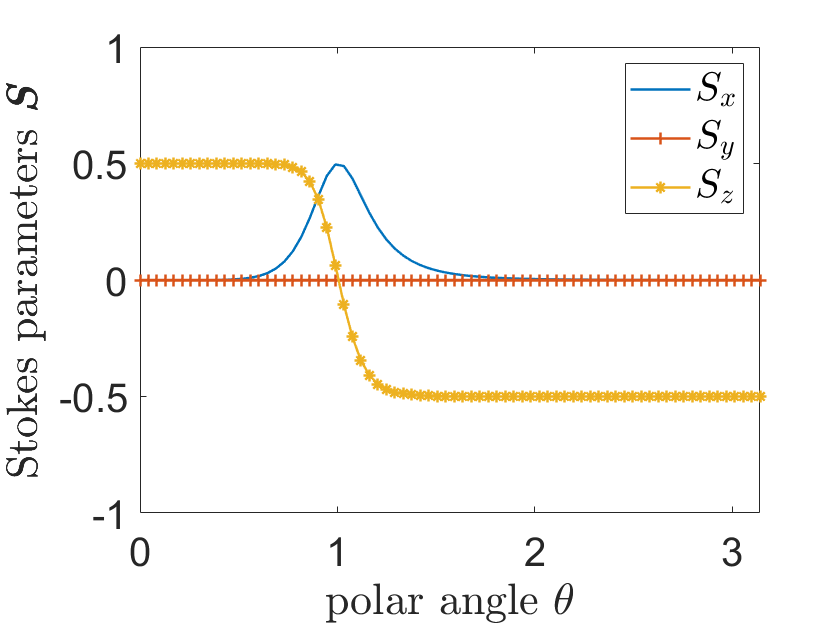}
        \label{figure53_DC_Sxyz_p=0.375}
    \end{minipage}
    }\subfigure[$p_2=0.75$]{
    \begin{minipage}[t]{0.33\linewidth}
        \centering
        \includegraphics[width=1.65in]{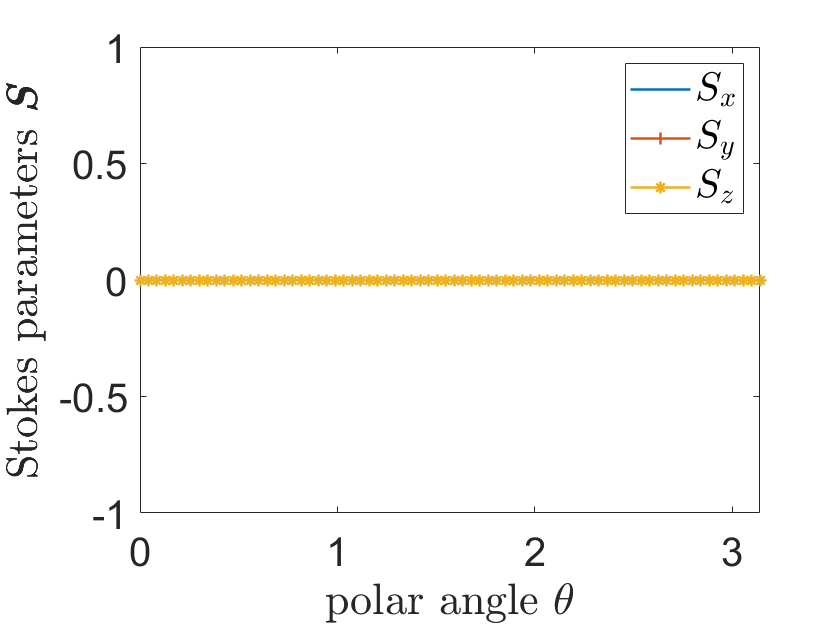}
        \label{figure53_DC_Sxyz_p=0.75}
    \end{minipage}
    }
    \caption{The Stokes parameters $(S_x,S_y,S_z)(\mathbf{r})$ vary with different decoherence strength ($p_2=0, p_2=0.375, p_2=0.75$) of depolarizing channel. We choose a slice of $S_y=0$ as an example.}
    \label{fig:Bloch S vector DC}
\end{figure*}

\begin{figure*}[ht!]
    \centering
    \subfigure[$\eta=1$]{
    \begin{minipage}[t]{0.33\linewidth}
        \centering
        \includegraphics[width=1.65in]{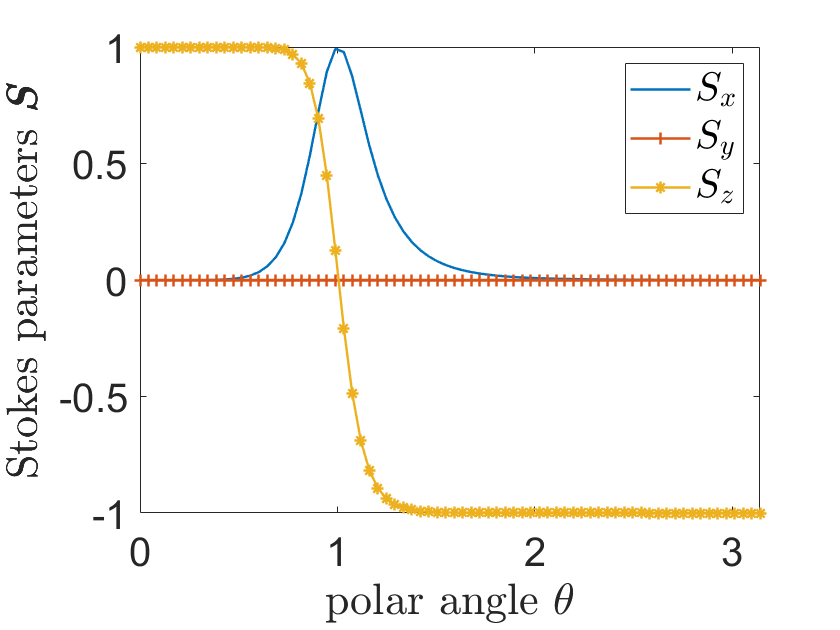}
        \label{figure51_ADC_Sxyz_eta=1}
    \end{minipage}
    }\subfigure[$\eta=0.75$]{
    \begin{minipage}[t]{0.33\linewidth}
        \centering
        \includegraphics[width=1.65in]{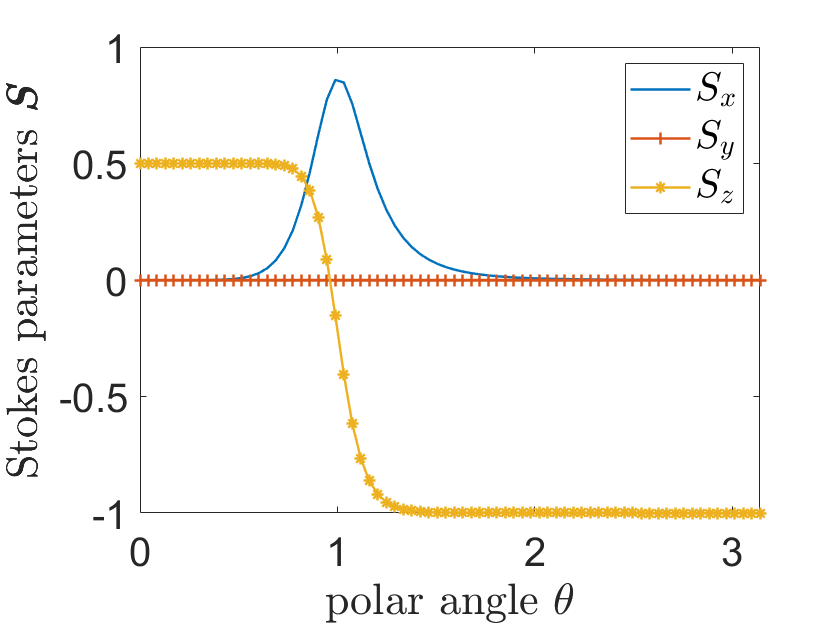}
        \label{figure51_ADC_Sxyz_eta=0.75}
    \end{minipage}
    }\subfigure[$\eta=0.5$]{
    \begin{minipage}[t]{0.33\linewidth}
        \centering
        \includegraphics[width=1.65in]{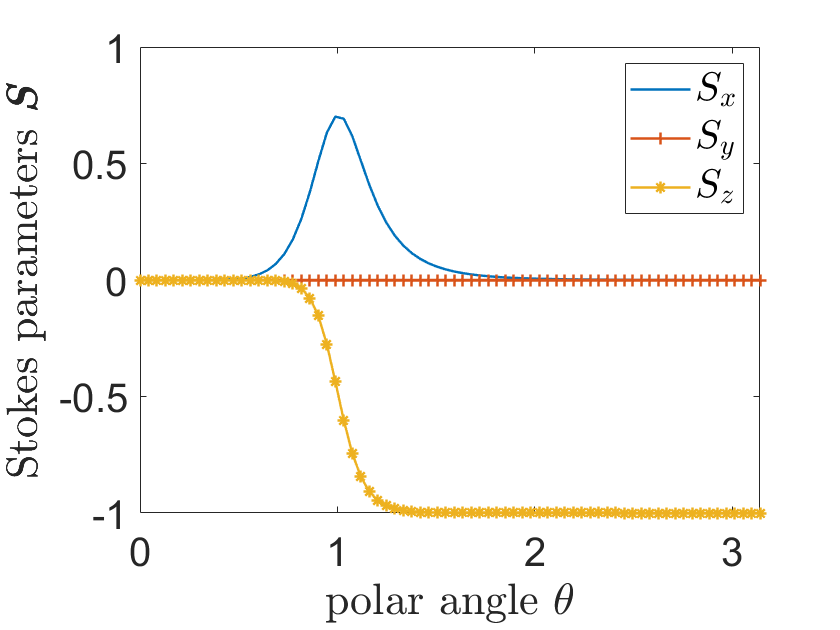}
        \label{figure51_ADC_Sxyz_eta=0.5}
    \end{minipage}
    }
    \caption{The Stokes parameters $(S_x,S_y,S_z)(\mathbf{r})$ vary with different decoherence strength ($\eta=1, \eta=0.75, \eta=0.5$) of amplitude damping channel. We choose a slice of $S_y=0$ as an example.}
    \label{fig:Bloch S vector ADC}
\end{figure*}

An ordinary arbitrary two-level state is usually denoted by a density matrix $\rho$ which can be expanded using the identity $\mathbb{I}$ and three Pauli matrices. For a 2D density matrix, it can be expressed by $\rho = \frac{1}{2}(\mathbb{I}+\mathbf{n}\cdot\vec{\sigma})$, where the vector $\mathbf{n} = (n_x,n_y,n_z) = |\mathbf{n}|(\sin\theta \cos \phi,\sin\theta\sin \phi,\cos\theta)$, ($\theta \in [0,\pi], \phi \in [0,2\pi]$), is called the Bloch vector pointing in the spin direction and $\vec{\sigma} = (\sigma_x, \sigma_y, \sigma_z)$. Actually, the 3D Stokes parameters $\mathbf{S}(\mathbf{r})$ equivalently represent the spin vector direction in the Bloch sphere. We numerically demonstrate the variations of the Stokes parameters with different decoherence strength. Taking a slice of azimuthal angle $\phi=0$ (i.e. $S_y = 0$) as an example, we observe the curves in Fig. \ref{fig:Bloch S vector PDC}, \ref{fig:Bloch S vector DC}, \ref{fig:Bloch S vector ADC}. In Fig. \ref{fig:Bloch S vector PDC}, the $z$ component of the Stokes parameters, $S_z$, remains invariant yet the $x$ component of the Stokes parameters $S_x$ shrinks and eventually $S_x=0$, resulting in complete decoherence. In Fig. \ref{fig:Bloch S vector DC}, as the decoherence strength increases, both $S_x$ and $S_z$ decrease until they reach zero. In Fig. \ref{fig:Bloch S vector ADC}, $S_x$ gradually decreases with the increase of decoherence strength of amplitude damping channel, while $S_z$ becomes unable to reach the upper hemisphere when the damping factor $\eta\leq \frac{1}{2}$.

\section*{\expandafter{\romannumeral6}. Entanglement sudden death (ESD)}
The interaction between two entangled qubits and the surrounding noisy environment may provoke the dissipation or/and dephasing effects and even the loss of quantum correlation. Quantum correlation can be quantified by the concurrence $C$ which varies between $C=0$ of classical correlation and $C=1$ of maximal entanglement. The expression of concurrence is $C(\rho) = \max\{0, \sqrt{\lambda_1}-\sqrt{\lambda_2}-\sqrt{\lambda_3}-\sqrt{\lambda_4}\}$ and $\lambda_i (i=1,2,3,4)$ are the positive eigenvalues in decreasing order of the operator $R = \rho(\sigma_y\otimes\sigma_y)\rho^\ast(\sigma_y\otimes\sigma_y)$ \cite{almeida2007environment}. We generally use the term `decoherence' to describe the decay and loss of quantum correlation or entanglement and several studies indicate that local decoherence alone can induce ESD \cite{yashodamma2014effectiveness,almeida2007environment,bavontaweepanya2018effect}. ESD is a phenomenon that entanglement of a system vanishes in a finite time and it is different from the asymptotic decay in a infinite time. For three decoherence channels, there are distinct features. \par
In phase damping channel, the concurrence conforms to $C(\rho_{out}) = 2|a||b|(1-p_1)$. When $p_1=1$, the density matrix corresponds to the maximally mixed state and entanglement completely vanishes.\par
In depolarizing channel, the final density matrix is in Eq. (\ref{rhoo_dc}) and the upper bound of $p_2$ is $\frac{3}{4}$. The depolarizing channel with $p_2=3/4$ can map any input to the maximally mixed state. The concurrence is $C(\rho_{out}) = \max\{2|a||b|(1-2p_2),0\}$. When $p_2 = 1/2$, the concurrence is zero, which reveals the occurrence of ESD. \par
In amplitude damping channel, from Eq. (\ref{rhoo_adc}), the qubit (or two-level system) always stays at the ground state if $\eta=0$. The concurrence in amplitude damping channel is $C(\rho_{out}) = 2|a||b|\sqrt{\eta}$, which indicates there is no ESD as entanglement does not completely vanish until $\eta=0$.\par
Obviously, depolarizing noise can cause finite-time entanglement decay yet phasing noise and amplitude noise have no ESD.

\section*{\expandafter{\romannumeral7}. The skyrmion numbers in loss}
When we consider the loss of two polarization components of the optical skyrmion field, there are two types of possible cases, i.e., the equal loss parameters and unequal loss parameters. The unnormalized state is expressed as
\begin{equation}
    \ket{\Psi} = aT_a\ket{l_1}\ket{0}+bT_b\ket{l_2}\ket{1},
\end{equation}
where $T_a\in [0,1]$ and $T_b\in [0,1]$ are the loss parameters of two components, respectively. $T_a=1$ and $T_b=1$ indicate no loss and $T_a=0$ (or $T_b=0$) means the skyrmion number of the field is zero. Similarly, we can calculate the corresponding skyrmion numbers by Eq. (\ref{skyrme number}). \par
If $T_a=T_b\neq 0$, the normalized state is 
\begin{equation}
    \ket{\Psi}_{\text{Loss}}^{\text{equal}} = c_1 aT_a\ket{l_1}\ket{0}+c_1 bT_b\ket{l_2}\ket{1}.
    \label{state loss equal}
\end{equation}
The output density matrix is 
\begin{align}
    \rho_{\text{Loss}}^{\text{equal}} &= |c_1|^2 [|a|^2T_a^2\ket{l_1}\bra{l_1}\otimes\ket{0}\bra{0}+ab^\ast T_a T_b\ket{l_1}\bra{l_2}\otimes\ket{0}\bra{1} \notag \\ 
    &+ a^\ast b T_a T_b\ket{l_2}\bra{l_1}\otimes\ket{1}\bra{0} + |b|^2T_b^2\ket{l_2}\bra{l_2}\otimes\ket{1}\bra{1}] \notag \\
    &= |c_1|^2T_a^2 [|a|^2\ket{l_1}\bra{l_1}\otimes\ket{0}\bra{0}+ab^\ast \ket{l_1}\bra{l_2}\otimes\ket{0}\bra{1} \notag \\
    &+ a^\ast b \ket{l_2}\bra{l_1}\otimes\ket{1}\bra{0} + |b|^2 \ket{l_2}\bra{l_2}\otimes\ket{1}\bra{1}],
    \label{rho loss equal}
\end{align}
where $c_1 = \frac{1}{T_a}$ is the normalized coefficient. Comparing Eq. (\ref{rho loss equal}) with Eq. (\ref{intial rho}), they differ only by a factor $|c_1|^2T_a^2=1$ that has no impact on computing the locally normalized Stokes parameters in Eq. (\ref{initial Sx}-\ref{initial Sz}). Thus, the skyrmion number remains constant when the two polarization components have equal loss parameters unless $T_a=T_b=0$.\par
If $T_a\neq T_b$ and $T_a\neq 0, T_b\neq 0$, the normalized state is 
\begin{equation}
    \ket{\Psi}_{\text{Loss}}^{\text{unequal}} = c_2 aT_a\ket{l_1}\ket{0}+c_2 bT_b\ket{l_2}\ket{1}.
    \label{state loss unequal}
\end{equation}
the output density matrix is 
\begin{align}
    \rho_{\text{Loss}}^{\text{unequal}} &= |c_2|^2 [|a|^2T_a^2\ket{l_1}\bra{l_1}\otimes\ket{0}\bra{0}+ab^\ast T_a T_b\ket{l_1}\bra{l_2}\otimes\ket{0}\bra{1} \notag \\ 
    &+ a^\ast b T_a T_b\ket{l_2}\bra{l_1}\otimes\ket{1}\bra{0} + |b|^2T_b^2\ket{l_2}\bra{l_2}\otimes\ket{1}\bra{1}]\notag \\
    &= \sum _{pqst=1}^2 \mu''''_{pqst}\ket{l_p}\ket{l_q}\otimes\ket{e_s}\bra{e_t},
    \label{rho loss unequal}
\end{align}
where $c_2 = \sqrt{\frac{1}{T_a^2|a|^2+T_b^2|b|^2}}$ is the normalized coefficient. 
We can rewrite the normalized state in Eq. (\ref{state loss unequal}) using new coefficients $a'$ and $b'$, i.e., 
\begin{equation}
    \ket{\Psi}_{\text{Loss}}^{\text{equal}} = a'\ket{l_1}\ket{0}+b'\ket{l_2}\ket{1},
    \label{state loss unequal new}
\end{equation}
where $a'=\frac{T_a a}{\sqrt{T_a^2|a|^2+T_b^2|b|^2}}$ , $b'=\frac{T_b b}{\sqrt{T_a^2 |a|^2+T_b^2 |b|^2}}$ and $|a'|^2+|b'|^2=1$. The form of Eq. (\ref{state loss unequal new}) is equivalent to that of the state $\ket{\Psi} = a\ket{l_1}\ket{0}+b\ket{l_2}\ket{1}$.
$\ket{e_1}=\ket{0}$, $\ket{e_2}=\ket{1}$, $\mu''''_{1111} = |c_2|^2T_a^2|a|^2=|a'|^2$, $\mu''''_{1212}=|c_2|^2T_aT_bab^\ast = a' b'^\ast$, $\mu''''_{2121}=|c_2|^2T_aT_ba^\ast b = a'^\ast b'$, $\mu''''_{2222}=|c_2|^2T_b^2|b|^2=|b'|^2$, and other coefficients are zero.
 Thus the skyrmion number in this case exhibits the stability against the loss parameters according to the descriptions in Sec.~\uppercase\expandafter{\romannumeral4} (the skyrmion number remains invariant unless the entanglement vanishes ($a'=0$ or $b'=0$), as described in Ref. \cite{ornelas2024non}). Here, we restrict our discussion to the influence of intensity loss in the two orthogonal components on the skyrmion number. However, it is worth emphasizing that even when the skyrmion field, already subjected to such intensity loss, further propagates through the aforementioned three decoherence channels, the skyrmion number still retains certain topological resilience.

\section*{\expandafter{\romannumeral8}. Generation of inhomogeneous yet continuous decoherence channels}
\begin{figure*}[ht!]
    \centering
    \subfigure[$\epsilon=4$]{
    \begin{minipage}[t]{0.33\linewidth}
        \centering
        \includegraphics[width=1.75in]{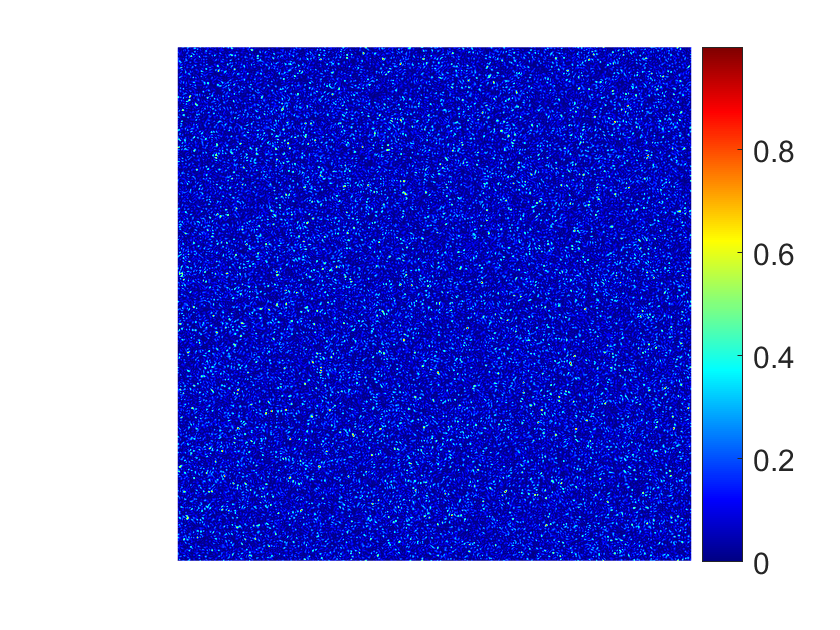}
        \label{PDC_4}
    \end{minipage}
    }\subfigure[$\epsilon=32$]{
    \begin{minipage}[t]{0.33\linewidth}
        \centering
        \includegraphics[width=1.75in]{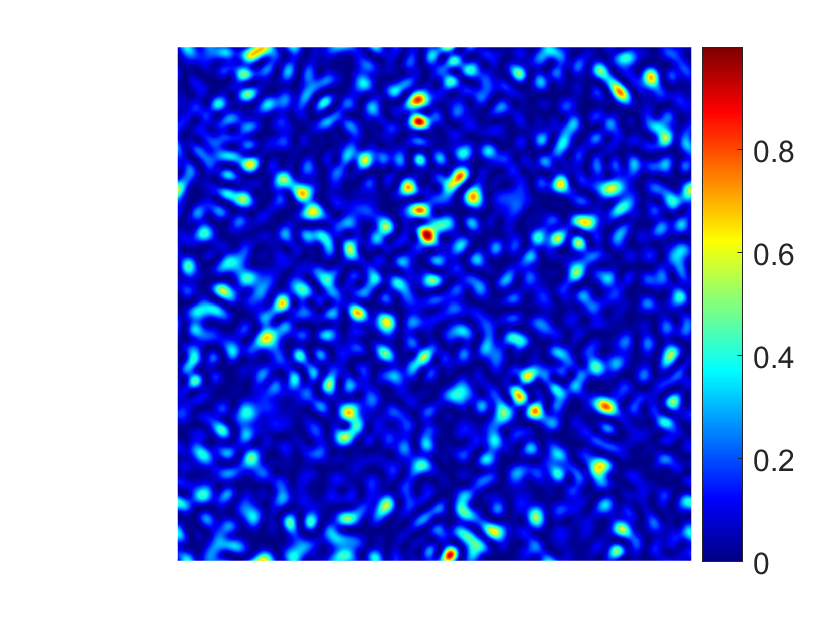}
        \label{PDC_32}
    \end{minipage}
    }\subfigure[$\epsilon=256$]{
    \begin{minipage}[t]{0.33\linewidth}
        \centering
        \includegraphics[width=1.75in]{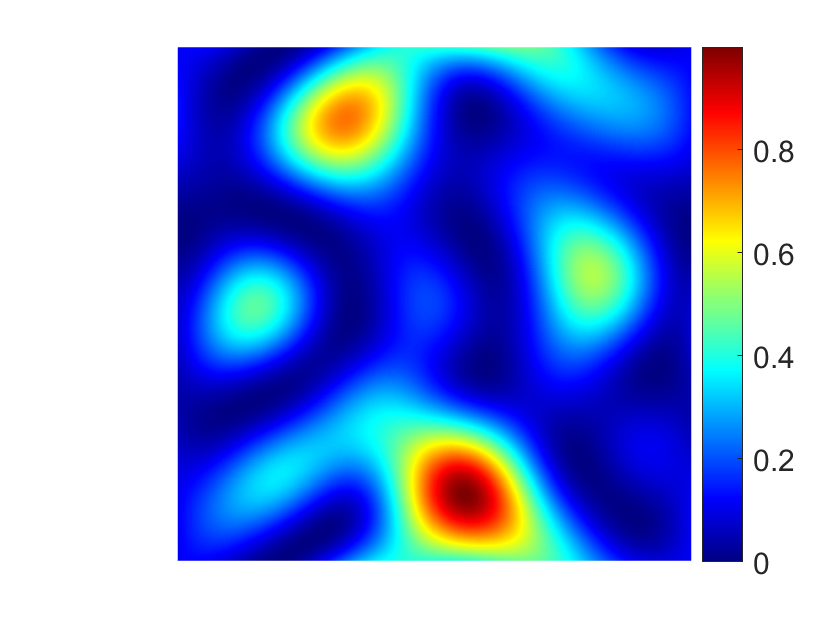}
        \label{PDC_256}
    \end{minipage}
    }
    \caption{Phase damping channel. The size of grid is $1024\times1024$, and we create three spatial distributions of the damping factor $p_1$ with different correlation lengths ($\epsilon=4, 32, 256$) as examples. $p_1(\mathbf{r})\in [0,1).$}
    \label{fig:inhomogeneous PDC}
\end{figure*}

\begin{figure*}[ht!]
    \centering
    \subfigure[$\epsilon=4$]{
    \begin{minipage}[t]{0.33\linewidth}
        \centering
        \includegraphics[width=1.75in]{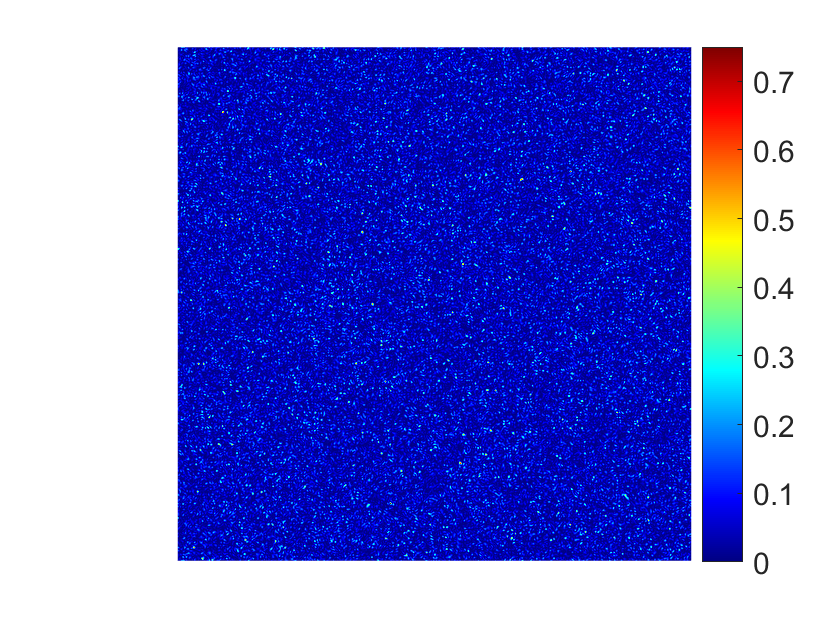}
        \label{DC_4}
    \end{minipage}
    }\subfigure[$\epsilon=32$]{
    \begin{minipage}[t]{0.33\linewidth}
        \centering
        \includegraphics[width=1.75in]{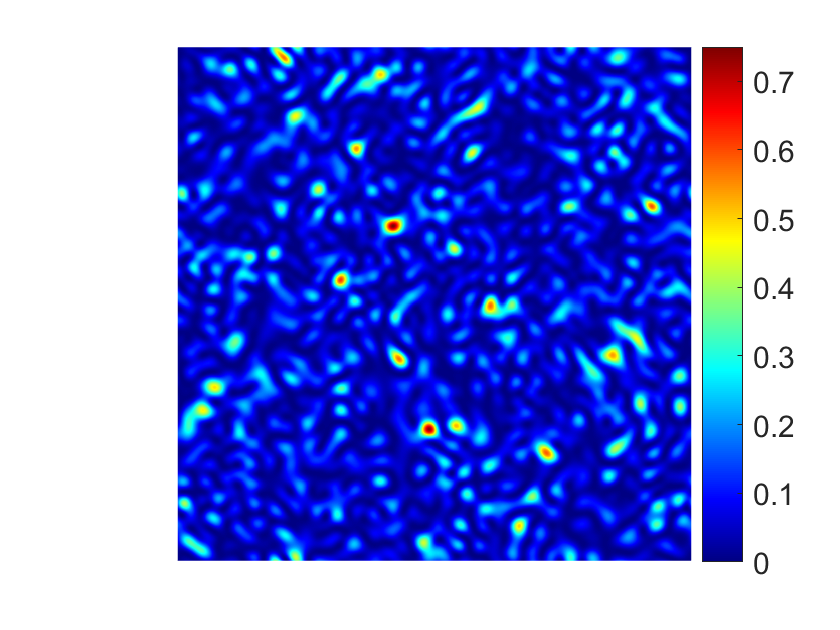}
        \label{DC_32}
    \end{minipage}
    }\subfigure[$\epsilon=256$]{
    \begin{minipage}[t]{0.33\linewidth}
        \centering
        \includegraphics[width=1.75in]{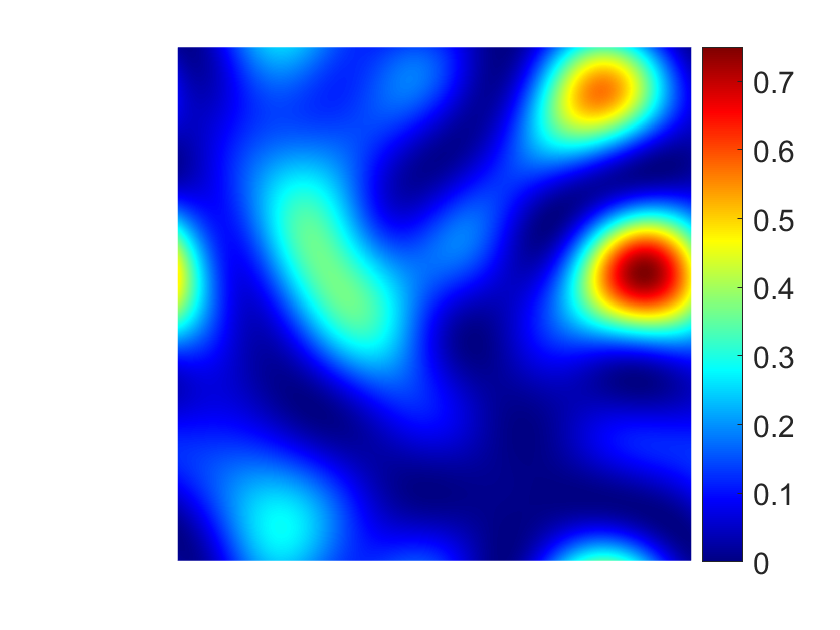}
        \label{DC_256}
    \end{minipage}
    }
    \caption{Depolarizing channel. The size of grid is $1024\times1024$, and we create three spatial distributions of the damping factor $p_2$ with different correlation lengths ($\epsilon=4, 32, 256$) as examples. $p_2(\mathbf{r})\in [0,\frac{3}{4}).$}
    \label{fig:inhomogeneous DC}
\end{figure*}

\begin{figure*}[ht!]
    \centering
    \subfigure[$\epsilon=4$]{
    \begin{minipage}[t]{0.33\linewidth}
        \centering
        \includegraphics[width=1.75in]{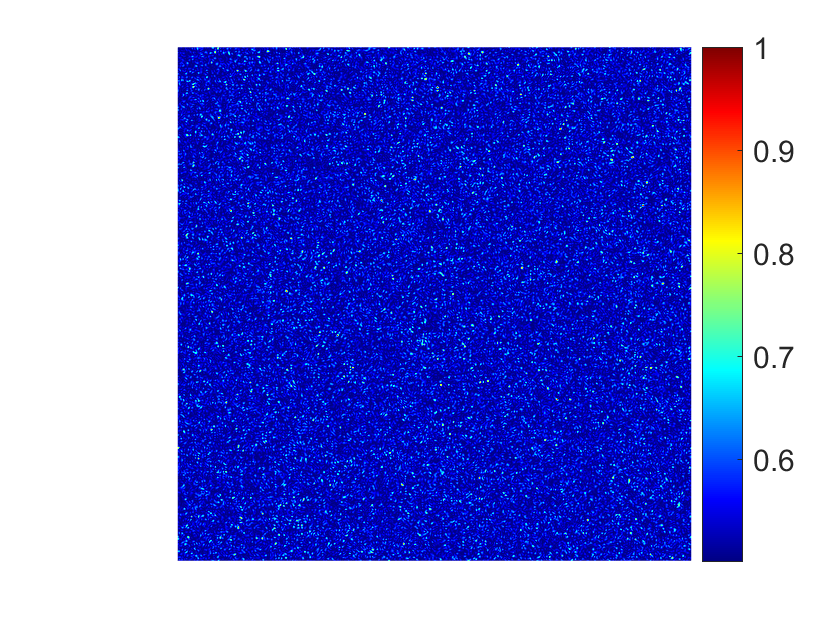}
        \label{ADC_4}
    \end{minipage}
    }\subfigure[$\epsilon=32$]{
    \begin{minipage}[t]{0.33\linewidth}
        \centering
        \includegraphics[width=1.75in]{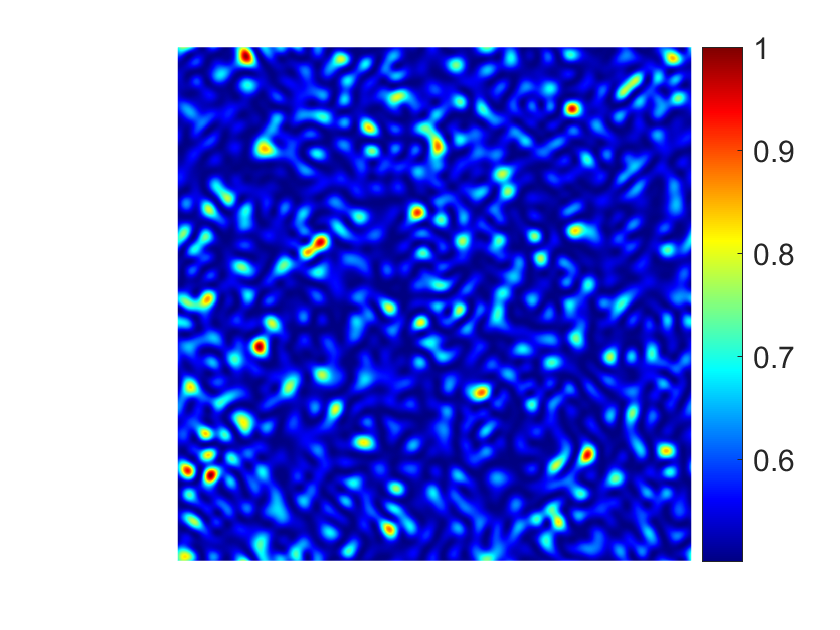}
        \label{ADC_32}
    \end{minipage}
    }\subfigure[$\epsilon=256$]{
    \begin{minipage}[t]{0.33\linewidth}
        \centering
        \includegraphics[width=1.75in]{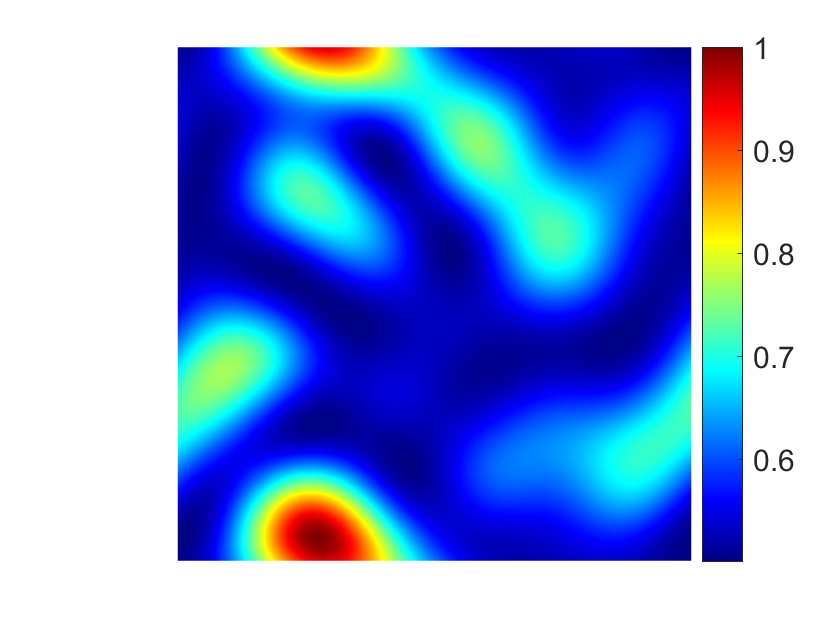}
        \label{ADC_256}
    \end{minipage}
    }
    \caption{Amplitude damping channel. The size of grid is $1024\times1024$, and we create three spatial distributions of the damping factor $\eta$ with different correlation lengths ($\epsilon=4, 32, 256$) as examples. $\eta(\mathbf{r})\in (\frac{1}{2},1].$}
    \label{fig:inhomogeneous ADC}
\end{figure*}

To generate a decoherence channel with the method of a spatially distributed yet continuous damping factor, we control the correlation length $\epsilon$ of distribution and use Fourier transformation \cite{goodman2007speckle}. Considering a screen size of $N\times N$ (with $N=1024$), we first create a Gaussian complex random matrix of size $M\times M$ (with $M=N/\epsilon$), filling the elements of $(1:M,1:M)$. Then, we apply Fourier transformation to the entire matrix and finally obtain the modulus square of the matrix as the desired result. The correlation function of non-uniformly distributed decoherence noise obtained by this method is $\sinc$ function. For example, in Fig. \ref{fig:inhomogeneous PDC}, \ref{fig:inhomogeneous DC}, \ref{fig:inhomogeneous ADC}, we demonstrate several inhomogeneous distributions of the damping factors in three local decoherence scenarios with different correlation lengths. In this simulation, the wavelength is $780$ nm (it should be noted that the selection of the wavelength does not affect this model and simulation), the input Gaussian beam's radius is $1.5$ mm, the spatial grid size is $1024\times1024$ (the corresponding screen size is $12.8~\mathrm{mm}\times12.8~\mathrm{mm}$), the initial coefficients are $a=b=\frac{1}{\sqrt{2}}$ and the two input OAM modes are $l_1=8$ and $l_1=0$ with the same radial index $p=0$. Here, we choose the horizontal polarization and vertical polarization as the two mutually orthogonal polarizations.

\section*{\expandafter{\romannumeral9}. Comparison between bit flip channel and phase damping channel}
It is well known that the noise of the phase damping channel is from the $\sigma_z$ operator and the noise of the bit flip channel is from the $\sigma_x$ operator. When the input state is expressed in the diagonal basis, i.e, 
\begin{equation}
    \ket{\Psi(\mathbf{r})}=a\ket{l_1}\ket{0}+b\ket{l_2}\ket{1}=\left (\frac{a}{\sqrt{2}}\ket{l_1}+\frac{b}{\sqrt{2}}\ket{l_2}\right)\ket{+}+\left(\frac{a}{\sqrt{2}}\ket{l_1}-\frac{b}{\sqrt{2}}\ket{l_2}\right)\ket{-},
    \label{diagonal state}
\end{equation}
where $\ket{\pm}=\frac{1}{\sqrt{2}}(\ket{0}\pm \ket{1})$, we have two formulas $\sigma_x\ket{+} = \sigma_x(\ket{0}+\ket{1})/\sqrt{2}=(\ket{0}+\ket{1})/\sqrt{2}=\ket{+}$ and $\sigma_x\ket{-}=\sigma_x(\ket{0}-\ket{1})/\sqrt{2}=(\ket{1}-\ket{0})/\sqrt{2}=-\ket{-}$ which is equivalent to the action of the $\sigma_z$ operator. Thus, the bit flip channel's effect is equivalent to that of a phase damping channel, which acts on the state in the diagonal basis. To explain this more clearly, we calculate the output density matrices and Stokes parameters in two cases.\par
First of all, we consider a phase damping channel and input a state under diagonal basis in Eq.~(\ref{diagonal state}). Here, our new basis is $\{\ket{\Tilde{0}}=\ket{+},\ket{\Tilde{1}}=\ket{-}\}$ and Eq.~(\ref{diagonal state}) is transformed into 
\begin{equation}
    \ket{\Psi(\mathbf{r})}=\left(\frac{a}{\sqrt{2}}\ket{l_1}+\frac{b}{\sqrt{2}}\ket{l_2}\right)\ket{\Tilde{0}}+\left(\frac{a}{\sqrt{2}}\ket{l_1}-\frac{b}{\sqrt{2}}\ket{l_2}\right)\ket{\Tilde{1}}.
    \label{diagonal state new}
\end{equation}
According to Eq.~(\ref{pdc_rho}), the corresponding output density matrix is 
\begin{align}
    \Tilde{\rho}_{\text{PDC}} = \sum _{pqst=1}^2 \Tilde{\mu}''''_{pqst} \ket{l_p}\bra{l_q}\otimes \ket{\Tilde{e_s}}\bra{\Tilde{e_t}},
\end{align}
where $\ket{\Tilde{e_1}}=\ket{\Tilde{0}}$, $\ket{\Tilde{e_2}}=\ket{\Tilde{1}}$, $\Tilde{\mu}''''_{11\Tilde{1}\Tilde{1}} = |a|^2/2$, $\Tilde{\mu}''''_{12\Tilde{1}\Tilde{1}}=ab^\ast/2$, $\Tilde{\mu}''''_{21\Tilde{1}\Tilde{1}}=a^\ast b/2$, $\Tilde{\mu}''''_{22\Tilde{1}\Tilde{1}}=|b|^2/2$, $\Tilde{\mu}''''_{11\Tilde{2}\Tilde{2}} = |a|^2/2$, $\Tilde{\mu}''''_{12\Tilde{2}\Tilde{2}}=-ab^\ast/2$, $\Tilde{\mu}''''_{21\Tilde{2}\Tilde{2}}=-a^\ast b/2$, $\Tilde{\mu}''''_{22\Tilde{2}\Tilde{2}}=|b|^2/2$, $\Tilde{\mu}''''_{11\Tilde{1}\Tilde{2}} = (1-p_1')|a|^2/2$, $\Tilde{\mu}''''_{12\Tilde{1}\Tilde{2}}=-(1-p_1')ab^\ast/2$, $\Tilde{\mu}''''_{21\Tilde{1}\Tilde{2}}=(1-p_1')a^\ast b/2$, $\Tilde{\mu}''''_{22\Tilde{1}\Tilde{2}}=-(1-p_1')|b|^2/2$, $\Tilde{\mu}''''_{11\Tilde{2}\Tilde{1}} = (1-p_1')|a|^2/2$, $\Tilde{\mu}''''_{12\Tilde{2}\Tilde{1}}=(1-p_1')ab^\ast/2$, $\Tilde{\mu}''''_{21\Tilde{2}\Tilde{1}}=-(1-p_1')a^\ast b/2$, $\Tilde{\mu}''''_{22\Tilde{2}\Tilde{1}}=-(1-p_1')|b|^2/2$. The parameter $p_1'\in [0,1]$ is the damping factor and $p_1'=0$ indicates no damping. To calculate the corresponding Stokes parameters, we need to convert the diagonal basis back into the initial basis ($\{\ket{0}, \ket{1}\}$) and the output density matrix in the initial basis is 
\begin{align}
    \rho'_{\text{PDC}} = \sum _{pqst=1}^2 \mu''''_{pqst} \ket{l_p}\bra{l_q}\otimes \ket{e_s}\bra{e_t},\label{pdc rho initial basis}
\end{align}
where $\ket{e_1}=\ket{0}$, $\ket{e_2}=\ket{1}$, $\mu''''_{1111} = (1-p_1'/2)|a|^2$, $\mu''''_{1212}=(1-p_1'/2)ab^\ast$, $\mu''''_{2121}=(1-p_1'/2)a^\ast b$, $\mu''''_{2222}=(1-p_1'/2)|b|^2$, $\mu''''_{1122} = (p_1'/2)|a|^2$, $\mu''''_{1221}=(p_1'/2)ab^\ast$, $\mu''''_{2112}=(p_1'/2)a^\ast b$, $\mu''''_{2211}=(p_1'/2)|b|^2$, and other coefficients are zero. The unnormalized Stokes parameters are 
\begin{align}
    S_x^{'\text{PDC}}(\mathbf{r}) &= \frac{2\Re[ab^\ast \psi_{l_1}(\mathbf{r})\psi_{l_2}^\ast(\mathbf{r})]}{|a|^2|\psi_{l_1}(\mathbf{r})|^2+|b|^2|\psi_{l_2}(\mathbf{r})|^2},\label{Sx pdc'}\\
    S_y^{'\text{PDC}}(\mathbf{r}) &= (1-p_1')\frac{-2\Im[ab^\ast \psi_{l_1}(\mathbf{r})\psi_{l_2}^\ast(\mathbf{r})]}{|a|^2|\psi_{l_1}(\mathbf{r})|^2+|b|^2|\psi_{l_2}(\mathbf{r})|^2},\label{Sy pdc'}\\
    S_z^{'\text{PDC}}(\mathbf{r}) &= (1-p_1')\frac{|a|^2|\psi_{l_1}(\mathbf{r})|^2-|b|^2|\psi_{l_2}(\mathbf{r})|^2}{|a|^2|\psi_{l_1}(\mathbf{r})|^2+|b|^2|\psi_{l_2}(\mathbf{r})|^2}.\label{Sz pdc'}
\end{align}
\par
Secondly, we consider a bit flip channel and the output density matrix has the form
\begin{equation}
    \rho_{out} = \left(1-\frac{p_3}{2}\right)\rho + \frac{p_3}{2}(\mathbb{I}\otimes \sigma_x)\rho (\mathbb{I}\otimes \sigma_x),
    \label{bit flip rho}
\end{equation}
where $p_3\in [0,1]$ is the flipping factor of the bit flip channel and $p_3=0$ indicates no flipping. Thus, we can obtain the output density matrix, i.e., 
\begin{align}
    \rho_{\text{Flip}} = \sum _{pqst=1}^2 \mu'''''_{pqst} \ket{l_p}\bra{l_q}\otimes \ket{e_s}\bra{e_t},\label{bitflip rho initial basis}
\end{align}
where $\ket{e_1}=\ket{0}$, $\ket{e_2}=\ket{1}$, $\mu'''''_{1111} = (1-p_3/2)|a|^2$, $\mu'''''_{1212}=(1-p_3/2)ab^\ast$, $\mu'''''_{2121}=(1-p_3/2)a^\ast b$, $\mu'''''_{2222}=(1-p_3/2)|b|^2$, $\mu'''''_{1122} = (p_3/2)|a|^2$, $\mu'''''_{1221}=(p_3/2)ab^\ast$, $\mu'''''_{2112}=(p_3/2)a^\ast b$, $\mu'''''_{2211}=(p_3/2)|b|^2$, and other coefficients are zero. We can see that Eq.~(\ref{bitflip rho initial basis}) is the same as Eq.~(\ref{pdc rho initial basis}). Furthermore, the corresponding Stokes parameters in the bit flip channel are the same as Eq.~(\ref{Sx pdc'})-(\ref{Sz pdc'}). Therefore, these two cases are equivalent.\par
\begin{figure}
    \centering
    \includegraphics[width=3in]{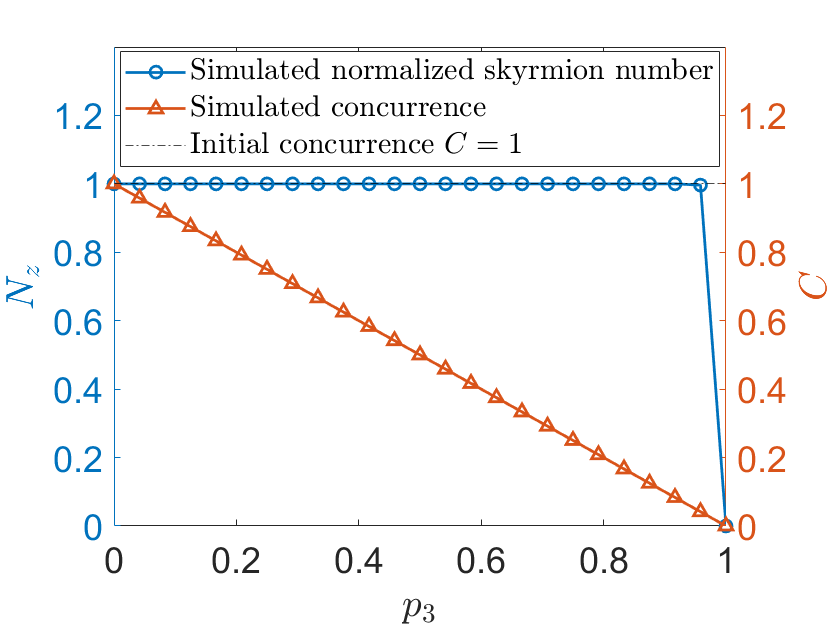}
    \caption{The simulated normalized skyrmion numbers and concurrence values in the bit flip channel. The initial spatial modes are $l_1=8$ and $l_2=0$. Coefficients $a_0=1/2$, $a=1/\sqrt{2}$ and $b=1/\sqrt{2}$ correspond to an initial concurrence of $C=1$.}
    \label{fig:BitFlip_Nz}
\end{figure}

Additionally, for the bit flip channel (or the phase damping channel in the diagonal basis), we numerically demonstrate that the skyrmion number remains stable when $0\leq p_3<1$ (or $0\leq p_1'<1$). When $p_3=1$ (or $p_1'=1$), i.e., maximal flipping noise (or completely phase damping), the skyrmion number is trivial, as shown in Fig.~\ref{fig:BitFlip_Nz}. Meanwhile, the concurrence decreases linearly with the flipping factor $p_3$.
If the flipping factor is an inhomogeneous distribution in the $x-y$ plane ($p_3(x,y)$ or $p_3(\mathbf{r})$), there is a similar result as the inhomogeneous phase damping channel. As shown in Tab.~\ref{tab:bit flip vary damping}, the skyrmion number is still stable.
\begin{table}[htbp!]
    \centering
    \setlength{\abovecaptionskip}{0.17cm}
    \setlength{\belowcaptionskip}{-0.46cm}
    \begin{tabular}{c|c}
    \hline\hline
    \diagbox[]{$\epsilon$}{$N_z^{\text{sim}}$/$N_z^{\text{the}}$}{Type} & Bit Flip Channel\\
         \hline
       1 & 0.9990 \\
       2 & 0.9999 \\
       4 & 1.0000 \\
       8 & 1.0000 \\
       16 & 1.0000 \\
       32 & 1.0000 \\
       64 & 1.0000 \\
       128 & 1.0000 \\
       256 & 0.9999 \\
       512 & 0.9978 \\
         \hline\hline
    \end{tabular}
    \caption{Normalized skyrmion numbers after averaging 50 realizations in the inhomogeneous bit flip channel with different correlation lengths $\epsilon$ in units of a single grid point. $N_z^{\text{sim}}$ is the simulated skyrmion number and $N_z^{\text{the}}$ is the theoretical skyrmion number. For the bit flip channel, the flipping factor $p_3(\mathbf{r})\in [0,1)$}
    \label{tab:bit flip vary damping}
\end{table}